%% file: disc06corr.tex
\documentclass[11pt]{article}
\usepackage{smallsec}

\usepackage{chicagor}
\usepackage{amssymb}
\usepackage{times}
\usepackage{hyperref}
\usepackage{graphicx}

\newcommand{\podc}[1]{#1}
\newcommand{\discin}[1]{\commentout{#1}}
\newcommand{\discout}[1]{#1}


\input{defn}

\newcommand{\eqref}[1]{(\ref{e:#1})}         %

\newenvironment{bogustabbing}{\begin{tabbing}\=\hspace{2em}\=\=\=\kill}%
{\end{tabbing}}

\newenvironment{program*}{%
\begin{bogustabbing}}{\end{bogustabbing}}

\renewcommand{\Box}{\mathbin{\vcenter{\hrule
    \hbox{\vrule \kern .6em
          \vbox to .6em{}\vrule}\hrule}}\hspace{.17ex}}
\newcommand{\intension}[1]{[\![ #1 ]\!]}

\newcommand{\val}{{\mathit{val}}}

\newcommand{\send}{{\mathit{send}}}

\newcommand{\msg}{{\mathit{msg}}}

\newcommand{\Pg}{\mathit{Pg}}

\newcommand{\Names}{{\bf N}}
\newcommand{\Nbr}{{\bf Nbr}}
\newcommand{\leftinput}{{\mathit{left\_input}}}

\newcommand{\done}{{\mathit{done}}}

\newcommand{\lefta}{L}
\newcommand{\righta}{R}
\newcommand{\Pgkb}{{\sf Pg}_{{\mathit kb}}}

\newcommand{\newinfo}{{\mathit{new\_info}}}
\newcommand{\inewinfo}{{\mathit{cont(new\_info)}}}

\newcommand{\somenewinfo}{{\mathit{some\_new\_info}}}
\newcommand{\maxid}{{\mathit{maxid}}}
\newcommand{\mmax}{{\mathit{max}}}

\newcommand{\id}{{\mathit{id}}}
\newcommand{\name}{{{\mathbf{n}}}}
\newenvironment{oldthm}[1]{\par\noindent{\bf Theorem #1:} \em \noindent}{\par}
\newenvironment{oldlem}[1]{\par\noindent{\bf Lemma #1:} \em \noindent}{\par}
\newenvironment{oldcor}[1]{\par\noindent{\bf Corollary #1:} \em \noindent}{\par}
\newenvironment{oldpro}[1]{\par\noindent{\bf Proposition #1:} \em \noindent}{\par}
\newcommand{\othm}[1]{\begin{oldthm}{\ref{#1}}}
\newcommand{\eothm}{\end{oldthm} \medskip}
\newcommand{\olem}[1]{\begin{oldlem}{\ref{#1}}}
\newcommand{\eolem}{\end{oldlem} \medskip}
\newcommand{\ocor}[1]{\begin{oldcor}{\ref{#1}}}
\newcommand{\eocor}{\end{oldcor} \medskip}
\newcommand{\opro}[1]{\begin{oldpro}{\ref{#1}}}
\newcommand{\eopro}{\end{oldpro} \medskip}
\newcommand{\RCond}{>}
\newcommand{\doact}{{\mathit{do}}}
\newcommand{\sfa}{{{\sf act}}}
\newcommand{\Rrep}{{\bf R}}
\newcommand{\close}{{\tt close}}
\newcommand{\cis}{\J}
\newcommand{\rkgen}{\sigma}
\newcommand{\ecis}{\tilde{\cal I}}
\newcommand{\Calls}{{\mathit{Calls}}}
\newcommand{\diam}{{\mathit{diam}}}
\newcommand{\Isys}{{\bf I}}
\newcommand{\lta}{\mbox{{$<${\hskip -6.8pt}$<$}}}
\newcommand{\closest}{{\tt closest}}
\newcommand{\ogen}{o}
\newcommand{\rgen}{\sigma}
\newcommand{\Rp}{\cR^+}
\newcommand{\rank}{\kappa}
\newcommand{\mini}{{\sf min}_i}
\newcommand{\Pgcb}{{\sf Pg}_{{\it cb}}}
\newcommand{\rgend}{\sigma^*}
\newcommand{\minj}{{\sf min}_j}
\newcommand{\Pgcbb}{\Pgcb}

\newcommand{\noop}{{\sf skip}}
\newcommand{\ainfo}{{\mathit{has\_all\_info}}}
\newcommand{\heardainfo}{{\mathit{heard\_from\_all}}}

\begin{document}
\title{A Knowledge-Based Analysis of Global Function Computation%
\thanks{Work supported in part by NSF under grants   
CTC-0208535, ITR-0325453, and IIS-0534064, by ONR under grant
N00014-02-1-0455, by the DoD Multidisciplinary University Research   
Initiative (MURI) program administered by the ONR under   
grants N00014-01-1-0795 and N00014-04-1-0725, and by AFOSR under grants
F49620-02-1-0101 and FA9550-05-1-0055.}}   
\author{Joseph Y. Halpern\\
Cornell University\\
Ithaca, NY 14853\\
halpern@cs.cornell.edu
\and
Sabina Petride\\
Cornell University\\
Ithaca, NY 14853\\
petride@cs.cornell.edu}

\date{ }

\maketitle

\begin{abstract}
Consider a distributed system $N$ in which each agent has an input 
value and each communication link has
a weight.  Given a global function, that is, a function $f$ whose value
depends on the whole network, the goal is for every agent to eventually 
compute the value $f(N)$. We
call this problem {\em global function computation}.
Various solutions for instances of this  problem, such as  
Boolean function computation, leader election, 
(minimum) spanning tree construction, and
network determination,  have been proposed, 
each under particular assumptions about 
what processors know about the system and how this knowledge can be 
acquired. 
We give a necessary and sufficient condition for the problem to be 
solvable
that generalizes a number of well-known results
\cite{attiya88,YK96a,YK99}.
We then
provide a {\em knowledge-based (kb) program\/} 
(like those of Fagin, Halpern, Moses, and Vardi \citeyear{FHMV,FHMV94}) that solves
global  
function computation whenever possible.  Finally, we improve the message
overhead inherent in our initial kb program by 
giving a {\em counterfactual belief-based program} 
\cite{HM98a} that also solves the global function computation
whenever
possible, 
but where agents send messages only when they believe it is
necessary to do so.
The latter program is shown to be implemented by a 
number of well-known  algorithms for solving leader election.
\end{abstract}

\section{Introduction}
Consider a distributed system $N$ in which each agent has an input 
value and each communication link has
a weight.  Given a global function, that is, a function $f$ whose value
depends on the whole network, the goal is for every agent to eventually 
compute the value $f(N)$. We
call this problem {\em global function computation}.
Many distributed protocols involve computing some global function of 
the network. 
This problem is typically straightforward if the network is known. For 
example, if the goal is to compute the
spanning tree of the network, one can simply apply one of the 
well-known algorithms proposed by Kruskal or Prim.
However, in a distributed setting, agents may have only local 
information, which makes the problem more difficult. 
For example, the 
algorithm proposed by Gallager, Humblet and Spira 
\citeyear{gallager83} is known 
for its complexity.%
\footnote{Gallager, Humblet, and Spira's algorithm does not actually
solve the minimum spanning tree as we have defined it, since agents do
not compute the minimum spanning tree, but only learn relevant
information about it, such as which of its edges lead in the direction
of the root.}
Moreover, the algorithm does not work for all
networks,
although
it is
guaranteed to work correctly when agents have distinct inputs and no two 
edges have identical weights.

Computing shortest paths between nodes in a network is another instance
of global function computation that has been studied extensively
\cite{ford62,bellman58}.
The well-known {\em leader election problem\/} \cite{Lyn97}
can also be viewed as an instance of 
global computation in all systems
where agents have distinct inputs: the leader is the agent with 
the largest (or smallest) input.
The difficulty in solving global function computation depends on what 
processors know. 
For example, when processors  know their identifiers (names) and all ids are 
unique, 
several solutions for the leader election problem have been proposed, 
both in the synchronous and asynchronous settings
\cite{CR79,L77,peterson82}.
On the other hand, Angluin \citeyear{angluin80}, and Johnson and 
Schneider \citeyear{johnson85} proved that 
it is impossible to deterministically elect a leader if agents may share
names.
In a similar vein,
Attiya, Snir and Warmuth \citeyear{attiya88} prove that there is 
no deterministic algorithm that computes a non-constant Boolean global 
function in a ring of unknown and arbitrarily large size if agents'
names are 
not necessarily unique. 
Attiya, Gorbach, and Moran \citeyear{AGM02} characterize what can be
computed in what they call \emph{totally anonymous shared memory
systems}, where access to shared memory is anonymous.

We aim to better understand what agents need to know to 
compute a global function.
We do this using the framework of {\em knowledge-based (kb) programs}, 
proposed by Fagin, Halpern, Moses and Vardi 
\citeyear{FHMV,FHMV94}.
Intuitively, in a  kb program,  an agent's actions may 
depend on his knowledge. 
To say that the agent with identity $i$ knows some fact $\varphi$ we 
simply write $K_i \varphi$. For example,
if agent $i$ sends a message $\msg$ to agent $j$ only if he does not 
know that $j$ already has the message,
then  the agent is following a kb program 
that can be
written as 
$${\tt \bf if}~ K_i(has_j(\msg)) ~{\tt {\bf then} ~\noop ~{\bf 
else} ~ \send(\msg)}.$$
Knowledge-based programs abstract away from particular details of
implementation and  
generalize classes of standard programs. They provide a high-level framework 
for the design and specification of distributed protocols.
They have been applied to a number of problems, 
such as {\em atomic commitment} \cite{Had}, {\em distributed
commitment} \cite{Maz},
Byzantine agreement \cite{DM,HMW}, sequence transmission 
\cite{HZ}, and  analyzing the TCP protocol \cite{SV02}.

We first characterize when global function computation is 
solvable, i.e., for which networks $N$ and global
functions $f$ agents can eventually learn $f(N)$. As we said earlier,
whether or not agents can learn $f(N)$ depends
on what they initially know about $N$. We model what agents initially 
know as a set ${\cal N}$ of networks; the 
intuition is that ${\cal N}$ is the set of all networks such that it is 
common knowledge that $N$ belongs to ${\cal N}$.
For example, if it is commonly known that the network is a ring, ${\cal 
N}$ is the set of all rings; this corresponds
to the setting considered by Attiya, Snir and Warmuth  \citeyear{attiya88}. 
If, in addition, the size $n$ of $N$ is common knowledge, then ${\cal 
N}$ is the (smaller) set of all rings of size $n$. 
Yamashita and Kameda \citeyear{YK96a} focus on three 
different types of sets ${\cal N}$: 
(1) for a given $n$, the set of all networks of size $n$, 
(2) for a fixed $d$, 
the set of all networks of diameter 
at most $d$, and (3) for a graph $G$, the set of  networks 
whose underlying graph is $G$, for all possible
labelings of nodes and edges.
In general, the more that is initially known, the smaller ${\cal N}$ is. Our 
problem can be rephrased as follows:
given $N$ and $f$, for which sets ${\cal N}$ is it possible for 
all
agents 
in $N$ to eventually learn $f(N)$?

For simplicity, we assume that the network is finite and connected,
that communication is reliable, and that no agent fails.
Consider the following simple protocol, run by each agent in the 
network: 
agents start by sending what they
initially know to all of their neighbors; agents wait until they 
receive information from all their
neighbors; and then agents transmit all they know on all outgoing 
links.
This is a {\em full-information protocol}, since agents send to their
neighbors everything they know. 
Clearly with the full-information protocol all agents will eventually 
know all available information about the network.
Intuitively, if $f(N)$ can be computed at all, %
then it can be  computed when agents run this full-information 
protocol. 
However, there are cases when this protocol fails; 
no matter how long agents run the protocol, they will never learn 
$f(N)$.  
This can happen because
\begin{enumerate}
\item 
although the agents actually have all the information they could
possibly get, and this information suffices to compute the value of $f$,
the agents do not know this;
\item although the agents have all the information they could
possibly get (and perhaps even know this), the information does not
suffice to compute the function value.
\end{enumerate}
In Section~\ref{sec:solvable}, we illustrate these situations with 
simple examples.
We show that there is a natural way of capturing what agents know in 
terms of {\em bisimilarity relations}
\cite{milner89}, and
use bisimilarity to characterize exactly when global function computation
is solvable.
\commentout{Roughly speaking, agent $i$ in network $N$ is {\em
$k$-bisimilar} to  
agent $i'$ in $N'$, written as
$(N,i)\sim_k (N',i')$, if after 
hearing all the information from agents at most $k$-distance apart, 
$i$ thinks he may be $i'$ in $N'$ and vice-versa.
We can then  characterize the exact conditions under which global 
computation is solvable:
$f(N)$ can be computed given ${\cal N}$ if there exists a value 
$k_{{\cal N},N,f}$ such that, 
for any agent $i$ in $N$,
for any networks $N'\in {\cal N}$ and agents $i'$ in $N'$,
if  $(N,i) \sim_{k_{{\cal N},N,f}} (N',i')$, then $f(N')=f(N)$.
In particular, if it is common knowledge that agents have distinct 
names, %
then we can simply take $k_{{\cal N},N,f}=|N|$; additionally, in this 
case,
we prove that when agents have all the information they also {\em know} 
that they have all the information.   
We can naturally generalize these results to the more general setting 
in which the initial information is not
necessarily common knowledge, as for example when agents may initially 
know 
different features of the network.
}
We show that this characterization provides  a 
significant
generalization of 
results of 
Attiya, Snir, and Warmuth \citeyear{attiya88}
and Yamashita and Kameda \citeyear{YK99}.
We then show that the simple program where each agent
just forwards all the new information it obtains about the network solves
the global function computation problem
whenever possible.
It is perhaps obvious that, if anything works at all, this program
works.  We show that the program terminates with each agent
knowing the global function value iff
the condition that we have identified holds.

Our program, while correct, is typically not optimal in terms of the number of messages sent.
Generally speaking, the problem is that agents may send information to
agents who already 
know it or will get it via another route. For example, consider an oriented ring. A simple
strategy of always sending information to the right is just as effective as sending information in 
both directions.
Thus, roughly speaking, we want to change the program so that an agent sends whatever information he learns
to a neighbor only if he does not know that the neighbor will eventually learn it anyway.

Since agents decide which actions to perform based on what they
know, this will be a kb
program. 
While the intuition behind this kb program is
quite straightforward, there are subtleties involved in formalizing it.
One problem is that, 
in describing kb programs, it has been
assumed that names are commonly known.  However, if the network size is
unknown, then the names of all the agents in the network cannot be
commonly known.  Things get even more complicated if we assume that
identifiers are not unique.  For example, if identifiers are not unique,
it does not make sense to write ``agent $i$ knows $\varphi$'';
$K_i \varphi$ 
is not well defined if more than one agent can have the id $i$.
We deal with these problems using techniques introduced by Grove and
Halpern \citeyear{Grove95,GroveH2}.
Observe that it makes perfect sense to talk
about each agent acting based on his own knowledge 
by saying ``if {\em I\/} know $\phi$, then \ldots''.
$I$ here represents  the name each agent uses to 
refer to himself.
This deals with self-reference; by using relative names appropriately,
we can also handle the problem of how an agent refers to other agents.

A second problem arises in expressing 
the fact that an agent should  send
information to a 
neighbor only if the neighbor will not eventually learn it anyway.
As shown by Halpern and Moses \citeyear{HM98a} 
the most obvious way of expressing it does not work;
to capture this intuition
correctly we must use {\em counterfactuals}.  These are statements of
the form $\phi \RCond  \psi$, which are read ``if $\phi$ then $\psi$'',
but the ``if ... then'' is not treated as a standard material
implication.  In particular, the formula is not necessarily true if
$\phi$ is false.  
In Section~\ref{sec:k prog}, 
we provide a kb program that uses counterfactuals which
solves the global function computation problem
whenever possible, while considerably reducing  communication overhead.

As a reality check, for the special case of leader election in networks 
with distinct ids, we show in 
Section~\ref{sec:election}  that the  
kb program is essentially
implemented by the 
protocols of Lann, Chang and Roberts \cite{L77,CR79}, and 
Peterson \citeyear{peterson82}, which all work in rings
(under slightly different assumptions),
and by the 
optimal flooding protocol \cite{Lyn97} in networks of bounded diameter.
Thus, the kb program with counterfactuals shows the
underlying commonality of all these programs
and captures the key intuition behind their design.  

\podc{
The rest of this paper is organized as follows.  In
Section~\ref{sec:solvable}, we give our characterization of when global
function computation is possible.  In Section~\ref{sec:s k progs} we
describe the kb program for global function computation,
and show how to optimize it so as to minimize messages.
In Section~\ref{sec:election}, we
show
that the program essentially implements some standard solutions to
leader election in a ring.
}
\discout{
We remark that
to define kb programs with
counterfactuals requires a lot of technical machinery, which can
sometimes obscure the essential simplicity of the ideas.  
Thus, we defer}
\discin{For space reasons, we defer}
the detailed formal definitions 
and the proofs of results to the
\discout{
appendix, giving only the essential ideas in the main part of the paper.}
\discin{full paper.}

\section{Characterizing when global function computation is solvable}\label{sec:solvable} 
We model a  network as a directed, simple (no self-loops), connected, 
finite graph, where both nodes and edges
are labeled. Each node represents an agent; its label is the agent's 
input, possibly together with the 
agent's name (identifier). Edges represent communication links; edge
labels usually denote the cost of message
transmission along links.
Communication is reliable, meaning that every message sent is 
eventually delivered and no 
messages are  duplicated or corrupted.

We  assume that initially agents know their {\em local information}, 
i.e., their own input value,
the number of outgoing links,
and the weights associated with these links.
However, agents do not necessarily know the weights on non-local edges, 
or any
 topological characteristics of the network, such as size, upper bound 
on the diameter, or the underlying graph.
Additionally, agents may not know the identity of the agents they can 
directly communicate with, or
if they share their names with other agents. In order to uniquely 
identify agents in a network $N$ of size $n$, we label
agents with ``external names'' $1$, $\dots$, $n$.  Agents do not
necessarily know these external names; we use them for our convenience
when reasoning about the system.
In particular, we assume that the global function $f$ does not depend
on these external names; $f(N)=f(N')$ for any two networks $N$ and $N'$ 
that differ only in the way that nodes are labeled.

Throughout the paper we use 
the following notation:
We write $V(N)$ for the set of agents in $N$ and $E(N)$ for the set of 
edges.
For each $i\in V(N)$, let $Out_N(i)$ be  the set of $i$'s neighbors on
outgoing links,
so that
$Out_N(i)=\lbrace j \in V(N) \: | \: (i,j)\in E(N)\rbrace$; let $In_N(i)$ 
be the set of $i$'s neighbors on incoming links, so that
$In_N(i)=\lbrace j \in V(N) \: | \: (j,i) \in 
E(N))\rbrace$; let $in_N(i)$ denote $i$'s input value.
Finally, if $e$ is an edge in
$E(N)$, let $w_N(e)$ denote $e$'s label.

We want to understand, 
for a given network $N$ and global function $f$, when it is possible 
for agents to eventually know $f(N)$. 
This depends on what agents know about $N$.
As mentioned in the introduction,
the general (and unstated) assumption in the literature is that, 
besides their local information,
whatever agents know initially about the network is {\em common 
knowledge}. 
We start our analysis by making the same assumption,
and characterize the initial common knowledge as a 
set $\N$ of networks.

In this section, we assume  
that agents are following a full-information protocol. 
We think of the protocol as proceeding in {\em rounds}:
in each round agents send to all neighbors messages 
describing all the information they have; messages are stamped
with the round number;
round $k$ for agent $i$
starts after he has received all round $k-1$ messages from his neighbors
(since message delivery is reliable, this is guaranteed to happen). 
The round-based version of the full-information protocol 
makes sense both in synchronous and  asynchronous settings, 
and for any assumptions about the order in which messages are
delivered. 

\commentout{, as during the same 
round agents may send and receive message
at different times. 
It follows from our logical division of runs into rounds that, at the 
beginning of round $k$, agents
have received all the information about agents at $k$ distance apart, 
and know all  the 
information about agents at most $k$ distance apart.
}

\begin{figure}[htb]
\centerline{\includegraphics[width=18pc,height=8pc]{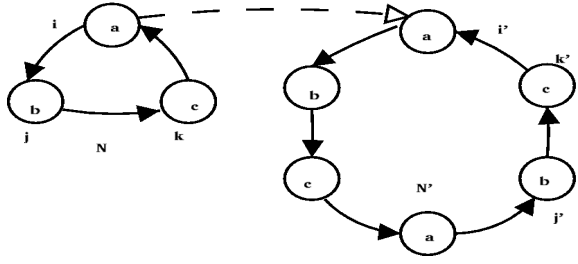}}
\caption{Two indistinguishable networks.
\label{fig:ring2}}
\end{figure}

Intuitively, the full-information protocol reduces uncertainty. 
For example, suppose that $\N$ consists of all 
unidirectional 3-node rings, and let $N$ be a three node ring 
in which agents have inputs $a$, $b$, and $c$, and all edges have the same 
weight $w$.
Let $i$ be the 
external name of the
agent with 
input $a$.
Initially, $i$ considers possible 
all 3-nodes rings in which the weight on his outgoing edge is $w$ 
and his input is $a$.
After the first round, $i$
learns from his incoming neighbor, 
who has external name $j$,
that $j$'s incoming edge also has
weight $w$, and that $j$ has input $c$.
Agent $j$ learns in the first round 
that his incoming neighbor 
has input $b$ and that his incoming edge also has
weight $w$.  
Agent $j$
communicates this information to $i$ in round 2. 
At the end of round $2$, $i$ knows everything about the network $N$, 
as do the other two  agents. Moreover, he knows exactly what the network
is. 
But this depends on the fact that $i$ knows that the ring has size $3$.
\begin{figure}[htb]
\begin{tabbing}
\quad \=\includegraphics[width=9pc,height=7pc]{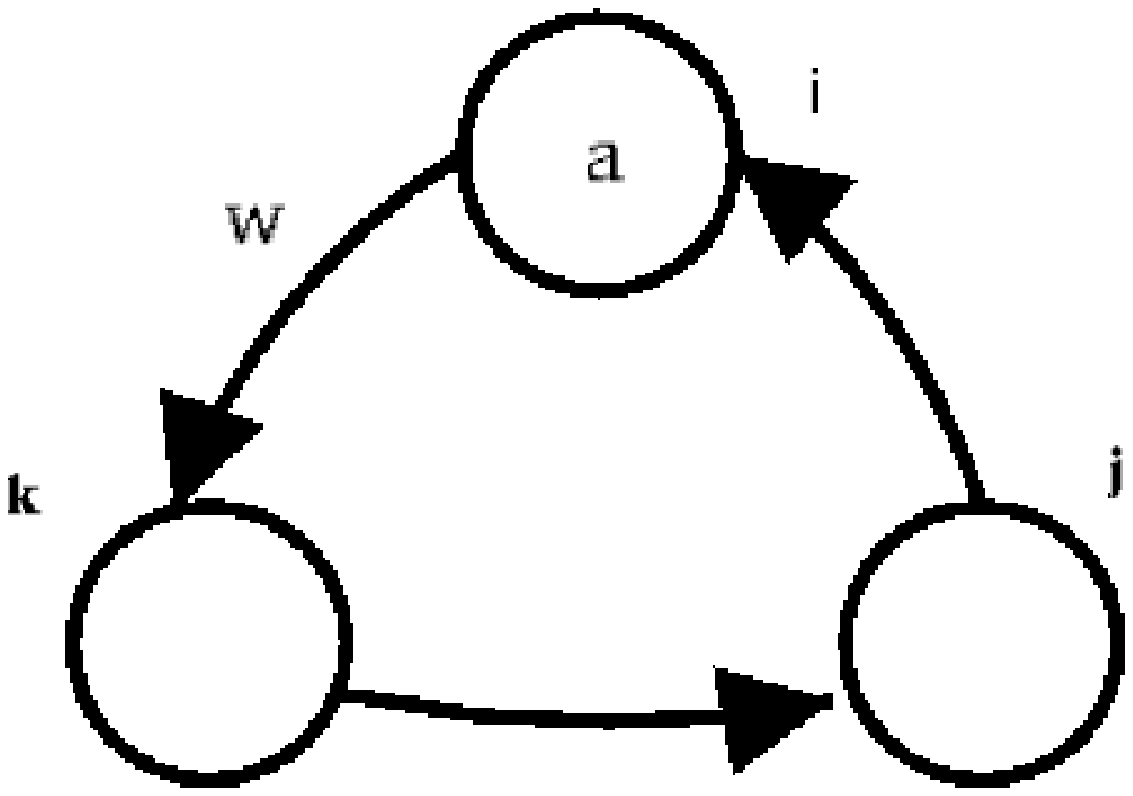} 
\quad \=\includegraphics[width=9pc,height=7pc]{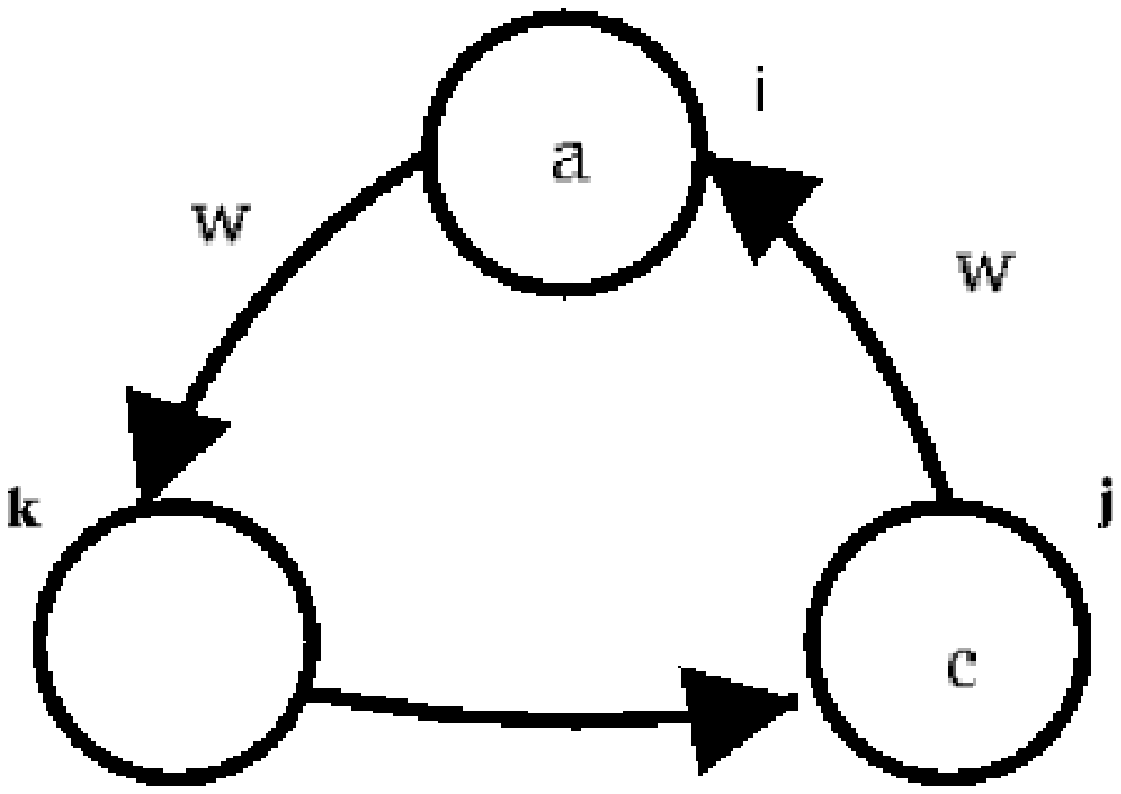} 
\quad \=\includegraphics[width=9pc,height=7pc]{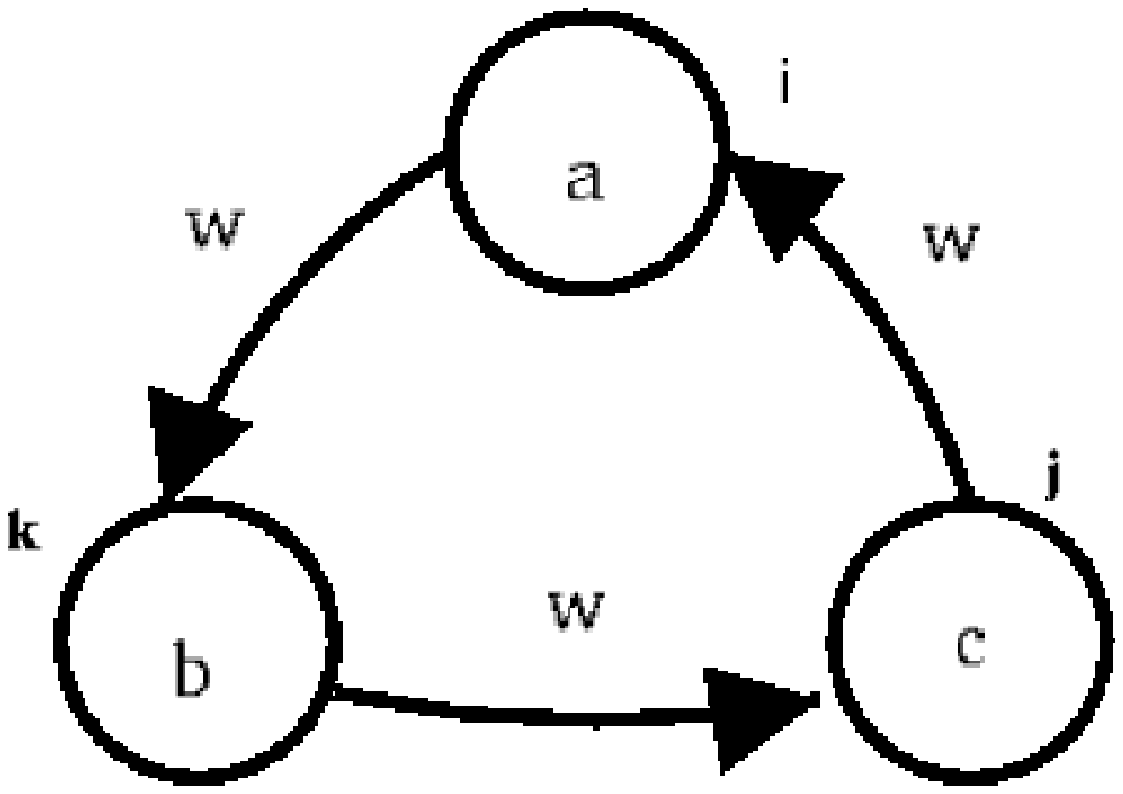}\\
\>\mbox{\ } \quad\quad\quad \mbox{Round ~0} \>\mbox{\ }\quad\quad\quad
\mbox{Round ~1}  \>\mbox{\ }\quad\quad\quad\quad \mbox{Round~2} 
\end{tabbing}
\caption{How $i$'s information changes with the full-information protocol.
\label{fig:rounds}}
\end{figure}
Now consider the same network $N$, but suppose that agents do not know the 
ring size, i.e., 
 ${\cal N}$ is the set of all unidirectional rings, of all possible 
sizes and for
all input and weight distributions. 
Again, at the end of round 2, agent $i$ has all
the information that he could possibly get, 
as do the other two agents. 
However, at no point are agents able to distinguish the network $N$  
from a 6-node
ring $N'$ in which agents look just like the agents on the 3-node ring 
(see Figure~\ref{fig:ring2}).
 Consider  the pair of agents $i$ in $N$ and $i'$ in $N'$.
It is easy to check that these agents get exactly the same messages in
every round of the full-information protocol.  Thus, they have no way of
distinguishing which is the true situation.
If the function $f$ has different values on $N$ and $N'$, then the
agents cannot compute $f(N)$.
On the other hand,
if ${\cal N}$ consists only of networks where inputs 
are distinct, then $i$ realizes 
at the end of round 2
that he must be $k$'s neighbor, and
then he knows the network configuration.

We want to characterize
when agent $i$ in network $N$ thinks he could be agent $i'$ in network $N'$.
Intuitively, 
at round $k$, $i$ thinks it possible that he could be $i'$ 
if there is a bijection 
$\mu$
that maps $i$'s incoming neighbors 
to $i'$'s incoming neighbors
such that, at the previous round $k-1$, each  incoming neighbor $j$
of $i$ thought that he could be 
$\mu(j)$.

\dfn%
\label{def:k-bisim}
Given networks $N$ and $N'$ and agents $i\in V(N)$ and $i' \in 
V(N')$, $i$ and $i'$ are 
{\em $0$-bisimilar}, written  $(N,i)\sim_0 (N',i')$, iff 
\begin{itemize}
\item $in_N(i)=in_{N'}(i')$;
\item there is a bijection $f^{out}:Out_N(i)\longrightarrow 
Out_{N'}(i')$ that preserves edge-labels; that is,
for all $j\in Out_N(i)$, we have $w_N(i,j)=w_{N'}(i',f^{out}(j))$.
\end{itemize}

For $k >0$, $i$ and $i'$ are {\em $k$-bisimilar}, written 
$(N,i)\sim_k (N',i')$, iff
\begin{itemize}
\item $(N,i)\sim_0 (N',i')$, and 
\item there is a  bijection $f^{in}:In_N(i)\longrightarrow In_{N'}(i')$ 
such that for all $j\in In_N(i)$
   \begin{itemize}
   \item $w_N(j,i)=w_{N'}(f^{in}(j),i')$, 
   \item the $(j,i)$ edge is bidirectional iff the $(f^{in}(j),i')$ edge
   is bidirectional, and  
   \item $(N,j)\sim_{k-1} (N',f^{in}(j))$.
   \end{itemize}
\end{itemize}
\end{definition}
Note that $\sim_k$ is an equivalence relation on the set
of pairs $(N,i)$ with $i\in V(N)$, and that
$\sim_{k+1}$ is a refinement of $\sim_k$.

The following lemma relates bisimilarity and the
full-information protocol:

\lem\label{lem:same info k rounds}
The following are equivalent:
\begin{itemize}
\item[(a)] $(N,i) \sim_k (N',i')$.
\item[(b)] Agents $i \in V(N)$ and $i' \in V(N')$ have the same initial
local information and 
receive the same messages
in 
each of
the first $k$ rounds of the full-information protocol.
\item[(c)] If the system is synchronous, then $i$ and $i'$ 
have the same initial local information and
receive the same messages in 
each of the first $k$ rounds of every 
deterministic
protocol.
\end{itemize}
\elem
\discout{
\prf
We first prove that (a) implies (c).
Let $P$ be an arbitrary deterministic
protocol. The proof proceeds by induction, with the base
case following from the definition of $\sim_0$.
Suppose that, if $(N,i)\sim_k (N',i')$, then
$i$ and $i'$ start with the same local information and receive same 
information in each of the first $k$ rounds of protocol $P$ 
and that $(N,i)\sim_{k+1} (N',i')$. Then $(N,i)\sim_k (N',i')$,
and there exists a bijection $f^{in}:In_N(i)\longrightarrow In_{N'}(i)$
such that $(N,j)\sim_{k} (N',f^{in}(j))$ for all $j\in In_N(i)$.
{F}rom the inductive hypothesis, it follows that $i$ and $i'$ have 
the same initial information and receive the same messages in the first $k$
rounds of $P$; 
similarly, for each $j$ incoming neighbor of $i$, $j$ and $f^{in}(j)$ have
same initial information and receive same messages in each of the first $k$
rounds of $P$. 
Hence, $j$ and $f^{in}(j)$ have the same local state at time $k$
and, since $P$ is deterministic, $j$ sends $i$ the same
messages as $f^{in}(j)$ 
sends to $i'$. Thus, $i$ and
$i'$ receive same 
messages in round $k+1$ of protocol $P$.

To prove that (c) implies (b), it suffices to notice that 
the full-information protocol is a special case of a deterministic
protocol and that,
given how we have defined rounds in an asynchronous setting, $i$
receives the same messages in round $k$ of the full-information protocol
in both the synchronous and asynchronous case.

Finally, we prove that (b) implies (a) 
by induction on $k$. 
For $k=0$, it is clear from 
Definition~\ref{def:k-bisim} that 
$(N,i)\sim_0 (N',i')$ exactly when $i$ and $i'$ have the same initial 
local information. 
For the inductive step, 
\commentout{
First suppose
that $(N,i) \sim_{k+1}(N',i')$. Then 
exists a bijection  
$f^{in}:In_N(i)\longrightarrow In_{N'}(i')$  that preserves 
edge-weights and orientation such that
for all $j\in In_N(i)$, $(N,j) \sim_k (N',f^{in}(j))$. 
By the induction hypothesis, 
$j$ and $f^{in}(j)$ have  the same initial local information and received
the same messages up to round $k$.  
Thus, 
following the full-information protocol,
in round $k+1$, $j$ sends $i$ the same messages as 
$f^{in}(j)$ sends to $i'$. 
That means that $i$ and $i'$ receive the same  messages in round $k+1$.

For the converse,
} %
suppose that $i$ and $i'$ have the same initial local
information and receive the same messages at each round $k' \le k+1$.
We can then construct a mapping, say 
$f^{in}$, from $In_N(i)$ to $In_{N'}(i')$ such that for all 
$j\in In_N(i)$, the information that $i$ 
receives from $j$ is the same
as the information that $i'$ receives from $f^{in}(j)$
in each of the first $k+1$ rounds.
Since $j$ is following a full-information protocol, it follows that $j$
must have the same initial local information as $j'$ and that $j$ and
$j'$ receive the same messages in each of the first $k$ rounds.
By the induction hypothesis, 
$(N,j) \sim_k (N',f^{in}(j))$. 
Since
part of $i$'s 
information from $j$ is also the weight
of edge $(j,i)$, 
$f^{in}$ must preserve edge-weights. 
Thus, 
$(N,i)\sim_{k+1} (N',i')$.  
\eprf
}%

Intuitively, if the function $f$ can be computed on $N$, then it can be
computed using a full-information protocol.
The value of $f$ can be computed when 
$f$ takes on the same value at all networks that the agents
consider possible. 
The round at which this happens may depend on the network $N$, 
the function $f$, and what it is initially known.
Moreover, if it does not happen, then $f$ is not computable.
Using Lemma~\ref{lem:same info k rounds}, we can characterize if and when
it
 happens.

\thm%
\label{thm:solvability}
The global function $f$ can be computed on networks in $\N$
iff, for all networks $N \in \N$, there exists a constant
$k_{{\cal N},N,f}$, such that, for all networks $N' \in \N$, all $i \in
V(N)$, and all $i' \in V(N')$, if 
$(N,i) \sim_{k_{{\cal N},N,f}} (N',i')$ then
$f(N')=f(N)$. \ethm
\prf
First suppose that the condition in the statement of the theorem 
holds.
At the beginning of each round $k$, each agent $i$ in the network
proceeds as follows.
If $i$ received the value of $f$ in the previous round, then $i$
forwards the value to all of its neighbors and terminates; otherwise,
$i$ computes $f$'s value on all the networks $N'$ such that
there exists an $i'$  such that agent $i'$ would have received the same
messages in the first $k-1$ rounds in network $N'$ as $i$ actually
received.  (By Lemma~\ref{lem:same info k rounds}, these are just the
pairs $(N',i')$ such that $(N',i') \sim_{k-1} (N,i)$.)
If all the values are equal, then $i$ sends the value to all his neighbors
and terminates; otherwise, $i$ sends whatever new information he has received
about the network to all his neighbors.
Let $k_i$ be the first round with the property
that for all $N'\in \N$ 
and $i'$ in $N'$, if $(N,i)\sim_{k_i} (N',i')$, then $f(N')=f(N)$.
(By assumption, such
a $k_i$ exists and it is 
at most $k_{{\cal N},N,f}$.) 
It is easy to see that, by round $k_i$, $i$ learns the value of $f(N)$,
since either $i$ gets the same messages that it gets in the
full-information protocol up to round $k_i$ or it gets the function value.
Thus, $i$ terminates by the end of round $k_i+1$ at the
latest, after sending the value of $f$, and  
the protocol terminates in at most $k_{{\cal N},N,f}+1$ rounds. 
Clearly all agents learn $f(N)$ according to this protocol.

Now suppose that the condition in the theorem does not hold and, 
by way of contradiction,
that the value of $f$ can be computed by some protocol $P$ on all the
networks in $\N$.  
There must exist some network $N$ for which the condition in the theorem
fails.
Consider a run where all messages are delivered synchronously.
There must be some round $k$ such that 
all agents in $N$ have computed the function value by round $k$.  Since
the condition fails, there must exist a network $N' \in \N$ and agents
$i \in V(N)$ and $i' \in V(N')$ such that $(N,i) \sim_k  
(N',i')$ and $f(N) \ne f(N')$.  
By Lemma~\ref{lem:same info k rounds}, $i$ and $i'$ have the same
initial information and receive the same messages in the first $k$
rounds of protocol $P$.
Thus, they must output the same value for the function at round $k$. But since
$f(N) \ne f(N')$, one of these answers must be wrong, contradicting our
assumption that $P$ computes the value of $f$ in all networks in $\N$.
\eprf

Intuitively, $k_{{\cal N}, N, f}$ is a round at which each agent $i$
knows that $f$ takes on the same value at all the networks $i$ considers
possible at 
that round.  Since we are implicitly assuming that agents do not forget,
the set of networks that agent $i$ considers possible never grows.
Thus, if \discout{the function} $f$ takes on the same value at all the networks
that agent $i$ considers possible at round $k$, then $f$ will take on
the same value at all networks that $i$ considers possible at round $k'
> k$,  
so every agent knows
the value of $f(N)$ in round
$k_{{\cal N}, N, f}$. 
In some cases, we can provide a useful upper bound on
$k_{{\cal N}, N, f}$.  For example,
if $\N$ consists only of
networks with distinct identifiers, or, more
generally, of networks in which no two agents are {\em locally} the same, 
i.e., $(N,i)\not \sim_0 (N,j)$ for all $i \neq j$, then
we can take $k_{{\cal N},N,f}=\diam(N)+1$, where $\diam(N)$ is the
diameter of $N$.

\thm\label{cor:non-anon}
If initially it is common knowledge that no two agents are 
locally the same, then all global functions
can be computed; indeed, we can take 
$k_{{\cal N},N,f}=\mbox{$\diam(N)+1$}$. 
\ethm

\prf
Since $f(N) = f(N')$ if $N$ and $N'$ are isomorphic, it suffices to show
that $(N,i) \sim_{\diam(N)+1}  (N',i')$ implies that $N$ and $N'$ are
isomorphic for 
all $N, N' \in \N$.  
First observe that, by an easy induction on $k$, if there is a path of
length $k \le \diam(N)$ from $i$ to $j$ in $N$, then there must exist a
node $j' \in V(N')$ such that there is a path from $i'$ to $j'$ of
length $k$ and $(N,j) \sim_{\diam(N) + 1 -k} (N',j')$.  Moreover, note
that $j'$ must be unique, since if $(N,j) \sim_{\diam(N) + 1 -k}
(N',j'')$, then $j$, $j'$, and $j''$ must be locally the same and, by
assumption, no distinct agents in $N'$ are locally the same.
Define a map $h$ from $N$ to $N'$ by taking $h(j) = j'$.  
This map is 1-1, since if $h(j_1) = h(j_2)$, then $j_1$ and $j_2$ must be
locally the same, and hence identical.  

Let $N''$ be the subgraph of
$N'$ consisting of all nodes of distance at most $\diam(N)$ from $i'$.
An identical argument shows that there is a 1-1 map $h'$ from $N''$ to $N$
such that $j'$ and $h'(j')$ are locally the same for all $j' \in
V(N'')$.  The function $h'$ is the inverse of $h$, since $h(h'(j'))$ and
$j'$ are locally the same, and hence identical, for all $j' \in V(N)$.  
Finally, we must have that $h$ is a graph isomorphism from $N$ to $N''$,
since the fact $j$ and $h(j)$ are locally the same guarantee that they
have the same labels, and if $(j_1,j_2) \in E(N)$, then $(h(j), h(j'))
\in E(N'')$ and the two edges have the same label.

It remains to show that $N' = N''$.  Suppose not.  Then there is a node
$j_1 \in V(N')$ of distance $\diam(N) + 1$ from $i'$.  Let $j_2 \in V(N)$
be such that $j_1$ is an outgoing neighbor of $j_2$ and the distance from
$i'$ to $j_2$ is $\diam(N)$.  By construction, $j_2 \in V(N'')$; by our
previous argument, there is a node $j_3 \in V(N)$ such that
$(N,j_3) \sim_1 (N',j_2)$.  Since $j_2$ and $j_3$ are locally the same,
they must have the same number of outgoing links, say $m$.  That means
that there are $m$ nodes in $N$ that have $j_3$ as an incoming neighbor,
say $i_1, \ldots, i_m$.  Thus, each of $h(i_1), \ldots, h(i_m)$, all of
which are in $N''$, must
have $j_3$ as an incoming neighbor.  But $j_3$ has only $m$ outgoing
edges, and one of them goes to $j_2$, which is not in $N''$.  This is a
contradiction.  
\eprf

Attiya, Snir, and Warmuth \citeyear{attiya88} prove an analogue of
Lemma~\ref{lem:same info k rounds} in their setting (where all networks
are rings) and use it to prove a number of impossibility results.  In
our language, these impossibility results all show that there does not
exist a $k$ such that $(N,i) \sim_k (N',i')$ implies $f(N) = f(N')$ for
the functions $f$ of interest, and thus are instances of
Theorem~\ref{thm:solvability}.%
\footnote{We remark that Attiya, Snir, and Warmuth allow their global
functions to depend on external names given to agents in the network.
This essentially amounts to assuming that the agent's names are part of
their input.} 
Yamashita and Kameda characterize when global functions can be computed
in undirected  
networks (which have no weights associated with the edges), 
assuming that an upper bound  on the size of the network is known.
They define
a notion of {\em view} and show that two agents have the same information
whenever their views are {\em similar}
in a precise technical sense; $f(N)$ is computable 
iff for all networks $N'$  such that 
agents in $N$ and $N'$ have similar views, $f(N')=f(N)$.
Their notion of similarity
is essentially our notion of bisimilarity restricted to undirected
networks with no edge labels.
Thus, their result is a special case of Theorem~\ref{thm:solvability}
for the case that ${\cal N}$ consists of undirected networks 
with no edge labels
of size at
most $n^*$ for some fixed constant $n^*$; they show that 
$k_{{\cal N},N,f}$ can be taken to be $n^*$ in that case.
Not only does our result generalize theirs, but our characterization is
arguably much cleaner.

Theorem \ref{cor:non-anon}  sheds light on why 
the well-known protocol for minimum spanning tree construction 
proposed by Gallager, 
Humblet, and Spira \citeyear{gallager83} can deal both with
systems with distinct ids (provided that there is a commonly-known
ordering on ids) and for  networks with identical ids 
but distinct edge-weights. 
These are just instances of situations where 
it is common knowledge that no two agents are locally the same.

\commentout{
But is common knowledge of the fact that no two agents are  locally the 
same a necessary condition 
for any function to be computed? The answer is no, but it is necessary 
that  this is
{\em distributed knowledge} among the agents
in the network, meaning that if agents would gather all they initially 
know about the network, they 
would know they are locally distinct. To see why this is the case, 
consider a more general setting in
which it is not required that all that agents initially know about the 
network, besides their local
neighborhood, is common knowledge. In this setting we model the initial 
knowledge of any agent $i$
in $N$ as a set of networks ${{\cal N}_i}\subseteq Net$, with $N\in 
{\cal N}_i$,
 and place no conditions on the relation between ${\cal N}_i$
and ${\cal N}_j$ for different agents $i$ and $j$. 

Similar proofs as for lemma \ref{lem:same info k rounds} and theorem 
We can prove results similar 
Similar proofs as for lemma \ref{lem:same info k rounds} and theorem 
to Theorem~\ref{thm:solvability} 
even if the set of possible networks is not common knowledge.
To present these results, for 
given network $N$ and $i,j \in V(N)$, 
we write $d_N(j,i)$ as the {\em nodal distance} in $N$ from $j$ to $i$, 
i.e., the number of edges along the shortest
{\em directed} path in $N$ from $j$ to $i$. The set of all agents in 
$N$ at most $k$-distance apart from $i$
is denoted as $V_{N,k}(i)$.
We can show that the distributed knowledge among all agents
at most $k$-distance apart from $i$ precisely characterizes $i$'s 
knowledge at round $k$ of any execution
of the full-information protocol:

\lem\label{lem:same info k round gen}
Let $N$ be an arbitrary network and $\lbrace{\cal N}_i\rbrace_{i\in 
V(N)}$ the initial knowledge of agents in $N$.
Let $r_N^{info}$ be an arbitrary run of the full-information protocol 
on $N$ wrt 
$\lbrace{\cal N}_i\rbrace_{i\in V(N)}$.
Then for any agent $i \in V(N)$,  the set of networks $i$ considers 
possible at the beginning of
round $k$ in $r_N^{info}$ is
$$\lbrace N'\in \cap_{j\in V_{N,k}(i)}{\cal N}_j\: | \: \exists i'\in 
V(N'). (N,i)\sim_k(N',i') \rbrace. $$
\elem
}
\commentout{
Similar results can be proved in a more general setting in which it is 
not required that 
the set $\N$ of possible networks be common knowledge.
Suppose that each agent $i$ initially considers the 
set ${\cal N}_i$
of networks possible, and the actually network $N$ is in $\inter_{i \in
V(N)} \N_i$ (so that all agents consider the actual network possible).

We assume that agents
running the full-information protocol 
now also communicate 
information that they have about the set of possible networks.
Thus, at the first step, agent $i$ sends $\N_i$ to his neighbors and, at
later steps, agent $i$ sends whatever information new information he has
learned about the set of possible networks. 
In this setting, we can simply replace $\N$ by $\inter_{i \in V(N)}
\N_i$ in Theorem~\ref{thm:solvability}.

\thm\label{pro:solvability gen}
The problem of computing the global function $f$ on network $N$ with
respect to the 
initial knowledge 
$\lbrace{\cal N}_i\rbrace_{i \in V(N)}$
is solvable if and only if the following condition is satisfied:
$$
\begin{array}{l}
\exists k^*.
 \: \forall i \in V(N). 
\forall N'\in \cap_{j\in V(N)}{\cal N}_j.
\forall  i' \in V(N'). 
(N',i')\sim_{k^*}  (N,i) \rimp
(f(N')=f(N)).  
\end{array}
$$
\ethm
}

\commentout{
The fact that networks are connected and finite graphs ensures that, by 
running the full-information protocol,
eventually agents have all information, in other words each agent $i$ 
eventually receives all the 
initial information of agents in the network. 
}
\commentout{To make this formal we define $diam(N)$ as the diameter of 
$N$, i.e., the maximum of the nodal distances
$d_N(j,i)$ for any pair of agents $i$ and $j$ in $N$; we also 
abbreviate
$\lbrace N'\in \cap_{j\in V_{N,k}(i_1)}{\cal N}_j\: | \: \exists i'\in 
V(N'). (N,i_1)\sim_k (N',i') \rbrace$
as ${\cal N}_{N,k}(i)$.

\lem\label{lem:all info}
Let $N$ be an arbitrary network and $\lbrace{\cal N}_i\rbrace_{i\in 
V(N)}$ the initial knowledge of agents in $N$.
Let $r_N^{info}$ be an arbitrary run of the full-information protocol 
on $N$ wrt 
$\lbrace{\cal N}_i\rbrace_{i\in V(N)}$.
Then for any round $k> diam(N)$,  for any two agents $i$ and $j$, at 
the beginning of round $k$ in run
$r_N^{info}$ the set of networks $i$
considers possible is identical to the set of networks $j$ considers 
possible, and does not change in time: 
$$ {\cal N}_{N,k}(i)={\cal N}_{N,k}(j)={\cal N}_{N,diam(N)}(i).$$
\elem
}

\commentout{
This means that, for computing some function $f$, initially  having 
distributed knowledge of some feature 
of the network is as good as ensuring that  initially everybody knows 
the feature.  
\rem\label{rem:distrib}
The problem of computing the global function $f$ on network $N$ is 
solvable wrt initial knowledge
$\lbrace{\cal N}_i\rbrace_{i \in V(N)}$ iff it is solvable wrt the 
initial knowledge  
$\lbrace \cap_{j \in V(N)} {\cal N}_j\rbrace_{i \in V(N)}$
\erem

\cor\label{cor:non-anon gen}
If initially it is distributed knowledge that no two agents are {\em 
locally} the same, 
then {\em any} function can be computed.
Otherwise, there are networks $N$ and functions $f$ that cannot be 
computed on $N$.
\ecor
}

\commentout{
Roughly speaking, the information that each agent is computing during 
the protocol
is represented as an infinite tree, called {\em view}. To distinguish 
among their neighbors, agents
arbitrarily label their links; for example, an agent with degree $n$ 
labels his links from $1$ to $n$. 
Two agents have the same information by round $k$ exactly when their 
views are {\em similar}, 
i.e., isomorphic up to a labeling. 
$f$ is proved computable if it takes same value on all networks that 
generate similar views.
We can show that the notion of similar views amounts to $k$-bisimilarity, and subsequently
restate the above condition for computing $f$ as in theorem 
\ref{thm:solvability}. 

As explained by Kranakis, Krizanc, and Berg \citeyear{boolcomp94}, 
Yamashita and Kameda's model
of views can be better understood as a continuation of the work in the 
'80s of Angluin \citeyear{angluin80}
and Fisher, Lynch, and Merritt \citeyear{easyimpos85} of characterizing 
classes of networks in which agents
behave similarly in terms of {\em graph coverings}. Here, a graph $H$ 
is a {\em covering} of graph $G$ if 
$H$ looks locally like $G$, i.e., there is a surjection $f:V(H) 
\longrightarrow V(G)$ such that for any
node $i$ in $V(H)$, $f$  is an isomorphism between $i$'s neighborhood 
in $H$ and $f(i)$'s neighborhood in $G$.
A {\em universal covering} of $G$ is an infinite (unless $G$ is a tree) 
tree covering of $G$, 
and the universal covering of any graph $G$ is unique up to an 
isomorphism 
(see Angluin's work \citeyear{angluin80}). The notion of universal 
coverings
is relevant to distributed computing because any two networks with same 
universal covering behave identically.
There is a natural extension of the notion of graph covering to that of 
{\em network covering} if we require node and edge labels to be 
preserved. 
Following Norris  \citeyear{norris95},
for any network $N$ and agent $i$ in $N$, we denote the universal cover 
of a graph $N$ 
rooted at $i$ as $U_N(i)$; 
$U_N(i)$ truncated up to depth $k$ is denoted as $U_N^k(i)$. The 
intuition is that, for
{\em undirected} networks, $i$ in $N$ and $i'$ in $N'$ have same 
information by round $k$ of a full-information
protocol exactly when their corresponding universal covers are 
isomorphic. 
We conjecture that $(N,i)\sim_k(N',i')$
wrt the set of all undirected networks iff $U_N^k(i)$ and 
$U_{N'}^k(i')$ are isomorphic.
However, this is not always the case for {\em directed} networks,  as a 
simple example based on 
\cite{norris95} proves (see Section~\ref{sec:proofs-solvable}).  

\rem\label{rem:bisim and covers}
There exist $N$ and $N'$ {\em directed} networks such that there are 
agents $i\in V(N)$ and $i'\in V(N')$ and 
values $k$ such that $(N,i)\sim_k (N',i')$, but $U_N^k(i)$ and  
$U_{N'}^k(i')$ are not isomorphic.
\erem
We see this result as sustaining the idea that, for the purpose of 
characterizing solvable cases for
global function computation, the notions of bisimilarity and graph 
coverings
complement each other; analysis based on the theory of universal 
coverings proved 
interesting connections between network topology and computable 
functions, while
bisimilarity relations allow natural and simple general 
characterization of the required
knowledge for computing global functions.

}
\section{A standard program for global function computation}\label{sec:s k progs}
\subsection{Standard programs with shared names}\label{sec:k prog}
A standard {\em program} $\Pg$ has the
form
$$
\begin{array}{l}
{\bf if }~ t_1 ~{\bf then} ~{\sfa_1}\\
{\bf if }~ t_2 ~{\bf then} ~{\sfa_2}\\
\dots,\\
\end{array}
$$
where the $t_j$s are standard tests (possibly involving temporal operators 
such as $\Diamond$), and the $\sfa_j$s are actions.
The intended interpretation is that agent $i$ runs this
program forever.  At each point in time, $i$ nondeterministically
executes one of the actions $\sfa_j$ such 
that the test 
$t_j$
is satisfied; if no such action exists,
$i$ does nothing.
We sometime use obvious abbreviations like ${\bf if} \ldots {\bf then}
\ldots {\bf else}$.
\commentout{
Even in a standard program, there are issues of naming if we work in
networks where the names of agents are not common knowledge (see
\cite{MosesRoth,GroveH2}). 
}%
Following Grove and Halpern \cite{Grove95,GroveH2} (GH from now on),
we distinguish between agents and their names.
We assume that programs mention only names, not agents (since in general
the programmer will have access only to the names, which can be viewed
as denoting roles).
We use $\Names$ to denote the set of all possible names and assume that
one of the names is $I$. 
In the semantics, we associate with each name  the agent who has that name.
\commentout{
With each \emph{run} (or execution) of the program, we
associate the set of agents that exist in that run. For simplicity, 
we assume that the set of agents is constant over the run;  that is,
we are not allowing agents to enter the system or leave the system.
However, different sets of agent may be associated with different runs.
}%
We assume that each agent has a way of naming his neighbors, and gives
each of his neighbors different names.  However, two different agents
may use the same name for different neighbors.  For example, in 
a ring, each agent may name his neighbors $\lefta$ and $\righta$;
in an arbitrary network, an agent whose outdegree is $d$ may
refer to his outgoing neighbors as $1$, $2$, ..., $d$.
We allow actions in a program to depend on names, so the
meaning of an action may depend on which agent is running it. 
For example, in our program for global function computation, 
if $i$ uses name $\name$  
to refer to his neighbor $j$, we write $i$'s action of sending message
$\msg$ to $j$ as $\send_{\name}(\msg)$. 
Similarly, 
if $A$ is a set of names, then we take $\send_A(\msg)$ to be the action
of sending $\msg$ to each of the agents in $A$ (and not sending anything
to any other agents).  Let 
$\Nbr$ denote the neighbors of an agent, so that $\send_{\Nbr}(\msg)$ is
the action of sending $\msg$ to all of an agent's neighbors.

We assume that message delivery is handled by the channel (and is
not under the control of the agents).
In the program, we use a primitive proposition $\somenewinfo$ that we
interpret as 
true for agent $i$ 
iff $i$ has received some new information; 
in our setting, that means that 
$i$ has learned about another agent in the network and his
input, has learned the weight labeling some edges, or has learned that
there are no further agents in the network.  (Note that in the latter
case, $i$ can 
also compute the function value.  For example, in doing leader election
on a unidirectional ring, if $i$ gets its id back after sending 
it around the network, then $i$ knows that it has heard from all agents
in the network, and can then compute which agent has the highest id.)
Note that $\somenewinfo$ is a 
proposition whose truth is relative to an agent.  
As already pointed out by GH, once we
work in a setting with relative names, then both propositions and names
\discout{need to be interpreted relative to an agent; we make this more
precise 
in the next section.%
}
\discin{need to be interpreted relative to an agent.
}
In the program, the action
$\send_{\name}(\newinfo)$ has the effect of $i$ sending $\name$ 
whatever new information $i$ learned.

With this background, we can describe the 
program for global function computation, 
which we  call  
$\Pg^{GC}$; each agent runs the program
\podc{$$
\begin{array}{l}
}
$${\bf if}~ \somenewinfo~{\bf then}~
\send_{\Nbr}(\newinfo); \mathit{receive},
$$
\podc{
\end{array} 
$$}
where the $\mathit{receive}$ action updates the agent's state by
receiving any messages that are waiting to be delivered.
As written, $\Pg^{GC}$ does not terminate; however, we can easily modify
it so that it terminates if agents learn the function value.  (They will
send at most one message after learning the function value.)
We would like to prove that $\Pg^{GC}$ solves the global function
computation problem.  To do this, we need to give precise semantics
to programs; that is the subject of the next section.

\subsection{Protocols, systems, and contexts}\label{sec:sys}
We interpret programs in the {\em runs and systems} framework of
Fagin et 
al. \citeyear{FHMV}, adapted 
to allow for names.
We start with a  possibly infinite set $\A$ of agents.
At each point in time, only finitely 
many agents are present.
Each of these agents $i$ is in some local state $l_i$. 
The \emph{global state} of the system at a particular point is a tuple $s$ 
consisting of the local states of the agents that exist at that point.
Besides the agents, it is also convenient to assume that there is 
an {\em environment state}, which keeps track of
everything relevant to the system not included in the agents' states.
In our setting, the environment state simply describes the network.

A {\em run\/} is a function from time (which we take here to range over
the natural numbers) to global states.  Intuitively, a run describes the
evolution of the system over time. 
With each 
run, we
associate the set of agents that exist in that run. For simplicity, 
we assume that the set of agents is constant over the run;  that is,
we are not allowing agents to enter the system or leave the system.
However, different sets of agent may be associated with different runs.
\commentout{
 For simplicity, 
in this paper
we assume that the
same set of agents is associated with the global state at each point in
a run.  Intuitively, agents do not enter or leave the system during the
run.
}%
 (While this is appropriate in our setting, it is clearly not
appropriate in general.  We can easily extend the framework presented
here to allow agents to enter or leave the system.)  Let $\A(r)$ denote the
agents present in run $r$.  A pair $(r,m)$ consisting of a run $r$ and
time $m$ is called a {\em point}.  If 
$i \in \A(r)$, we use $r_i(m)$ to denote agent $i$'s 
local state at the point $(r,m)$.
A {\em system\/} $\R$ consists of a set of runs.

In a \emph{system for global function computation}, 
each agent's initial local information is encoded in the
agent's local state; it must be consistent with the environment.  For
example, if according to the environment the network is a bidirectional
ring, each agent must 
have
two outgoing edges according to its local state.
We 
assume that agents have {\em perfect recall}, so that they
keep track in their local states of everything that they have heard and
when they heard it.  
This means that, in particular, the local state of an agent encodes
whether the agent has obtained new information about the network in a
given round $k$.

We are particularly interested in systems generated by protocols.
A protocol $P_i$ for agent $i$ is a function from 
$i$'s local states to nonempty sets of actions that $i$ 
may perform.
If the protocol is deterministic, then  
$P_i(\ell)$ is a singleton for  each local state $\ell$. 
A {\em joint protocol} is a tuple $P=\{P_i: i \in \A\}$, which
consists of one protocol for each agent.

We can associate with each joint protocol $P$ a system, given
a {\em context}.  
A context describes the environment's protocol, the initial states, the
effect of actions, and the association of names with agents.  Since names
are relative to agents, we do the association using a \emph{naming
function} $\mu: \G \times \A \times \Names
\rightarrow \A$, where $\G$ is the set of global states.
Intuitively, $\mu(g,i,\name) = j$
if  agent $i$ assigns name $\name$ to agent $j$ at the global state $g$.
Thus, we take a 
context $\gamma$ to be a tuple $(P_e,\Gz,\tau,\mu)$,
where $P_e$ is a protocol for the environment,
$\Gz$ is a set of initial global states, 
$\tau$ is a {\em transition function}, and $\mu$ is a naming
function.%
\footnote{Fagin et al.~\citeyear{FHMV} also have a component of the
context that describes the set of ``allowable'' runs.  This plays a role
when considering issues like fairness, but does not play a role in this
paper, so we omit it for simplicity.
Since they do not consider names, they do not have a component $\mu$ in
their contexts.}
The environment is viewed as running a protocol just like the agents;
its protocol is used to capture, for example, when messages are
delivered in an asynchronous system.  The transition function $\tau$ 
and naming function $\mu$ determine a mapping denoted $\tau_\mu$
associating with each {\em joint action\/}
(a tuple consisting of an action for the environment and one
for each of the agents)
a {\em global state transformer}, that is, a mapping from global states 
to global states.  
Note that we need the naming function since actions may involve names.
For the simple programs considered in this paper, the
transition function will be almost immediate from the description of the
global states.

We focus in this paper on a family of contexts that we call 
{\em contexts for global function computation}.  Intuitively, the
systems that represent programs in a context for global function
computation are systems for global function computation. 
A context $\gamma^{GC} =
(P_e,\Gz,\tau,\mu)$ for global function computation 
has the following features:
\begin{itemize}
\item The environment's protocol $P_e$ controls message delivery and is
such that all messages are eventually delivered, and no messages are
duplicated or corrupted.
\item The initial global states 
are such that
the environment's state records the network $N$ 
and agent $i$'s local state records agent $i$'s initial local
information;
we use $N_r$ to denote the network in a run $r$ (as encoded by the
initial global state in $r$). 
\item The transition function $\tau_\mu$ is such that the agents keep track
of all messages sent and delivered and the set of agents does not change
over time.  That is, 
if $s$ is a global state, $\sfa$ is a joint action, and $s' =
\tau_{\mu}(\sfa)(s)$, then $\A(s) = \A(s')$ and agent $i$'s local state in
$s'$ is the result of appending all messages that $i$ sent and received
as a result of action $\sfa$ to $i$'s local state in $s$.
We assume that $\tau_\mu$ is such that the action
$\send_{\name}(\newinfo)$ has the
appropriate effect, i.e., if  
$\send_{\name}(\newinfo)$ is agent $i$'s component of a joint action
$\sfa$ and agent $i$ gives agent $j$ name $\name$ in the global state
$s$ (note here we need the assumption that the naming function $\mu$
depends only on the global state)
and $s' = \tau_{\mu}(\sfa)(s)$, then in $s'$, $j$'s local state records
the
fact that $j$ has received the information from $i$.
\end{itemize}
In the following, we will denote the set of all networks encoded in the initial
global states of a context $\gamma^{GC}$ for global function computation as 
$\N(\gamma^{GC})$.

A run $r$ is consistent with a joint protocol $P$ if it could have been
generated when running $P$.  Formally,
run $r$ is {\em consistent with joint
protocol $P$ in context $\gamma$\/} if
its  initial global state $r(0)$ is one of
the initial global states~$\Gz$ 
given in~$\gamma$, 
and for all~$m$, the transition
from global state $r(m)$ to $r(m+1)$ is the result of performing 
one of the joint actions specified by~$P$ according to the agents in
$r$, and the environment protocol $P_e$
(given in~$\gamma$) in the global state $r(m)$.  
That is, if $P = \{P_i: i \in \A\}$ and $P_e$ is the environment's
protocol in context $\gamma$, then $r(0) \in \Gz$, and if
$r(m) = (\ell_e, \{\ell_i: i \in \A(r)\})$, then 
there must be a joint action $(\sfa_e, \{\sfa_i: i \in r(\A)\})$
such that $\sfa_e \in P_e(\ell_e)$, $\sfa_i \in P_i(\ell_i)$ for $i \in
r(\A)$, 
and $r(m+1) = \tau_\mu(\sfa_e,\{\sfa_i: i \in r(\A)\})(r(m))$ (so
that $r(m+1)$ is the result of applying the joint action
$(\sfa_e,\{\sfa_i: i \in \A\})$ to $r(m)$.
For future reference, we will say that a run~$r$ is {\em consistent
with $\gamma$} if $r$ is consistent with {\em some} joint protocol~$P$
in~$\gamma$. 
A system $\R$ {\em represents\/} a joint protocol $P$ in a context
$\gamma$
if it consists of all runs consistent with $P$ in $\gamma$.
We use $\Rrep(P,\gamma)$ to denote the system representing~$P$ in 
context~$\gamma$.
We want to associate with a program a protocol.  To do
this, we need to interpret the tests in the program.  
In doing so, we need to consider the fact that tests in the programs we consider here 
may contain names. 
This is the case for example of leader election programs in a ring network,
where an agent may send a message only if his identifier is larger than his left neightbor's. 
We can write this as $\id_I > \id_{\lefta}$, and clearly this test holds for the agent 
with maximum id, but does not hold for the agent with minimum id. 
This is why we need to  interpret the tests in a program relative to an agent 
and with respect to a naming function $\mu$ that resolves names relative to the agent.
Given a set $\Phi$ of primitive propositions, let an
\commentout{
{\em interpretation\/} be a function 
$\pi: \G \times \A \times \Phi \rightarrow \{\true,
\false\}$.
Intuitively, $\pi(g,i,p) = \true$ if $p$ is true at the global state $g$
relative to agent $i$.  Of course, we can easily extend this truth
assignment to arbitrary propositional formulas.
}
{\em interpretation\/} $\pi$ be a mapping that associates with each naming 
function $\mu$ a function
$\pi_{\mu}: \G \times \A \times \Phi \rightarrow \{\true, \false\}$.
Intuitively, $\pi_{\mu}(g,i,p) = \true$ if $p$ is true at the global state $g$
relative to agent $i$. 
Furthermore, we need to ensure that the interpretation
is consistent, in the sense that if $\id_I > \id_{\lefta}$ is interpreted as true
in a global state $g$ with respect to agent $i$, and $i$'s left neighbor refers 
to $i$ as his right neighbor, then $\id_{\righta} > \id_I$ is taken as true in same
global state, this time when interpreted relative to $i$'s left neighbor.
To formalize this, we take $\Phi'$ to be the set of all propositions in $\Phi$ with
relative names replaced by ``external names'' $1$, $\dots$, $n$, and take
functions $\pi' : \G \times \Phi' \rightarrow \{\true,\false\}$ to be 
{\em objective interpretation functions\/}.
We say that $\pi_{\mu}$ is {\em consistent\/} if there exists an objective 
interpretation $\pi'$ such that, 
for all global states $g$, agents $i$ and 
tests $p$ in $\Phi$, $\pi_{\mu}(g, i,p)=\true$ if and only if $\pi'(g,p')=\true$,
where $p'$ is just like $p$, except that all names $\name$ are replaced by
the external name $\mu(g,i,\name)$. In the following, we will focus only on 
contexts $\gamma$ and interpretations $\pi$ such that $\pi_{\mu}$ (for $\mu$ the naming
function in $\gamma$) is consistent.
Of course, we can extend $\pi_{\mu}$ 
to arbitrary propositional formulas, in the standard way; for example,
we take $\pi_{\mu}(g,i,\neg \varphi)=\true$ iff $\pi_{\mu}(g,i,\varphi)=\false$,
$\pi_{\mu}(g,i,\varphi\wedge \psi)=\true$ iff $\pi_{\mu}(g,i,\varphi)=\true$
and  $\pi_{\mu}(g,i,\psi)=\true$, etc.

An interpretation is 
\emph{local}
(for program $\Pg$ and in context $\gamma$) if
the tests $\phi$ in $\Pg$ depend only on the
local state, in the sense that if $\ell$ is agent $i$'s local state in the
global state $g$ and also agent $j$'s local state in the global state
$g'$, 
then $\pi_{\mu}(g,i,\phi) = \true$ iff $\pi_{\mu}(g',j,\phi) = \true$.  In
this case, we write $\pi_{\mu}(\ell,\phi) = \true$.
Given an interpretation $\pi$ that is 
local,
we can associate with a program $\Pg$ for agent $i$ a protocol
$\Pg^{\pi_{\mu}}$.
We
define $\Pg^{\pi_{\mu}}(\ell) = \{\sfa_j \: | \: \pi_{\mu}(\ell,t_j) = \true\}$
if there exist tests $t_j$ such that $\pi_{\mu}(\ell,t_j) = \true$, and take
$\Pg^{\pi_{\mu}}(\ell) = \noop$ otherwise.
Define $\Isys(\Pg,\gamma,\pi) = \Rrep(\Pg^{\pi_{\mu}},\gamma)$, for $\mu$ the naming
function in context $\gamma$.

An {\em interpreted context for global function computation\/}
is a pair $(\gamma,\pi)$, where $\gamma$ is a context for global
function computation and $\pi_{\mu}$ interprets $\somenewinfo$ appropriately
(so that $\pi_{\mu}(g,i,$ $\somenewinfo)$ $ = \true$ if $i$ received some
new information about the network in $g$ and has not sent a message
since receiving that information).

For the purpose of global function computation, we often talk about agents \emph{knowing} 
a fact about the network, some piece of information, or the function value, and how this
knowledge changes during a run of a protocol like ${(\Pg^{GC})}^{\tau_{\mu}}$.
Intuitively, this says that, regradless of the agent's uncertainity about the network, and 
in general about the global state he is in, $\phi$ holds. 
$i$'s uncertainity about the global world  comes from two sources: $i$'s uncertainty
about the local states of other agents, and $i$'s uncertainity about his own identity and 
the identities of the other agents he can refer to by certain names. More precesily,
when in some local state $\ell=r_i(m)$, $i$ cannot distinguish between the global world 
$r(m)$ and any global world $r'(m')$ such that there exists an agent $i'$ with same local state
as $i$, i.e., $r'_{i'}(m')=\ell$. In the following, we will a tuple $(r,m,i)$ a 
\emph{situation}, and we will say that situations $(r,m,i)$ and $(r',m',i')$  are
indistinguishable to agent $i$ if $i$ thinks possible he is $i'$ in $r'(m')$, i.e.,
$r_i(m)=r'_{i'}(m')$. 
We define an \emph{extended interpreted system} to be a tuple
$\I = (\R,\pi,\mu)$, where $\R$ is a system, $\pi$ is an interpretation,
and $\mu$ is a naming function. 
We say that fact $\phi$ holds at situation $(r,m,i)$ and with respect
to interpreted system $\I$, denoted as $(\I, r,m,i)\models \varphi$, precisely when 
$\pi_{\mu}(r(m), i,\phi)=\true$.
We can now formalize the fact that $i$ knows $\phi$  at point $(r,m)$ as
the condition that $\phi$ holds at all situations intistinguishable to $i$ from 
$(r,m,i)$, i.e., $(\I,r',m',i')\models \phi$ for all situations $(r',m',i')$ in $\I$ with
$r'_{i'}(m')=r_i(m)$.

Program $\Pg$ solves
the global function computation  problem for function $f$ 
in the interpreted context $(\gamma^{GC},\pi)$ if and only if, in all runs 
$r$ of $\Isys(\Pg,\gamma^{GC},\pi)$, eventually all agents in $\A(r)$
know the value $f(N_r)$.
That is, for all such runs $r$, there exists a time $m$ such that, for all agents 
$i$ in $\A(r)$, $f$ takes the same value $f(N_r)$ on all networks $i$ thinks possible when 
in local state $r_i(m)$, i.e., on all networks in runs $r'$ 
such that there exists a time $m'$ and an agent $i'$ with $r'_{i'}(m')=r_i(m)$.

\subsection{Proving the correctness of $\Pg^{GC}$}

\thm\label{thm:kb correct}
\commentout{
$\Pg^{GC}$  solves the global function computation problem whenever 
possible.
That is, if $\N$ and $f$ satisfy the condition in
Theorem~\ref{thm:solvability}, then
with $\Pg^{GC}$ eventually all agents know the function value;
otherwise, no agent ever knows the function value. 
}
If $f$ and $\N(\gamma^{GC})$ satisfy the condition
in Theorem~\ref{thm:solvability}, then
$\Pg^{GC}$ solves the global function computation problem 
for $f$ in all interpreted contexts $(\gamma^{GC},\pi)$
for global function computation. 
\ethm

\prf
\commentout{
Suppose that $\N$ and $f$ satisfy the condition in
Theorem~\ref{thm:solvability}.
Let $(\gamma,\pi)$ be an interpreted context for global function computation
such that in all 
initial states the network encoded in the environment state is in $\N$, 
let $\I=\Isys(\Pg^{GC,\pi},\gamma,\pi)$.
let $r$ be a run in $\I$,
and 
let
$N_r$ be the network associated with run~$r$.
}
Let $f$ be a global function and 
let
$(\gamma^{GC},\pi)$ be an interpreted 
system for global function computation such that $f$ and $\N(\gamma^{GC})$
satisfy the condition in Theorem~\ref{thm:solvability}. Let $r$ be a run in 
the system $\Isys(\Pg^{GC}, \gamma^{GC},\pi)$.

We first show that at some point in $r$, some
agent knows $f(N_r)$.  Suppose not.
Let $r'$ be the unique run of the full-information protocol 
starting with the same initial global state as $r$.
We show by induction on $k$ that there is a time $m_k$ such that, at
time $(r,m_k)$, all the agents in $\A(r)$ have at least as much information 
about the network as they do at the beginning of round $k$ in $r'$.
That is, for all agents $i$ in $\A(r)$, the set of networks $i$ considers possible
at time $m_k$ in $r$ (i.e., the set of all networks $N_{r''}$ for $r''$ run in 
$\Isys(\Pg^{GC}, \gamma^{GC},\pi)$ such that there exists a situation $(r'',m'',i'')$
with $r''_{i''}(m'')=r_i(m_k)$) is a subset of the set of networks $i$ considers possible
at the beginning of round $k$ in $r'$ (i.e., if $m'_k$ is the time in $r'$ when round $k$
begins, the set of networks $N_{r''}$ for $r''$ run of the full-information protocol
such that there exists a situation $(r'',m'',i'')$ with $r''_{i''}(m'')=r'_i(m'_k)$). 

The base case is immediate: we can take $m_1 = 0$ since, by assumption,
agents in $r$ and $r'$ start with the same initial states.
For the inductive step, 
suppose  that $i$ learns some
new information from $j$ in round $k$ of $r'$.
That means $j$ knew this information at the beginning of round $k$ in
$r'$ so, by the induction hypothesis,  $j$ must have known this
information by time $m_k$ in $r$.   
Thus, there is a time $m_k' \le m_k$ such that $j$ first learns this
information in run $r$ (where we take $m_k' = 0$ if $k=1$).
It follows from the semantics of $\Pg^{GC}$ that $j$
sends this information to $i$ at time $m_k'$ in $r$.   
Since we
have assumed that communication is reliable, $i$ learns it by some time
$m_k''$.  Since $i$ has only finitely many neighbors and there are only
finitely many pieces of information about the network, there must be a
time in $r$ by which $i$ learns all the information that it learns by
the beginning of round $k+1$ in $r'$.  And since there are only finitely
many agents in $\A(r)$, there must be a time $m_{k+1}$ by which all the
agents in $\A(r)$ learn all the information about the  network that they
know at the beginning of round $k+1$ in $r'$.  

By Theorem~\ref{thm:solvability}, there exists a round 
$k_{\N(\gamma^{GC}), N_r,f}$
such that, running the full-information protocol, 
for all networks 
$N'\in \N(\gamma^{GC})$,
all $i' \in V(N')$, and all $i \in V(N_r)$,
we have that  
$f(N_r) = f(N')$ if $(N_r,i) \sim_{k_{\N(\gamma^{GC}),N_r,f}} (N',i')$.
Suppose that $i$ is an agent in $N_r$, 
$r'$ is a run in $\Isys(\Pg^{GC},\gamma^{GC},\pi)$, and 
$i'$ is
an agent in $N_{r'}$ such that 
$r_i(m_{k_{\N(\gamma^{GC}),N_r,f}})=r'_{i'}(m')$.
A straightforward argument now shows that 
$(N_r,i) \sim_{k_{\N(\gamma^{GC}),N_r,f}} (N_{r'},i')$.  (Formally, we show by
induction on $k$ with a subinduction on $k'$ that if 
$k \le k_{\N(\gamma^{GC}),N_r,f}$, $k' \le k$, and $j$ is 
an agent at distance  $k'$
 from $i$ in $N_r$, then there exists an agent $j'$ of distance $k'$ from
$i'$ in $N_{r'}$ such that  $(N_r,i) \sim_{k-k'} (N_{r'},i')$, and
similarly switching the roles of $i$, $i'$, $N_r$, and $N_{r'}$.)
It follows that $i$ knows $f(N_r)$ by time $m_{k_{\N(\gamma^{GC}),N_r,f}}$ in
$r$, contradicting the assumption that no agent learns $f(N_r)$.

Suppose that $i$ is the first agent to learn the function value in $r$, 
and does so at time $m$ (or one of the first, if there are several
agents that learn the function value at time $m$).
We can now use the same argument as above to show that eventually all
agents learn the function value.  A formal proof proceeds by induction
on the distance of agent $j$ from $i$ in $N_r$; we omit details here.
\eprf

\section{Improving message overhead}

While sending only the new information that an agent learns at each step
reduces the size of messages, it does not preclude sending unnecessary
messages.  
One way of reducing communication is 
to have agent $i$ not send information to the
agent he names $\name$ if he {\em knows} that $\name$ already {\em knows}
the information.  
Since agent $i$ is acting based on what he knows, this is a {\em
knowledge-based (kb) program}. 
We now formalize this notion.

\commentout{
There is an obvious change to the second line of 
$\Pgkb^{GC}$ that captures this intuition.  Instead of saying
$${\tt for ~all~neighbors~\name}~{\bf do}~ \send_{\name}(\val(f)),$$ 
we say 
$${\tt for ~all~neighbors~\name}~{\bf do}~{\bf if}~K_I \exists v
K_{\name} (f=v)~{\bf then}~\noop {\bf else}~\send_{\name}(\val(f)).$$  
We want to similarly modify the fourth line, so that $i$ sends his new
information to $\name$ only if $i$ does not know that $\name$ already
knows it.  
}
\subsection{Knowledge-based  programs with shared names}\label{sec: kb pg syntax}
\commentout{
To capture this, assume first that there is a modal operator $K_{\name}$ in
the language 
for each name $\name \in \Names$.
When interpreted relative to agent $i$, 
$K_{\name} \varphi$ is read as 
\discout{
``the agent $i$ named $\name$ knows fact $\varphi$''
(see Appendix~\ref{sec:apx:kb-progs} for the formal semantics).
}
\discin{``the agent $i$ named $\name$ knows fact $\varphi$''.
}
}%
Consider a language with a modal operator $K_{\name}$ 
for each name $\name \in \Names$.
When interpreted relative to agent $i$,  $K_{\name} \varphi$ is read as 
``the agent $i$ names $\name$ knows $\varphi$''. 
A knowledge-based program $\Pg_{kb}$  has the form
$$
\begin{array}{l}
{\tt \bf if }~ t_1 \wedge k_1 ~{\tt \bf do} ~{\tt \sfa_1}\\
{\tt \bf if }~ t_2 \wedge k_2 ~{\tt \bf do} ~{\tt \sfa_2}\\
\quad \dots\\
\end{array}
$$
where $t_j$ and $\sfa_j$ are as for standard programs, and $k_j$
are knowledge tests (possibly involving belief and counterfactual tests, as we
will see later in the section).

Let $\inewinfo$ be a primitive proposition that 
characterizes the content of the message $\newinfo$.   For example, 
suppose that $N$ is a unidirectional ring, and $\newinfo$ says that $i$'s left
neighbor has input value $v_1$.  Then $\inewinfo$ is true at all points
where $i$'s left neighbor has input value $v_1$.  (Note that
$\inewinfo$ is a proposition whose truth is relative to an agent.)
Thus, it seems that the following kb program should solve
the global function computation problem, while decreasing the number of
messages: 
\begin{equation}\label{eq:kbprogram}
\begin{array}{l}
{\bf if}\ \somenewinfo\ {\bf then}\\
\quad {\bf for\ each}\ nonempty\ subset\ $A$\ of\ agents\ {\bf do}\\
\quad {\bf if}~ A = \{\name \in \Nbr: \neg K_I K_{\name}(\inewinfo)\}
~{\bf then}~\send_{A}(\newinfo); \mathit{receive}.
\end{array}
\end{equation}
There are, however, some subtleties involved giving semantics to this 
program; we consider these in the next section.  In the process, we
will see that there are number of ways that the message complexity of
the program can be further improved.

\subsection{Semantics of kb programs with shared names}\label{sec:apx:kb-progs}
\commentout{
We assume that there is a modal operator $K_{\name}$ in the language 
for each name $\name \in \Names$.
When interpreted relative to agent $i$,  $K_{\name} \varphi$ is read as 
``the agent $i$ named $\name$ knows fact $\varphi$''.
}
We can use the machinery 
that
we have developed to give semantics to
formulas such as $K_{\name}\phi$.  
\commentout{
Define an \emph{extended interpreted system} to be a tuple
$\I = (\R,\pi,\mu)$, where $\R$ is a system, $\pi$ is an interpretation,
and $\mu$ is a naming function. 
}
\commentout{
As mentioned before, the statement
$K_{\name}\varphi$ is subjective:    
it may be false or true, depending on the agent making the statement;
this means that we interpret $K_{\name}\varphi$ with respect to a point $(r,m)$ and an agent $i$ in $r$.
(We call a tuple $(r,m,i)$ a {\em situation}.)
Intuitively, if the agent 
that
$i$ names $\name$ is $j$, then $i$'s statement ``$\name$ knows $\varphi$'' is true
if $\varphi$ holds in all situations $j$ considers possible, i.e., in all situations where it has the same 
local state as in $(r,m)$. Notice that, since $j$ may be uncertain about his own identity, 
these are exactly the situations $(r',m',j')$ such that $j$ has the same local state in $(r,m)$ as $j'$ in $(r',m')$.
}
The statement $K_{\name}\varphi$ holds with respect to a situation $(r,m,i)$
and an interpreted system $\I$ precisely when the agent $j=\mu(r(m), i,\name)$ $i$ names $\name$ 
knows $\varphi$ when in local state $r_j(m)$, i.e., when $\varphi$ holds in all situations 
$(r',m',j')$ in $\I$ agent $j$ cannot distinguish from $(r,m,j)$.
We can then define
$$\begin{array}{lll}
(\I, r,m, i) \models K_\name \phi & \mbox{ iff, for all
$j, j'$ and points $(r',m')$ such that $\mu(r(m),i,\name) = j$}\\
&\mbox{and
$r_j(m) = r'_{j'}(m')$, we have
$(\I,r',m',j') \sat \phi$.}
\end{array}
$$
\commentout{
(For $\phi$ standard test, i.e., not involving knowledge operators $K_{\name}$, 
we take $(\I,r,m,i)\models \phi$ precisely when $\pi(r(m), i,\phi)=\true$. 
The satisfiability relation $\models$ can be extended to formulas build
up from both knowledge and standard tests in the standard way, e.g.,
$(\I,r,m,i)\models \phi\wedge \psi$ iff $(\I,r,m,i)\models \phi$ and
$(\I,r,m,i)\models \psi$, etc.) 
}

As observed by GH,
once we allow relative names, we must 
be careful about scoping.  For example, suppose that, in an oriented ring,
$i$'s left neighbor is $j$ and $j$'s left neighbor is $k$.  What does a 
formula such as $K_I K_{{\lefta}}(\leftinput =3)$ mean when it
is
interpreted relative to agent $i$?  Does it mean that $i$ knows that $j$
knows that $k$'s input is 3, or does it mean that $i$ knows that $j$
knows that $j$'s input is 3?  That is, do we interpret the ``left'' in 
$\leftinput$ relative to $i$ or relative to $i$'s left neighbor
$j$?  
\podc{
Similarly, to which agent does the second $L$ in 
$K_I K_{{\lefta}} K_{{\lefta}} \phi$ refer?  
That, of course, depends on the application.  Using a first-order
logic of naming, as in \cite{Grove95}, allows us to distinguish the two
interpretations readily.  In a propositional logic, we cannot do this.
In the propositional logic, GH 
assumed {\em innermost scoping\/}, so that the {\em left\/} in $\leftinput$
and the second  $\lefta$ in  $K_I K_{{\lefta}} K_{{\lefta}} \phi$ are
interpreted 
relative to the ``current'' agent considered when they are evaluated (which
is $j$).  
For the purpose of this paper,
}
in a formula such as 
$K_I K_{\name} ~\inewinfo$, we want to interpret $\inewinfo$ relative to
``$I$'',
the agent $i$ that sends the message, not with respect to the agent $j$
that is the interpretation of $\name$.
To capture this, we add limited quantification over names to
the language.  In particular, we allow formulas 
of the form 
$\exists \name' (\Calls(\name,I,\name') \land K_\name (\name'\mbox{'s}
\phi))$, which is interpreted as ``there exists a name $\name'$ such
that the agent $I$ names $\name$ gives name $\name'$ to the agent that
currently has name $I$ and $\name$ knows that $\phi$ interpreted
relative to $\name'$ holds".  
Thus, to emphasize the scoping,
instead of writing $K_I K_\name \inewinfo$, we write
$K_I (\exists \name' (\Calls(\name,I,\name') \land K_\name (\name'\mbox{'s}
\inewinfo)))$. 

We can now give semantics to kb programs.
We can  associate with a kb program $\Pgkb$ 
and an
extended
interpreted system $\I = (\R,\pi,\mu)$ a 
protocol for agent $i$ denoted  $(\Pgkb)_i^{\cal I}$.
Intuitively, we evaluate the standard tests in $\Pgkb$ 
according to~$\pi$ 
and $\mu$
and evaluate the knowledge tests according to $\I$.
Formally, for each local state $\ell$ of agent $i$, we define 
$(\Pgkb)_i^\I(\ell)$ to consist of all actions $\sfa_j$ such that the
test $t_j \land k_j$ holds with respect to a tuple $(r,m,i')$ in $\I$
such that $r_{i'}(m) = \ell$ (recall that protocols can be
nondeterministic); if there is no point in $\I$ where some agent has
local state $\ell$, then $(\Pgkb)_i^\I(\ell)$ performs the null action
(which leaves the state unchanged).

A 
joint
protocol $P$ is said to {\em implement} $\Pgkb$ in 
interpreted context $(\gamma,\pi)$ 
if, by interpreting $\Pgkb$ with respect to $\Isys(P,\gamma,\pi)$,
we get back protocol $P$; i.e., if, 
for each agent $i$, we have $P_i=(\Pgkb)^{{\cal I}(P,\gamma,\pi)}_i$.
Here we seem to be implicitly assuming that all agents run the same
kb program.  
This is certainly true for the programs we give for global function
computation, and actually does not result in any loss of generality.
For example, if names are commonly known,
the actions performed by agents can depend on tests of the form ``if
your name is $\name$ then \ldots''.  
Similarly, if we have a system where some agents are senders and others
are receivers, the roles of agents can be encoded in their local states,
and tests in the program can ensure that all agents act appropriately,
despite using the same program.

In certain cases we are interested in 
joint
protocols $P$ that
satisfy  
a condition slightly weaker than implementation, first defined by 
Halpern and Moses \citeyear{HM98a} (HM from now on).
Joint protocols $P$ and $P'$
are {\em equivalent in context $\gamma$},
denoted $P \approx_\gamma P'$,
if $P_i(\ell) = P'_i(\ell)$ for every local state $\ell=r_i(m)$ with
$r \in \Rrep(P,\gamma)$.
We remark that if $P \approx_\gamma P'$, then it easily follows that
$\Rrep(P,\gamma) = \Rrep(P',\gamma)$: we simply show by induction on $m$
that every prefix of a run in $\Rrep(P,\gamma)$ is a prefix of a run in
$\Rrep(P',\gamma)$, and vice versa.
$P$ {\em de facto implements}
$\Pgkb$ in context $\gamma$ if 
$P \approx_\gamma \Pgkb^{{\cal I}(P,\gamma,\pi)}$.
Arguably, de facto implementation suffices for most purposes, since all we
care about are the runs generated by the protocol.  We do not care about
the behavior of the protocol on local states that never arise
when we run the protocol.

\commentout{
While this essentially works, there are some subtleties in interpreting
this kb program.  As observed by GH,
once we allow relative names, we must 
be careful about scoping.  For example, suppose that, in an oriented ring,
$i$'s left neighbor is $j$ and $j$'s left neighbor is $k$.  What does a 
formula such as $K_I K_{{\lefta}}(\leftinput =3)$ mean when it
is
interpreted relative to agent $i$?  Does it mean that $i$ knows that $j$
knows that $k$'s input is 3, or does it mean that $i$ knows that $j$
knows that $j$'s input is 3?  That is, do we interpret the ``left'' in 
$\leftinput$ relative to $i$ or relative to $i$'s left neighbor
$j$?  
\podc{
Similarly, to which agent does the second $L$ in 
$K_I K_{{\lefta}} K_{{\lefta}} \phi$ refer?  
That, of course, depends on the application.  Using a first-order
logic of naming, as in \cite{Grove95}, allows us to distinguish the two
interpretations readily.  In a propositional logic, we cannot do this.
In the propositional logic, Grove and Halpern \citeyear{GroveH2} 
assumed {\em innermost scoping\/}, so that the {\em left\/} in $\leftinput$
and the second  $\lefta$ in  $K_I K_{{\lefta}} K_{{\lefta}} \phi$ are
interpreted 
relative to the ``current'' agent considered when they are evaluated (which
is $j$).  As we will see, in our interpretation, we want to interpret it
relative to $I$ (in this case, $i$).  
}
As we will see, in a formula such as 
$K_I K_{\name} ~\inewinfo$, we want to interpret $\inewinfo$ relative to
``$I$'',
the agent $i$ that sends the message, not with respect to the agent $j$
that is the interpretation of $\name$.
To capture this, we add limited quantification over names to
the language.  In particular, we allow formulas of the form 
$\exists \name' (\Calls(\name,I,\name') \land K_\name (\name'\mbox{'s}
\phi))$, which is interpreted as ``there exists a name $\name'$ such
that the agent $I$ names $\name$ gives name $\name'$ to the agent that
currently has name $I$ and $\name$ knows that $\phi$ interpreted
relative to $\name'$ holds".  
Then, instead of writing $K_I K_\name \inewinfo$, we write
$K_I (\exists \name' (\Calls(\name,I,\name') \land K_\name (\name'\mbox{'s}
\inewinfo)))$.
}%

The kb program $\Pg_{kb}$ solves the global function computation problem
for $f$ in the  
interpreted context $(\gamma^{GC},\pi)$ if, for all protocols $P$ that
de facto implement $\Pg_{kb}$ in $\gamma^{GC}$ and all runs $r$ in
$\R(P,\gamma)$,  
eventually all agents in $\A(r)$ know the value $f(N_r)$.

We can now show that the kb program (\ref{eq:kbprogram}) solves 
the global  function computation problem for all
functions $f$ and interpreted contexts $(\gamma^{GC},\pi)$ 
for global function computation
such that
$f$ and $\N(\gamma^{GC})$ satisfy the condition in
Theorem~\ref{thm:solvability}. 
Rather than proving this result, we focus on further
improving  
the message complexity of the kb program, 
and give a formal analysis of correctness only for the improved program.

\subsection{Avoiding redundant communication with counterfactual tests}
We can further reduce message complexity by not sending information  not
only if the recipient of the message already knows the information, but also
if he will {\em eventually\/} know the information.  It seems relatively 
straightforward to capture this: we simply 
add a $\Diamond$ operator 
to the kb program (\ref{eq:kbprogram} to get
$$\begin{array}{l}
{\bf if}\ \somenewinfo\ {\bf then}\\
\quad {\bf for\ each}\ nonempty\ subset\ $A$\ of\ agents\ {\bf do}\\
\quad {\bf if}~ A = \{\name \in \Nbr: 
\neg K_I \Diamond (\exists \name' (\Calls(\name,I,\name') \land K_\name (\name'\mbox{'s}
\inewinfo)))\}\\
\quad \quad {\bf then}~\send_{A}(\newinfo); \mathit{receive}.
\end{array}$$  
Unfortunately, this modification will not work: as
observed by HM,
once we add the $\Diamond$
operator, 
\commentout{  
the
resulting kb program
has no representation.   For suppose it is represented by a system $\I$,
that is, by interpreting the above program w.r.t. $\I$ we get back the
system $\I$. 
}
the resulting program has no implementation in the context $\gamma^{GC}$.
For suppose there exists a protocol $P$ that implements it, and let 
$\I = {\cal I}(P, \gamma^{GC}, \pi)$, that is, by interpreting the above program
w.r.t. $\I$, we get back the protocol $P$.
Does $i$ (the agent represented by $I$) send $\newinfo$ 
to $\name$
in $\I$?
\commentout{
If it does, then $\I$ can't represent the program because, 
in $\I$, 
$i$ knows that $\name$ will eventually know $\newinfo$,
so $i$ should not send $\newinfo$ to $\name$.  It follows that no agent
should send $\newinfo$ to $\name$ in $\I$.
}  
If $i$ sends its new information to $\name$ at time $m$ in a run $r$ of $\I$,
then, as communication
is reliable, eventually $\name$ will know $i$'s new information and $i$ knows that this
is the case, i.e.,
$(\I, r, m,i)\models K_I \Diamond (\exists \name' (\Calls(\name,I,\name') \land K_\name
(\name'\mbox{'s} \inewinfo)))$. 
As $P$ implements the above kb program and $\I =  {\cal I}(P, \gamma^{GC}, \pi)$,
it follows 
that $i$ does not send its new information to $\name$.
On the other
hand, if no one sends $\newinfo$ to $\name$, then $\name$ will not know it,
and $i$ should send it.  Roughly speaking, $i$ should send the
information iff $i$ does not send the information.
HM suggest the use of counterfactuals to
deal with this problem.  As we said in the introduction, a
counterfactual has the form $\phi \RCond \psi$, which is read as 
``if $\phi$ were the case then $\psi$''.
As is standard in the philosophy literature (see,
for example, \cite{Lewis73,Stalnaker68}), 
to give semantics to counterfactual statements, we assume that there is
a notion of \emph{closeness} defined on situations.  This allows us to
consider the situations closest to a given situation that have certain
properties.  
For example, if in a situation $(r,m,i)$ agent $i$ sends its new
information 
to neighbor $\name$, 
we would expect that 
the closest situations $(r',m,i)$ to $(r,m,i)$  
where $i$ does {\em not} send its
new information to $\name$ are such that,
in $r'$,
all agents use the same protocol in $r'$ as in $r$, excpet that, at time
$m$ in $r'$, $i$ sends its new information to all agents to which it
sends its new information at the point $(r,m)$ 
with the exception of $\name$. 
The counterfactual formula $\phi \RCond \psi$ 
is taken to be
true if, in the closest 
situations to the current situation
where $\phi$ is
true, $\psi$ is also true.  
\commentout{
\discout{
As the results in this paper can be understood intuitively without
presenting all the details, and to keep the presentation simple, we 
defer the review of HM's semantics for counterfactuals to 
Appendix~\ref{sec:apx:cbb-progs}.
In 
particular, we discuss 
in Appendix~\ref{sec:apx:cbb-progs} the HM
concrete interpretation of ``closest
worlds''.
}
}%

Once  we have counterfactuals, we must 
consider systems with runs that are not runs of the program.
These are runs where, for example, counter to fact, the
agent does not send a message (although the program says it should).  
Following HM, we can make these executions less likely relative to those
generated  
by running the program
by associating to each run a {\em rank}; the higher the rank, the less
likely the run.  We then require that the runs of the program  
be the only ones of minimal rank. 
Once we work with a system that includes runs other than those generated
by the program, agents may no longer \emph{know} that, for example, 
when the program says they should send a 
message to their  neighbor, they actually do so (since there could be an
run in the system not generated by the program, in which at some
point the agent has the same local 
state as in a run of the program, but it does not send a message).
Agents do know, however, that they send the message to their neighbor in
all runs  
of minimal rank, that is, in all the runs consistent with the program. 
By associating a rank with each run, we can talk about formulas $\phi$
that hold at all situations in runs of minimal rank among those
an agent $i$ cannot distinguish from the current situation.
If $\phi$ holds at all points in runs of minimal rank that $i$ considers
possible then we say that $i$ \emph{believes} $\phi$ (although $i$ may
not \emph{know} $\phi$.
\commentout{
\discout{
(See Appendix~\ref{sec:apx:cbb-progs} for details.)
}
}
We write $B_{\name}\varphi$ to denote that the agent named $\name$
believes $\phi$, although this is perhaps better read as ``the agent
named $\name$ knows that $\phi$ is (almost certainly) true''.  
We provide the formal semantics of belief and counterfactuals, which is
somewhat technical, in Appendix~\ref{sec:apx:cbb-progs}; we hope that
the intuitions we have provided will suffice for understanding what follows.

Using counterfactuals, we can modify the program to say 
that agent $i$ should send the information only if $i$ does not believe  
``if I do not send the information, then $\name$ will  eventually
learn it anyway''. 
To capture this, we use the proposition
$\doact(\send_{\name}(\newinfo))$, which is true if $i$ is about to send 
$\newinfo$ to $\name$.  
If there are only finitely many possible values of $f$, say $v_1, \ldots, v_k$,
then the formula $B_{\name}(f=v_1) \lor \ldots \lor B_{\name}(f=v_k)$ 
captures the fact that the agent with name $\name$ knows the
value of~$f$.  However, in general, we want to allow an unbounded number
of function values.   
For example, if agents have distinct numerical ids, 
we are trying to elect as leader the agent with the highest id, and
there is no bound on the size of the network, then the set of possible
values of $f$ is unbounded.  We deal with this problem by allowing 
limited quantification over values. 
In particular, we 
use formulas of the form $\exists v B_\name (f=v)$, 
which intuitively say that the agent with name $\name$
knows the value of~$f$.
Let $\Pgcb^{GC}$ denote the following modification of $\Pg^{GC}$:
$$
\begin{array}{l}
{\bf if}~\somenewinfo ~{\bf then}\\
\quad {\bf for\ each}\ nonempty\ subset\ $A$\ of\ agents\ {\bf do}\\
\quad {\bf if}~A = \{\name \in \Nbr: \neg
B_I[\neg\doact(\send_{\name}(\newinfo))\RCond  \\
\quad \quad \quad \quad \quad  \quad \Diamond(\exists \name'
(\Calls(\name,I,\name')  \land B_\name (\name'\mbox{'s} \inewinfo)) 
\lor \exists v   B_\name(f=v))]\}  \\
\quad {\bf then}~\send_A(\newinfo); 
\mathit{receive}.
\end{array}
$$
In this program, the agent $i$ representing $I$ 
sends $\name$
the new information if $i$ does not believe that $\name$ will eventually 
learn the new information or the function value in any case.
As shown in Appendix~\ref{sec: cb correct prf}, this
improved program still solves the global
function computation problem whenever possible.
\commentout{
Note that $i$ does not send 
$\name$ information 
about the network
not only if $i$ believes that $\name$ will eventually get the information
but also if $I$ believes that getting the information is unnecessary,
in that $\name$ will eventually compute the function value even without
it.
}
\thm\label{thm:cbb correct}
\commentout{
$\Pgcb^{GC}$  solves the global function computation problem 
whenever possible:
for all ${\cal N}$ and $f$ such that the condition in 
Theorem~\ref{thm:solvability}  is satisfied
and all protocols $P$ that de facto implement $\Pgcb^{GC}$, 
in every run $r$ of the system that represents $P$,
eventually all agents know  $f(N_r)$. 
}
If $f$ and $\N(\gamma^{GC})$ satisfy the condition
in Theorem~\ref{thm:solvability}, then
$\Pgcb^{GC}$ solves the global function computation problem 
for $f$ in all interpreted contexts $(\gamma^{GC},\pi)$
for global function computation. 
\ethm

\commentout{
Let $o$ be an order that respects protocols, $\rkgen$ a ranking generator
that is deviation compatible. Recall that we denote the context for global computation 
in which all networks encoded in initial global states belong to $\N$ as 
$\gamma^{GC,{\cal N}}$. Let $\zeta^{GC,{\cal N}}$ be the  extended
context $(\gamma^{GC,{\cal N}},\pi^{GC},\mu^{GC},o,\rkgen)$.
Let $\Pg$ be a protocol that de-facto implements $\Pgcbb^{GC}$ in $\zeta^{GC,{\cal N}}$.
For simplicity, we denote the extended counterfactual interpreted system
$({\cal I}^+(\gamma^{GC,{\cal N}}), \pi^{GC},\mu^{GC},o(\Pg),\rkgen(\Pg))$ as $\ecis$.
Let $r$ be a run of in $\Rrep(\Pg,\gamma^{GC,{\cal N}})$ corresponding to function $f$ and network $N\in \N$.
We will prove that at some point in run $r$  all agents in $N$ know the function value.

We first prove the following lemma:
\lem\label{lem:info spread cbb}
For any agent $i$ in $N$, formula $\varphi$, 
and name $\name$ that $i$ gives to one of his  outgoing neighbors,
if $(\ecis,r,m,i) \models B_I \varphi$,
then $$ (\ecis, r,m,i) \models \Diamond (\exists v  B_{\name} (f=v) \vee 
\exists \name^* \Calls(\name,I,\name^*)\wedge B_{\name} (\name^* ~\mbox{'s}~\varphi)).$$
In other words, all neighbors will eventually either know the function 
value or believe everything that $i$ believes.
\elem

\prf
Without loss of generality we can assume that $m$ is the first time $i$
believes $\varphi$
$$(\ecis,\sigma(Pg)),r,m,i) \models B_I \varphi \wedge \odot \neg B_I \varphi.$$
Let $j$ be the $i$'s neighbor named $\name$, i.e., $\mu^{GC}(r(m),\name,i)=j$.
There are two possible scenarios: either $i$ does not believe at time $m$ that, 
if he were not to send
$\varphi$ to $\name$, $\name$ would know either $\varphi$, or the function value.
\begin{itemize}
\item[(a)] Assume that 
$$(\ecis,,r,m,i)\models \neg B_I (\neg \doact(\send_{\name}(\varphi))\RCond 
\Diamond ((\exists \name^* \Calls(\name,I,\name^*)\wedge B_{\name}(\name^* ~\mbox{'s}~\varphi))
 \vee B_{\name}(f)))).$$
As $\Pg$ de facto implements $\Pgcbb^{GC}$, it must be that $i$ 
performs the action 
$\send_{\name}(\varphi)$. As communication is reliable, then eventually 
$j$ receives $\varphi$. At this point, $j$ also has a way of naming $i$, and let
$\name^*$ be one such name. Then
$(\ecis,r,m,j)\models \Diamond B_I (\name^* ~\mbox{'s}~\varphi)$
and so 
$$(\ecis,r,m,i)\models \Diamond (\exists \name^* \Calls(\name,I,\name^*)\wedge B_{\name}(\name^* ~\mbox{'s}~\varphi))$$

\item[(b)] Assume this time the contrary:
$$(\ecis,,r,m,i)\models B_I (\neg \doact(\send_{\name}(\varphi))\RCond 
\Diamond ((\exists \name^* \Calls(\name,I,\name^*)\wedge B_{\name}(\name^* ~\mbox{'s}~\varphi))
 \vee B_{\name}(f)))).$$
As $\rkgen$ is deviation-compatible, and $r$ is a run generated by 
running $\Pg$ in $\zeta^{GC,{\cal N}}$, Remark~\ref{rem:B true} ensures that 
indeed
$$(\ecis,r,m,i)\models \neg \doact(\send_{\name}(\varphi))\RCond  \Diamond ((\exists \name^* \Calls(\name,I,\name^*)\wedge B_{\name}(\name^* ~\mbox{'s}~\varphi))
 \vee B_{\name}(f)).$$

As $\Pg$ de-facto implements $\Pgcbb^{GC}$, it must be that $i$ 
does not send $\varphi$ to $\name$
at time $m$, and so
$$(\ecis,r,m,i)\models \neg \doact(\send_{\name}(\varphi)). $$
Then, the semantics of counterfactuals ensures that
$$(\ecis ,r,m,i)\models  \Diamond ((\exists \name^* \Calls(\name,I,\name^*)\wedge B_{\name}(\name^* ~
\mbox{'s}~\varphi)) \vee B_{\name}(f)).$$ 
\end{itemize} 
\eprf

Assume now by way of contradiction that in run $r$ no agent in $N$ learns the function value.
The above lemma ensures that, for any time $m$ and  agent $i$, eventually
all of $i$'s neighbors will believe whatever $i$ believes: if
$(\ecis,r,m,i)\models B_I \varphi$, then for any $j$ neighbor of $i$, there exists a name $\name^*$ that
$j$ assigns to $i$ such that $(\ecis,r,m,j)\models \Diamond B_I (\name^* ~\mbox{'s}~\varphi)$. 
This results says information is actually flooding agents in $N$; more precisely, we can
construct a run $r^{info}$ of the full-information protocol on $N$, such that for any round $k$,
agents in $r$ eventually have at least as much information as agents in $r^{info}$ by round $k$.
If we now consider the round number $k_{{\cal N},N,f}$ corresponding to the condition in 
Theorem~\ref{thm:solvability}, an argument similar to the one made in the proof of 
Theorem~\ref{thm:solvability} suffices to show that eventually all agents know the function the value.
We have reached a contradiction.
}%

\section{Case study: leader election}\label{sec:election}
In this section we 
focus on leader election.
If we take the function $f$ to describe a method for computing a
leader, and require that all agents eventually know who is chosen as
leader, this problem becomes an instance of global function computation.
We assume that agents have distinct identifiers 
(which is the context in which leader election has been studied in the
literature).  It follows from Corollary~\ref{cor:non-anon} that leader
election is solvable in this context; 
the only question is what the complexity is.
Although leader election is only one instance of the
global function computation problem, it is of particular interest,
since it has been studied so intensively in the literature.
We show that
a number of well-known protocols for leader election in the literature
essentially implement the program $\Pgcb^{GC}$.    
In particular, we consider 
a protocol combining ideas of Lann \citeyear{L77} and Chang and Roberts
\citeyear{CR79} 
(LCR from now on) presented by  Lynch \citeyear{Lyn97},
which works in unidirectional rings, and 
Peterson's \citeyear{peterson82} protocol P1 for unidirectional rings
and P2 for bidirectional rings.
We briefly sketch the LCR protocol and 
\discin{Peterson's protocol P2, closely following Lynch's \citeyear{Lyn97} 
treatment; we omit the description of P1 for space reasons.}
\discout{Peterson's protocols P1 and P2, closely following Lynch's
\citeyear{Lyn97} treatment.}
\commentout{
because there is a sense in which 
leader election is complete
for the problem of global function computation:
More precisely:

\pro\label{pro:leaderelectioncomplete} If leader election can be solved
in a network $N$ given common 
knowledge $\N$, then the global function problem can be solved for every
function $f$ in network $N$, given common knowledge $\N$.
\epro
We remark that the converse of
Proposition~\ref{pro:leaderelectioncomplete} does not hold in general.
For example, 
if $N$ is a ring in which all agents have the same input and all links have
same weights, and if $N$ is common knowledge, then any function can be
computed, but the symmetry of the networks makes leader election unsolvable.
} 
\commentout{
The optimal flooding protocol takes a parameter $d$ 
(intuitively, an
upper bound on the network diameter), and proceeds in rounds. 
Agents keep track of the maximum id they have seen.
Initially agents send their id  to
their neighbors, and in each round, if they have heard of a value 
larger than the current maximum, they
forward it on all outgoing links.  
After $d$ rounds, 
the agent whose id is the maximum it has 
seen declares itself leader.
It is easy to see that this protocol is correct in all networks whose
diameter is bounded by $d$, since by round $d$, all agents will have
heard about the maximum id in the network, and will know that the
leader is the agent whose id is the maximum.
}

The LCR protocol works in unidirectional rings, and does not assume a
bound on their size.
Each agent starts by sending its id along the ring; whenever it
receives a value, if the value is larger 
than
the maximum value seen so far, then the agent forwards it; if not, it
does nothing, except when it receives its own id.
If this id is $M$, the agent then sends the message ``the agent with id
$M$ is the leader'' to its neighbor.  Each agent who receives such a
message forwards it until it reaches the agent with id $M$ again.
The LCR protocol is correct because it 
ensures that the maximum id travels along the ring and is 
forwarded by each agent until 
some agent receives its own id back.  That agent then
knows that its id is larger than that of any other agent, and thus
becomes the leader. 

Peterson's protocol P2 
for bidirectional rings operates in phases.
In each phase, agents are designated as either
{\em active\/} or {\em  passive}.
Intuitively, the active agents are those still competing in the
election.  Once an agent becomes passive, it remains passive, but
continues to forward messages.
Initially all agents are active. 
In each phase, an active agent compares its id with the 
ids of the  closest active agent to its right and the closest active
agent to its left.
If its id is the largest of the three, it continues to be active;
otherwise, it becomes passive.
Just as with the LCR protocol,
when an agent receives back its own id, it
declares itself leader.  
Then if its id is $M$, it sends the message ``the agent with id
$M$ is the leader'', 
which is forwarded around the ring until
everyone knows who the leader is.

Peterson shows that,
at each phase, the number of active agents is at most half that of the
previous phase, and always includes the agent with the largest id.
It follows that, eventually, the only active agent is the one with the
largest id.
Peterson's protocol terminates when the agent that has the maximum id
discovers that it has the maximum id by receiving its own id back.
The message complexity of Peterson's protocol is thus $O(n
\log n)$, where $n$ is the number of agents.
Peterson's protocol P1 for unidirectional rings is similar.
Again, passive agents forward all messages they
receive, at each round at most half of the agents remain active, and 
the agent with the largest value becomes leader. There are, however, a number
of differences.  
Agents now have ``temporary'' $\id$s as well as their own $\id$s.  
It is perhaps better to think of an agent's $\id$ as being active if it has
an ``active temporary $\id$''.    (In the bidirectional case, we can
identify the temporary $\id$ with the actual $\id$, so an agent is
active iff its $\id$ is active.)  We take a temporary 
$\id$ to be active at phase $p+1$ if it is larger than the temporary $\id$s
that precede or follow it in phase $p$.  But since messages can
only be sent in one direction, the way to discover this is for an active
agent to forward its temporary $\id$ to the following two active agents.
An active agent can then tell if the preceding active agent's temporary
$\id$ was greater than the following and preceding active temporary
$\id$'s.  If so, it remains active, and takes as its temporary $\id$ 
what was the temporary $\id$ of the preceding active agent.  Otherwise,
the agent becomes passive.  It is not hard to check that an agent is
active in the bidirectional protocol iff its $\id$ is active in the
unidirectional protocol (i.e., iff its $\id$ is the temporary $\id$ of
an active agent in the unidirectional protocol).
When an agent receives its original value, then it declares
itself leader 
and sends a message describing the result of the election around the ring.
We remark that although they all work for rings, the LCR protocol is
quite different from P1 and P2.  In the LCR protocol,
agents forward their values along their unique outgoing link.
Eventually, the agent with the 
maximum input receives its own value and realizes that it has the maximum
value.  
In P1 and P2, 
agents are either
{\em active\/} or {\em passive}; in
each round, the number of active
agents is reduced, and eventually only the agent with the maximum value 
remains active.
Despite their differences, LCR, P1, and P2 all essentially implement 
$\Pgcb^{GC}$.  There are two reasons we write ``essentially'' here.
The first, rather trivial reason is that, 
when agents send information, they do not send all
the information they learn (even if the agent they are sending it to
will never learn this information).  
For example, in the LCR protocol,
if agent $i$ learns that its left neighbor has value 
$v$
and this is the
largest value that it has seen, it passes along 
$v$
without passing along
the fact that its left neighbor has this value.  
We can easily deal with this by modifying the protocols so that all the
agents send $\newinfo$ rather than whatever message they were supposed
to send.  However, this
modification does not suffice.  The reason is that the modified
protocols send some 
``unnecessary'' messages.  This is easiest to see in the case of LCR.
Suppose that $j$ is the processor with highest id.  When $j$ receives
the message with its id 
back and sends it around the ring again (this is essentially the message
saying that $j$ is the leader), in a full-information protocol, $j$'s
second message will include the id $j'$ of the processor just before $j$.
Thus, when $j'$ receives $j$'s second message, it will not need to
forward it to $j$.  If LCR$'$ is the modification of LCR where 
each process sends $\newinfo$ rather than 
the maximum id seen so far,
and the last message
in LCR is not sent, then we can show that LCR$'$ indeed de facto
implements $\Pgcb^{GC}$.  
The modifications to P2 
that are
needed to get a protocol P2$'$ that de facto
implements $\Pgcb^{GC}$ are 
\discin{similar in spirit, although}
somewhat more complicated.  
\discin{We leave details to the full paper.}
Each processor
$i$ running P2$'$ acts as it does in P2 
(modulo sending $\newinfo$)
until the point where it first gets a
complete picture of who is in the ring (and hence who the leader is).  
What happens next depends on whether $i$ is the first to find out who
the leader is or not and whether $i$ is active or not.  
We leave details to the
Appendix~\ref{sec:le proofs}.

\commentout{
Note that it is possible 
with P1 and P2 
that some
processor other than the leader will discover who the leader is before the
leader does (by getting the same id from the left and 
right).\footnote{Consider, for example, a ring with two agents with ids
$1$ and $2$.  It is easy to see that, with Peterson's algorithm, agent
$1$ will be the first one to learn the maximum id in the ring, 
and it will be passive when it does so.}  In this 
case, the message announcing the leader is just sent as far as necessary.%
}

\commentout{
[TO BE CHANGED]
\thm
\begin{itemize}
\item[(a)] 
Given parameter $d$,
the optimal flooding protocol  
de facto
implements $\Pgcb^{GC}$
in contexts where (i) all networks have diameter at most $d$ and 
(ii) all agents have distinct identifiers. 
\item[(b)] LCR$'$ de facto implements $\Pgcb^{GC}$ in contexts 
where  (i) all networks are unidirectional rings and (ii) agents have distinct identifiers.
\item[(c)] P1$'$ de facto implements
$\Pgcb^{GC}$ in contexts where (i) all networks are
unidirectional rings and (ii) agents 
have distinct identifiers.  
\item[(d)] P2$'$  de facto implements
$\Pgcb^{GC}$ in contexts where (i) all networks are bidirectional rings
and (ii) agents 
have distinct identifiers.  
\end{itemize}
\ethm
}

\thm\label{thm:le proofs}
The following all hold:
\begin{itemize}
\item[(a)]
Given parameter $d$,
the optimal flooding protocol  
\cite{Lyn97}
de facto
implements $\Pgcb^{GC}$
in contexts where (i) all networks have diameter at most $d$ and 
(ii) all agents have distinct identifiers. 
\item[(b)] LCR$'$ de facto implements $\Pgcb^{GC}$ 
in all contexts where 
(i) all networks are unidirectional rings and (ii) agents have distinct
identifiers.
\item[(c)] 
There exists a protocol P1$'$ that agrees with P1 up to the last
phase (except that it sends
$\newinfo$) and implements $\Pgcb^{GC}$
in all contexts where (i) all networks are
unidirectional rings and (ii) agents 
have distinct identifiers.
\item[(d)] 
There exists a protocol P2$'$ that agrees with P2 up to the last
phase (except that it sends $\newinfo$) 
and de facto implements $\Pgcb^{GC}$
in all contexts where (i) all networks are
bidirectional rings and (ii) agents 
have distinct identifiers.
\end{itemize}
\ethm
\commentout{
\thm
The general algorithm for computing idempotent Boolean functions in 
anonymous networks 
proposed by Kranakis et al. \citeyear{boolcomp94} implements 
$Pg^{B\RCond, GC}$.
\ethm
}
Theorem~\ref{thm:le proofs} brings out the underlying commonality of all
these protocols.  Moreover, it emphasizes the connection between
counterfactual reasoning and message optimality.  
Finally, it shows that reasoning at the kb level can be a
useful tool for improving the message complexity of protocols.  For
example, although P2$'$ has the same order of magnitude message
complexity as P2 ($O(n \log n)$), it typically sends $O(n)$ fewer messages.
While this improvement comes at the price of possibly
longer messages, it does suggest 
that this approach can result in nontrivial improvements.  Moreover, it
suggests that 
starting with a high-level 
kb program and then trying to
implement it using a standard program can be a useful design methodology.
Indeed, our hope is
that we will be able to synthesize standard programs by starting
with high-level kb specifications, synthesizing a
kb program that satisfies the specification, and then
instantiating the kb program as a standard program.  We
have some preliminary results along these lines that give us confidence
in the general approach \cite{BCHP05}; we hope that further work will
lend further credence to this approach.

\appendix
\commentout{
\section{Proofs for Section~\ref{sec:solvable}}
\olem{lem:same info k rounds}
The following are equivalent:
\begin{itemize}
\item[(a)] $(N,i) \sim_k (N',i')$.
\item[(b)] Agents $i \in V(N)$ and $i' \in V(N')$ have the same initial
local information and 
receive the same messages
in each of
the first $k$ rounds of the full-information protocol.
\item[(c)] If the system is synchronous, then $i$ and $i'$ 
have the same initial local information and
receive the same messages in 
each of the first $k$ rounds of every deterministic protocol.
\end{itemize}
\eolem

\prf
We first prove that (a) implies (c).
Let $P$ be an arbitrary deterministic
protocol. The proof proceeds by induction, with the base
case following from the definition of $\sim_0$.
Suppose that, if $(N,i)\sim_k (N',i')$, then
$i$ and $i'$ start with the same local information and receive same 
information in each of the first $k$ rounds of protocol $P$ 
and that $(N,i)\sim_{k+1} (N',i')$. Then $(N,i)\sim_k (N',i')$,
and there exists a bijection $f^{in}:In_N(i)\longrightarrow In_{N'}(i)$
such that $(N,j)\sim_{k} (N',f^{in}(j))$ for all $j\in In_N(i)$.
>From the inductive hypothesis, it follows that $i$ and $i'$ have 
the same initial information and receive the same messages in the first $k$
rounds of $P$; 
similarly, for each $j$ incoming neighbor of $i$, $j$ and $f^{in}(j)$ have
same initial information and receive same messages in each of the first $k$
rounds of $P$. 
Hence, $j$ and $f^{in}(j)$ have the same local state at time $k$
and, since $P$ is deterministic, $j$ sends $i$ the same
messages as $f^{in}(j)$ 
sends to $i'$. Thus, $i$ and
$i'$ receive same 
messages in round $k+1$ of protocol $P$.

To prove that (c) implies (b), it suffices to notice that 
the full-information protocol is a special case of a deterministic
protocol and that,
given how we have defined rounds in an asynchronous setting, $i$
receives the same messages in round $k$ of the full-information protocol
in both the synchronous and asynchronous case.

Finally, we prove that (b) implies (a) 
by induction on $k$. 
For $k=0$, it is clear from 
Definition~\ref{def:k-bisim} that 
$(N,i)\sim_0 (N',i')$ exactly when $i$ and $i'$ have the same initial 
local information. 
For the inductive step, 
suppose that $i$ and $i'$ have the same initial local
information and receive the same messages at each round $k' \le k+1$.
We can then construct a mapping, say 
$f^{in}$, from $In_N(i)$ to $In_{N'}(i')$ such that for all 
$j\in In_N(i)$, the information that $i$ 
receives from $j$ is the same
as the information that $i'$ receives from $f^{in}(j)$
in each of the first $k+1$ rounds.
Since $j$ is following a full-information protocol, it follows that $j$
must have the same initial local information as $j'$ and that $j$ and
$j'$ receive the same messages in each of the first $k$ rounds.
By the induction hypothesis, 
$(N,j) \sim_k (N',f^{in}(j))$. 
Since
part of $i$'s 
information from $j$ is also the weight
of edge $(j,i)$, 
$f^{in}$ must preserve edge-weights. 
Thus, 
$(N,i)\sim_{k+1} (N',i')$.  
\eprf

\othm{thm:solvability}
The global function $f$ can be computed on networks in $\N$
iff, for all networks $N \in \N$, there exists a constant
$k_{{\cal N},N,f}$, such that, for all networks $N' \in \N$, all $i \in
V(N)$, and all $i' \in V(N')$, if 
$(N,i) \sim_{k_{{\cal N},N,f}} (N',i')$ then
$f(N')=f(N)$. 
\eothm

\prf
First suppose that the condition in the statement of the theorem is
satisfied.  
At the beginning of each round $k$, each agent $i$ in the network
proceeds as follows.
If $i$ received the value of $f$ in the previous round, then $i$
forwards the value to all of its neighbors and terminates; otherwise,
$i$ computes $f$'s value on all the networks $N'$ such that
there exists an $i'$  such that agent $i'$ would have received the same
messages in the first $k-1$ rounds in network $N'$ as $i$ actually
received.  (By Lemma~\ref{lem:same info k rounds}, these are just the
pairs $(N',i')$ such that $(N',i') \sim_{k-1} (N,i)$.)
If all the values are equal, then $i$ sends the value to all his neighbors
and terminates; otherwise, $i$ sends whatever new information he has received
about the network to all his neighbors.
Let $k_i$ be the first round with the property
that for all $N'\in \N$ 
and $i'$ in $N'$, if $(N,i)\sim_{k_i} (N',i')$, then $f(N')=f(N)$.
(By assumption, such
a $k_i$ exists and it is 
at most $k_{{\cal N},N,f}$.) 
It is easy to see that, by round $k_i$, $i$ learns the value of $f(N)$,
since either $i$ gets the same messages that it gets in the
full-information protocol up to round $k_i$ or it gets the function value.
Thus, $i$ terminates by the end of round $k_i+1$ at the
latest, after sending the value of $f$, and  
the protocol terminates in at most $k_{{\cal N},N,f}+1$ rounds. 
Clearly all agents learn $f(N)$ according to this protocol.

Now suppose that the condition in the theorem does not hold and, 
by way of contradiction,
that the value of $f$ can be computed by some protocol $P$ on all the
networks in $\N$.  
There must exist some network $N$ for which the condition in the theorem fails.suffice to
For now, suppose that the system is synchronous.
There must be some round $k$ such that 
all agents in $N$ have computed the function value by round $k$.  Since
the condition fails, there must exist a network $N' \in \N$ and agents
$i \in V(N)$ and $i' \in V(N')$ such that $(N,i) \sim_k  
(N',i')$ and $f(N) \ne f(N')$.  
By Lemma~\ref{lem:same info k rounds}, $i$ and $i'$ have the same
initial information and receive the same messages in the first $k$
rounds of protocol $P$.
Thus, they must output the same value for the function at round $k$. But since
$f(N) \ne f(N')$, one of these answers must be wrong, contradicting our
assumption that $P$ computes the value of $f$ in all networks in $\N$.
The same argument works in asynchronous systems; we just restrict to runs
where the messages are delivered synchronously (which is always possible
in an asynchronous system).
\eprf
}%

\commentout{
\othm{cor:non-anon}
If initially it is common knowledge that no two agents are 
locally the same, then all global functions
can be computed; indeed, we can take 
$k_{{\cal N},N,f}=\mbox{$\diam(N)+1$}$. 
\eothm

\prf
Since $f(N) = f(N')$ if $N$ and $N'$ are isomorphic, it suffices to show
that $(N,i) \sim_{\diam(N)+1}  (N',i')$ implies that $N$ and $N'$ are
isomorphic for 
all $N, N' \in \N$.  
First observe that, by an easy induction on $k$, if there is a path of
length $k \le \diam(N)$ from $i$ to $j$ in $N$, then there must exist a
node $j' \in V(N')$ such that there is a path from $i'$ to $j'$ of
length $k$ and $(N,j) \sim_{\diam(N) + 1 -k} (N',j')$.  Moreover, note
that $j'$ must be unique, since if $(N,j) \sim_{\diam(N) + 1 -k}
(N',j'')$, then $j$, $j'$, and $j''$ must be locally the same and, by
assumption, no distinct agents in $N'$ are locally the same.
Define a map $h$ from $N$ to $N'$ by taking $h(j) = j'$.  
This map is 1-1, since if $h(j_1) = h(j_2)$, then $j_1$ and $j_2$ must be
locally the same, and hence identical.  

Let $N''$ be the subgraph of
$N'$ consisting of all nodes of distance at most $\diam(N)$ from $i'$.
An identical argument shows that there is a 1-1 map $h'$ from $N''$ to $N$
such that $j'$ and $h'(j')$ are locally the same for all $j' \in
V(N'')$.  The function $h'$ is the inverse of $h$, since $h(h'(j'))$ and
$j'$ are locally the same, and hence identical, for all $j' \in V(N)$.  
Finally, we must have that $h$ is a graph isomorphism from $N$ to $N''$,
since the fact $j$ and $h(j)$ are locally the same guarantee that they
have the same labels, and if $(j_1,j_2) \in E(N)$, then $(h(j), h(j'))
\in E(N'')$ and the two edges have the same label.

It remains to show that $N' = N''$.  Suppose not.  Then there is a node
$j_1 \in V(N')$ of distance $\diam(N) + 1$ from $i'$.  Let $j_2 \in V(N)$
be such that $j_1$ is an outgoing neighbor of $j_2$ and the distance from
$i'$ to $j_2$ is $\diam(N)$.  By construction, $j_2 \in V(N'')$; by our
previous argument, there is a node $j_3 \in V(N)$ such that
$(N,j_3) \sim_1 (N',j_2)$.  Since $j_2$ and $j_3$ are locally the same,
they must have the same number of outgoing links, say $m$.  That means
that there are $m$ nodes in $N$ that have $j_3$ as an incoming neighbor,
say $i_1, \ldots, i_m$.  Thus, each of $h(i_1), \ldots, h(i_m)$, all of
which are in $N''$, must
have $j_3$ as an incoming neighbor.  But $j_3$ has only $m$ outgoing
edges, and one of them goes to $j_2$, which is not in $N''$.  This is a
contradiction.  
\eprf
}

\commentout{
\section{Protocols, systems, and contexts}\label{sec:sys}
\commentout{
For ease of exposition, we tailor the syntax that we consider 
to
the
particular 
programs that we want to write.  In
particular, since quantification is used only in a limited way, we
permit only the minimum quantification necessary in the syntax.
Given a set $\Names$ of names and a set $\Phi$ of primitive
propositions, we consider the following syntax:   
$$\varphi \: ::= \: p \: | \: \neg \varphi \: | \: \varphi \wedge 
\varphi \: | \: K_\name \varphi \: | \:  \:  \exists v  K_\name (f=v) 
\: | \: \Diamond \varphi,
\quad p \in \Phi, \: \name \in \Names. $$
}
We interpret programs in the {\em runs and systems} framework of
Fagin et 
al. \citeyear{FHMV}, adapted 
to allow for names.
We start with a  possibly infinite set $\A$ of agents.
At each point in time, only finitely 
many agents are present.
Each of these agents $i$ is in some local state $l_i$. 
The \emph{global state} of the system at a particular point is a tuple $s$ 
consisting of the local states of the agents that exist at that point.
Besides the agents, it is also convenient to assume that there is 
an {\em environment state}, which keeps track of
everything relevant to the system not included in the agents' states.
In our setting, the environment state simply describes the network.
A {\em run\/} is a function from time (which we take here to range over
the natural numbers) to global states.  Intuitively, a run describes the
evolution of the system over time.  For simplicity, 
in this paper
we assume that the
same set of agents is associated with the global state at each point in
a run.  Intuitively, agents do not enter or leave the system during the
run.  (While this is appropriate in our setting, it is clearly not
appropriate in general.  We can easily extend the framework presented
here to allow agents to enter or leave the system.)  Let $\A(r)$ denote the
agents present in run $r$.  A pair $(r,m)$ consisting of a run $r$ and
time $m$ is called a {\em point}.  If 
$i \in \A(r)$, we use $r_i(m)$ to denote agent $i$'s 
local state at the point $(r,m)$.
A {\em system\/} $\R$ consists of a set of runs.  

\commentout{
Formulas are assigned meaning at each point 
$(r,m)$ of ${\cal R}$. 
To do this, we first need to associate a meaning to each primitive
proposition 
and to each name.  Both primitive propositions and names are in general
agent-relative (for example, the truth of $\somenewinfo$ depends on the
agent, as does the agent that $\lefta$ refers to).  Thus, we interpret
formulas in {\em extended interpreted systems}, which 
are tuples $\I = (\R,\pi,\mu)$, where 
$$\pi: \R \times \IN \times \A \times \Phi \rightarrow \{\true, \false\}
\mbox{ and } \mu: \R \times \IN \times \A \times \Names \rightarrow
\A.$$  
Intuitively,
$\pi(r,m,i,p) = \true$ if $p$ is 
true at the point $(r,m)$ relative to agent $i$, while 
$\mu(r(m),i,\name) = j$
if  agent $i$ assigns name $\name$ to agent $j$ at the point $(r,m)$.
We assume that $\pi$ and $\mu$ are actually functions of the global
state, so that if $r(m) = r'(m')$ and $\A(r) = \A(r')$, then 
$\pi(r,m,i,p) = \pi(r',m',i,p)$ and $\mu(r(m),i,\name) =
\mu(r'(m'),i,\name)$. 
Finally, we assume that $\mu(r,m,i,I) = i$; this guarantees that $I$ has
the appropriate interpretation.

We give semantics to all formulas in our language in extended
interpreted systems.  
Fagin et al.~\citeyear{FHMV} give semantics to formulas 
with respect to points $(r,m)$.  Because we allow primitive propositions
(and hence all formulas) to be relative to an agent, we give semantics
with respect to {\em situations}, that is, triples $(r,m,i)$ consisting
of a point $(r,m)$ and an agent $i \in \A(r)$.
There is one further subtlety.  Normally, a
formula $K_i \phi$ is true at a point $(r,m)$ if $\phi$ is true at all
the points that $i$ considers possible.  The points that $i$ considers
possible are all the points $(r',m')$ where $i$'s local state is the
same at $(r,m)$ and $(r',m')$.  This definition implicitly assumes that
$i$ knows who he is (or, equivalently, that $i$'s name is part of $i$'s
local state).  In our setting that is not the case.  In any anonymous
ring, if there are two agents with the same local state, then agent $i$
might not know which one he is.  Thus, if $i \in \A(r)$, then at the
point $(r,m)$, $i$ considers the point $(r',m')$ possible if there
exists some agent $j \in \A(r')$ such that $r_i(m) = r'_j(m')$.
With this background, we can define the satisfaction relation as
follows:

$$
\begin{array}{lll}
({\cal I}, r,m, i) \models p & \mbox{ iff } & \pi(r,m,i,p)=\true~ \mbox{for a
primitive proposition $p \in 
\Phi$}\\
({\cal I}, r,m, i) \models \neg \varphi & \mbox{ iff } &  ({\cal I}, r,m, i) 
\not\models \varphi\\
({\cal I}, r,m, i) \models \varphi \wedge \psi & \mbox{ iff } & ({\cal I}, r,m, 
i) \models \varphi ~\mbox{and}~ ({\cal I}, r,m, i) \models \psi\\
({\cal I}, r,m, i) \models \Diamond \varphi & \mbox{ iff } & \exists m'\ge m \: 
({\cal I}, r,m', i) \models \varphi  \\
({\cal I}, r,m, i) \models K_\name \phi & \mbox{ iff } & \mbox{for all
$j, j'$ and points $(r',m')$ such that $\mu(r,m,i,\name) = j$}\\
&&\mbox{and
$r_j(m) = r'_{j'}(m')$, we have
$(\I,r',m',j') \sat \phi$.}
\end{array}
$$

So far we have not interpreted the formula $\exists v K_\name (f=v)$.
This formula makes sense only in systems appropriate for computing the
function $f$.  We also want to ensure that the formula $\somenewinfo$ is
interpreted appropriately.  We take a {\em system for computing the
global function $f$\/} to be one where the environment state 
describes the network and associates with each agent a node in the
network.  Let $N_r$ be the network associated with run~$r$.  
To capture the common knowledge of the set $\N$ of possible networks,
there will be at
least one run in the network associated with each network $N \in \N$  
(as we shall see, in a synchronous system, there will be in fact exactly
one run associated with each network $N \in \N$), and no runs associated
with networks not in $\N$.   
Since each run is
associated with a network, we can also associate a value of $f$ with
true at a point $(r,m)$ according to agent $i$ if, at all points $i$
each run $r$, namely $f(N_r)$.   We take $\exists v  K_\name (f=v)$ to be
true at a point $(r,m)$ according to agent $i$ if, at all points that
the agent $j$ with name $\name$ according to $i$
considers possible, $f$ has the same value.  Formally,
$$
\begin{array}{lll}
({\cal I}, r,m, i) \models \exists v K_\name (f=v) &\mbox{iff} &\mbox{for 
all $j, j'$ and points $(r',m')$ such that $\mu(r,m,i,\name) = j$}\\
&&\mbox{and
$r_j(m) = r'_{j'}(m')$, we have
$f(N_r) = f(N_{r'})$}.
\end{array}
$$
(Of course, we could have a richer syntax that allowed arbitrary
first-order quantification over values.  We then would not have to give
separate semantics to $\exists v K_\name (f=v)$; it would follow from
the semantics of $K_\name \phi$ and $\exists v \phi$.)
}
In a \emph{system for global function computation}, 
each agent's initial local information is encoded in the
agent's local state; it must be consistent with the environment.  For
example, if according to the environment the network is a bidirectional
ring, each agent must 
have
two outgoing edges according to its local state.
We also
assume that agents have {\em perfect recall}, so that they
keep track in their local states of everything that they have heard and
when they heard it.  
This means that, in particular, the local state of an agent encodes
whether the agent has obtained new information about the network in a
given round $k$.

We are particularly interested in systems generated by protocols.
A protocol $P_i$ for agent $i$ is a function from 
$i$'s local states to nonempty sets of actions that $i$ 
may perform.
If the protocol is deterministic, then  
$P_i(\ell)$ is a singleton for  each local state $\ell$. 
A {\em joint protocol} is a tuple $P=\{P_i: i \in \A\}$, which
consists of one protocol for each agent.

We can associate with each joint protocol $P$ a system, given
a {\em context}.  
A context describes the environment's protocol, the initial states, the
effect of actions, and the association of names with agents.  Since names
are relative to agents, we do the association using a \emph{naming
function} $\mu: \G \times \A \times \Names
\rightarrow \A$, where $\G$ is the set of global states.
Intuitively, $\mu(g,i,\name) = j$
if  agent $i$ assigns name $\name$ to agent $j$ at the global state $g$.
Thus, we take a 
context $\gamma$ to be a tuple $(P_e,\Gz,\tau,\mu)$,
where $P_e$ is a protocol for the environment,
$\Gz$ is a set of initial global states, 
$\tau$ is a {\em transition function}, and $\mu$ is a naming
function.%
\footnote{Fagin et al.~\citeyear{FHMV} also have a component of the
context that describes the set of ``allowable'' runs.  This plays a role
when considering issues like fairness, but does not play a role in this
paper, so we omit it for simplicity.
Since they do not consider names, they do not have a component $\mu$ in
their contexts.}
The environment is viewed as running a protocol just like the agents;
its protocol is used to capture, for example, when messages are
delivered in an asynchronous system.  The transition function $\tau$ 
and naming function $\mu$ determine a mapping denoted $\tau_\mu$
associating with each {\em joint action\/}
(a tuple consisting of an action for the environment and one
for each of the agents)
a {\em global state transformer}, that is, a mapping from global states 
to global states.  
Note that we need the naming function since actions may involve names.
For the simple programs considered in this paper, the
transition function will be almost immediate from the description of the
global states.

We focus in this paper on a family of contexts that we call 
{\em contexts for global function computation}.  Intuitively, the
systems that represent programs in a context for global function
computation are systems for global function computation. 
A context $\gamma^{GC} =
(P_e,\Gz,\tau,\mu)$ for global function computation 
has the following features:
\begin{itemize}
\item The environment's protocol $P_e$ controls message delivery and is
such that all messages are eventually delivered, and no messages are
duplicated or corrupted.
\item The initial global states 
are such that
the environment's state records the network $N$ 
and agent $i$'s local state records agent $i$'s initial local
information;
we use $N_r$ to denote the network in a run $r$ (as encoded by the
initial global state in $r$). 
\item the transition function $\tau_\mu$ is such that the agents keep track
of all messages sent and delivered and the set of agents does not change
over time.  That is, 
if $s$ is a global state, $\sfa$ is a joint action, and $s' =
\tau_{\mu}(\sfa)(s)$, then $\A(s) = \A(s')$ and agent $i$'s local state in
$s'$ is the result of appending all messages that $i$ sent and received
as a result of action $\sfa$ to $i$'s local state in $s$.
We assume that $\tau_\mu$ is such that the action
$\send_{\name}(\newinfo)$ has the
appropriate effect, i.e., if  
$\send_{\name}(\newinfo)$ is agent $i$'s component of a joint action
$\sfa$ and agent $i$ gives agent $j$ name $\name$ in the global state
$s$ (note here we need the assumption that the naming function $\mu$
depends only on the global state)
and $s' = \tau_{\mu}(\sfa)(s)$, then in $s'$, $j$'s local state records
the
fact that $j$ has received the information from $i$.
\end{itemize}

A run $r$ is consistent with a joint protocol $P$ if it could have been
generated when running $P$.  Formally,
run $r$ is {\em consistent with joint
protocol $P$ in context $\gamma$\/} if
its  initial global state $r(0)$ is one of
the initial global states~$\Gz$ 
given in~$\gamma$, 
and for all~$m$, the transition
from global state $r(m)$ to $r(m+1)$ is the result of performing 
one of the joint actions specified by~$P$ according to the agents in
$r$, and the environment protocol $P_e$
(given in~$\gamma$) in the global state $r(m)$.  
That is, if $P = \{P_i: i \in \A\}$ and $P_e$ is the environment's
protocol in context $\gamma$, then $r(0) \in \Gz$, and if
$r(m) = (\ell_e, \{\ell_i: i \in \A(r)\})$, then 
there must be a joint action $(\sfa_e, \{\sfa_i: i \in r(\A)\})$
such that $\sfa_e \in P_e(\ell_e)$, $\sfa_i \in P_i(\ell_i)$ for $i \in
r(\A)$, 
and $r(m+1) = \tau_\mu(\sfa_e,\{\sfa_i: i \in r(\A)\})(r(m))$ (so
that $r(m+1)$ is the result of applying the joint action
$(\sfa_e,\{\sfa_i: i \in \A\})$ to $r(m)$.
A run~$r$ is {\em consistent
with $\gamma$} if $r$ is consistent with {\em some} joint protocol~$P$
in~$\gamma$. 
A system $\R$ {\em represents\/} a joint protocol $P$ in a context
$\gamma$
if it consists of all runs consistent with $P$ in $\gamma$.
We use $\Rrep(P,\gamma)$ to denote the system representing~$P$ in 
context~$\gamma$.
We want to associate with a program a protocol.  To do
this, we need to interpret the tests in the program.  Given a set
$\Phi$ of primitive propositions, let an
{\em interpretation\/} be a function 
$\pi: \G \times \A \times \Phi \rightarrow \{\true,
\false\}$.
Intuitively, $\pi(g,i,p) = \true$ if $p$ is true at the global state $g$
relative to agent $i$.  Of course, we can easily extend this truth
assignment to arbitrary propositional formulas.

An interpretation is 
\emph{local}
(for program $\Pg$) if
the tests $\phi$ in $\Pg$ depend only on the
local state, in the sense that if $\ell$ is agent $i$'s local state in the
global state $g$ and also agent $j$'s local state in the global state
$t'$, then $\pi(g,i,\phi) = \true$ iff $\pi(g',j,\phi) = \true$.  In
this case, we write $\pi(\ell,\phi) = \true$.
Given an interpretation $\pi$ that is 
local,
we can associate with a program $\Pg$ for agent $i$ a protocol
$\Pg^\pi$.
We
define $\Pg^\pi(\ell) = \{\sfa_j \: | \: \pi(\ell,t_j) = \true\}$
if there exist tests $t_j$ such that $\pi(\ell,t_j) = \true$, and take
$\Pg^\pi(\ell) = \noop$ otherwise.
Define $\Isys(\Pg,\gamma,\pi) = \Rrep(\Pg^\pi,\gamma)$.

An {\em interpreted context for global function computation\/}
is a pair $(\gamma,\pi)$, where $\gamma$ is a context for global
function computation and $\pi$ interprets $\somenewinfo$ appropriately
(so that $\pi(g,i,$ $\somenewinfo)$ $ = \true$ if $i$ received some
new information about the network in $g$ and has not sent a message
since receiving that information).
\commentout{
We can use this machinery we have developed to give semantics to
formulas such as $K_{\name}\phi$.  
Define an \emph{extended interpreted system} to be a tuple
$\I = (\R,\pi,\mu)$, where $\R$ is a system, $\pi$ is an interpretation,
and $\mu$ is a naming function.  We can then define
$$\begin{array}{lll}
({\cal I}, r,m, i) \models K_\name \phi & \mbox{ iff, for all
$j, j'$ and points $(r',m')$ such that $\mu(r(m),i,\name) = j$}\\
&\mbox{and
$r_j(m) = r'_{j'}(m')$, we have
$(\I,r',m',j') \sat \phi$.}
\end{array}
$$
}

\commentout{
We can associate with a kb program $\Pgkb$ and an
extended
interpreted system $\I = (\R,\pi)$ a 
protocol for agent $i$ denoted  $(\Pgkb)_i^{\cal I}$.
Intuitively, we evaluate the standard tests in $\Pgkb$ 
according to~$\pi$ 
and $\mu$
and evaluate the knowledge tests according to $\I$.
Formally, for each local state $\ell$ of agent $i$, we define 
$(\Pgkb)_i^\I(\ell)$ to consist of all actions $\sfa_j$ such that the
test $t_j \land k_j$ holds with respect to a tuple $(r,m,i')$ in $\I$
such that $r_{i'}(m) = \ell$ (recall that protocols can be
nondeterministic); if there is no point in $\I$ where some agent has
local state $\ell$, then $(\Pgkb)_i^\I(\ell)$ performs the null action
(which leaves the state unchanged).

A 
joint
protocol $P$ is said to {\em implement} $\Pgkb$ in 
interpreted context $(\gamma,\pi)$ 
if, by interpreting $\Pgkb$ with respect to $\Isys(P,\gamma,\pi)$,
we get back protocol $P$; i.e., if, 
for each agent $i$, we have $P_i=(\Pgkb)^{{\cal I}(P,\gamma,\pi)}_i$.
Here we seem to be implicitly assuming that all agents run the same
kb program.  While this is true in the case of the
programs for global function computation, it is certainly not true in
general.  In fact, there is no loss of generality in assuming that all
agents run the same program.  For example, if names are commonly known,
the actions performed by agents can depend on tests of the form ``if
your name is $\name$ then \ldots''.  
Similarly, if we have a system where some agents are senders and others
are receivers, the roles of agents can be encoded in their local states,
and tests in the program can ensure that all agents act appropriately,
despite using the same program.

In certain cases we are interested in 
joint
protocols $P$ that
satisfy  
a condition slightly weaker than implementation, first defined by 
HM.
Joint protocols $P$ and $P'$
are {\em equivalent in context $\gamma$},
denoted $P \approx_\gamma P'$,
$P_i(\ell) = P'_i(\ell)$ for every local state $\ell=r_i(m)$ with
$r \in \Rrep(P,\gamma)$.
We remark that if $P \approx_\gamma P'$, then it easily follows that
$\Rrep(P,\gamma) = \Rrep(P',\gamma)$: we simply show by induction on $m$
that every prefix of a run in $\Rrep(P,\gamma)$ is a prefix of a run in
$\Rrep(P',\gamma)$, and vice versa.
We say that $P$ {\em de facto implements}
$\Pgkb$ in context $\gamma$ if 
$P \approx_\gamma \Pgkb^{{\cal I}(P,\gamma,\pi)}$.
Arguably, de facto implementation suffices for most purposes, since all we
care about are the runs generated by the protocol.  We do not care about
the behavior of the protocol on local states that never arise.
}
\commentout{
For ${\cal N}$ set of networks, a context in which all networks encoded in 
Given a set $\N$ of networks, a context in which all networks encoded in 
}

\commentout{
\subsection{Scoping}\label{sec:scope}

We focus now on assigning meaning to statements of the form ``$\name$
knows $\inewinfo$", 
taking into account
the subtleties raised by scoping, as discussed in Section~\ref{sec:b prog}. 
Intuitively, from agent $i$'s point of view, agent $\name$ knows 
$\inewinfo$ exactly when $\name$ knows that $\inewinfo$ is true about $i$. 
Since $\name$ may not know who $i$ is, this amounts to saying that
$\name$ has a name for $i$, say  
$\name^*$, and $\name$ knows that $\inewinfo$ is true about $\name^*$. 
We define a primitive proposition $\mathit{calls}(\name, \name',
\name^*)$ to be true 
at a point $(r,m)$, from agent's $i$ point of view, exactly when the agent 
that
$i$ names $\name$ calls the agent 
that
$i$ names $\name'$ as $\name^*$.
In other words, assuming that $i$ names agent $j$ $\name$, and agent
$k$ $\name'$, then $j$ calls $k$ $\name^*$. This is formally translated
into the following rule: $({\cal I}, r,m,i)\models \mathit{calls}(\name, \name', \name^*) 
~\mbox{iff}~
\mbox{for}~\mu(r(m),i,\name)=j, ~\mbox{and}~ \mu(r(m),i,\name')=k, 
~ \mu(r(m),j,\name^*)=k$.

Consider for example that $\inewinfo$ is $(val_{\name^*}=x)$.
Then $i$'s statement ``agent $\name$ knows $\inewinfo$''
is correct if the following statement made by $j$, the agent that $i$ names $\name$,
is correct, assuming $j$ calls $i$  $\name'$:
``the value of the agent $\name'$ names $\name^*$ is $x$''. 
As discussed by Halpern and Grove,
we can refer to ``the agent $\name'$ names $\name^*$'' as the agent named $\name^* \circ \name'$.
Here $\circ$ is a binary operation on names, and the naming function defined on
combinations of names satisfies the following property:
\begin{itemize}
\item $\mu(r,m,\name_k \circ \dots \circ \name_1, j) \stackrel{def}{=} 
\mu(r,m,\name_k \circ \dots \circ \name_2, \mu(r,m,\name_1, j))$.
\end{itemize} 
For example, $\mu(r,m,\name^* \circ \name',j)=\mu(r,m,\name^*,\mu(r,m,\name',j))=
\mu(r,m,\name^*,i)$: indeed, the agent that $j$ names $\name^*\circ \name'$ is the agent that
$i$ names $\name^*$. 

For an arbitrary agent $i$, formula $\varphi$ and name $\name$, we write $(\name ~\mathit{'s}~\varphi)$
to refer to $i$ making the statement $\varphi$ about $\name$. As suggested by the above example, 
$(\name ~\mathit{'s}~\varphi)$ can be obtained from
$\varphi$ by recursively substituting any occurrence of a name $\name_k \circ \dots \circ \name_1$ with
$\name_k \circ \dots \circ \name_1 \circ \name$.
More formally:
\begin{itemize}
\item if $p\in \Phi$, then $(\name ~\mathit{'s}~p)$ is obtained by substituting any
name (or combination of names) $\name'$ with $\name' \circ \name$;
for example,  if $p=(val_{\name_2\circ \name_1}=x)$, 
then $\name ~\mathit{'s}~ p \stackrel{def}{=}( \val_{\name_2 \circ \name_1 \circ \name} =x)$
\item $\name ~\mathit{'s}~(\neg\varphi) \stackrel{def}{=} \neg (\name ~\mathit{'s}~\varphi)$
\item $\name ~\mathit{'s}~(\varphi_1 \wedge \varphi_2) \stackrel{def}{=} 
(\name ~\mathit{'s}~\varphi_1) \wedge (\name ~\mathit{'s}~\varphi_2)$
\item $\name ~\mathit{'s}~ (\odot \varphi)\stackrel{def}{=} \odot (\name ~\mathit{'s}~ \varphi)$, and
\item $\name ~\mathit{'s}~ (\Diamond \varphi) \stackrel{def}{=} \Diamond (\name ~\mathit{'s}~ \varphi)$.
\end{itemize}

We are now able to give the intended interpretation to the statement ``$\name$ knows $\inewinfo$".
We say that $i$'s statement ``$\name$ knows $\inewinfo$" is true at some point $(r,m)$ when the following
conditions are all satisfied:
   \begin{enumerate}
   \item $\mu(r,m,i,\name)$ is a singleton, and let $j=\mu(r,m,i,\name)$
   \item there exists a name $\name^*$ such that $({\cal I}, r,m,i)\models \mathit{calls}(\name,I,\name^*)$
   \item $({\cal I},r,m,i)\models K_{\name}(\name^* ~\mathit{'s}~\varphi)$.
   \end{enumerate}
This also suggests a compact syntax for our formula as 
$\exists \name^* \mathit{calls}(\name,I,\name^*)\wedge K_{\name} (\name^* ~\mathit{'s}~ \varphi)$.

There are a number of subtleties to be addressed. 
First, the need to quantify over names arises from the lack of {\em standard} names: given an
arbitrary network $N$, there is not always a unique way some agent $j$ may refer to another agent $i$.
Even if we assume, as Yamashita and Kameda \citeyear{YK99}, that an agent $j$ with degree $d$
names his neighbors $\#1$,\dots, $\#d$, the distribution of names can be done arbitrarily; 
furthermore, $i$ is not initially aware of $j$'s particular choice of distributing names.

Second, it could be that $j$ uses more than one name to refer to agent $i$, and if
$\name'$ and $\name^*$ are two such names, then whatever $j$ knows about 
$\name'$ he must also know about $\name^*$. Subsequently, we modify  condition 3 above 
as follows:
\begin{enumerate}
\item[3'] for all $\name^*$ such that $({\cal I}, r,m,i)\models \mathit{calls}(\name,I,\name^*)$,
          $({\cal I},r,m,i)\models K_{\name}(\name^* ~\mathit{'s}~\varphi)$.
\end{enumerate}

Third, it makes sense to impose  that
if two agents are indistinguishable, then they use the same names to talk about 
other agents; in other words, whenever 
 $r'_{j'}(m')=r_j(m)$, $j$ assigns some name $\name'$ to an agent iff
so does $j'$: for all names $\name'$, $\mu(r,m,j,\name')$ is defined (and 
singleton) iff $\mu (r',m',\name',j')$ is defined (and singleton). However, 
it may well be that $\mu(r,m,j,\name')$ and $\mu(r',m',j',\name')$ are 
different agents. 

Furthermore, as initially agents may not know the number of incoming links, 
it could be that at the time $i$ makes the statement ``$\name$ knows $\inewinfo$",
$i$'s neighbor $j$ is not aware of $i$'s existence; subsequently, $j$ does not have a name for $i$, and condition 1 above is not satisfied. However, $j$ will find out that he has an incoming neighbor
the moment he receives information from $i$. 
It also makes sense to assume that, once an agent $j$ has a way of naming his incoming
neighbor $i$, he will always keep at least one such name to refer to $i$; in other words,
for any run $r$, if at time $m$ there exists $\name'$ such that $\mu(r,m,j,\name')=i$, then
there exists a name $\name^*$ such that $\mu(r,m',j,\name^*)=i$ for all $m'\ge m$.
This suggests that the correct interpretation of 
$i$'s statement ``$\name$ knows $\inewinfo$" is ``eventually $\name$ will have a way of referring to $i$,
and then $\name$ will know $\inewinfo$ about $i$":
$\Diamond (\exists \name^* \mathit{calls}(\name,I,\name^*)\wedge K_{\name} (\name^* ~\mathit{'s}~ \inewinfo))$.

Without getting into details, we denote $\mu^{GC}$ a naming function that satisfies
the properties discussed above, and write $\gamma^{GC}$ to denote a context with naming
function $\mu^{GC}$.

\commentout{
Recall that a context $\gamma$ is a tuple $({\cal G}_0, P_e, \tau, 
\Psi)$ of environment protocol $P_e$, initial global states ${\cal G}_0$, 
transition function $\tau$ and set of admissible runs $\Psi$. We impose 
that ${\cal G}_0$ be such that for any global initial state $s=(s_e, 
(s_i)_{i \in {\cal A}(s)})$, the environment's state records the network 
$N$ and the function $f$ to be computed, and $s_i$s contain information 
about initial local inputs, edges, and weights. More specifically, if 
$n$ is the size of the graph encoded in $s_e$, then ${\cal A}(s)=\lbrace 
1,2, \dots,n\rbrace$ and for each $i$ from 1 to $n$, $s_i$ contains 
$i$'s input value, the set of edges adjacent to $i$, their orientation and 
weights, as specified by the network encoded in $s_e$. 
$s_i$s also encode information about edge labeling in case of multiple 
links with same weight.

The protocol of the environment $P_e$ should be such that no messages 
are duplicated or corrupted and the only admissible runs (i.e., runs in 
$\Psi$) are those in which communication is reliable. Technically, the 
environment $\gamma$ is such defined that the systems generated in 
$\gamma$ are {\em message-passing systems}, meaning that the events of 
sending and receiving messages are all recorded in the local states. Further 
more, we assume  that the local states of agents at time $m+1$ encode 
at least as much information as at time $m$, which ensures that once 
agents know a fact, they do not forget it; formally, this is the {\em 
perfect-recall} semantics for message-passing systems, as discussed by 
Fagin et al. \citeyear{FHMV}

As specified in the introduction, the set of agents is fixed, as no 
agents may leave or join the network. Further more, the implied assumption 
is that this fact is common knowledge among the agents in the system. 
We can formally capture this assumption by constraining the transition 
function $\tau$ to maintain the set of agents, i.e., for any state $s$ 
and joint action $a$, if $s'=\tau(a)(s)$, then ${\cal A}(s')={\cal 
A}(s)$. 

As the network is simple, we must constrain naming functions such that 
no agent uses the same name $\#n$ for different agents; we write this 
formally as $\mu(s,\#n)(i)\neq \mu(s,\#n')(i)$  and $i \not \in 
\mu(s,\#n)(i)$ for all initial global states $s$, agents $i \in {\cal A}(s)$ 
and $\#n$ and $\#n'$ names as described above. Additionally, just by 
letting time pass, the identity of the agents named as $\#n$ by some agent 
does not change; we impose that for any state $s$, joint action $a$, if 
$s'=\tau(a)(s)$ and $i$ is some agent in $s'$, then for any name of the 
form $\#n$, $\mu(s',\#n)(i)=\mu(s,\#n)(i)$. Other names may change; for 
example, in the case of leader election, it makes sense to have a 
special name $leader$ or $candidates$, and as expected, at different points 
along a run these names may refer to different agents.

Naming  functions and contexts with the above properties are denoted as 
$\mu^{GC}$, respectively $\gamma^{GC}$, as in global computation. 
Additional assumptions about the level of knowledge in the system can be 
expressed by adding corresponding constraints to $\gamma^{GC}$. 
In general, we can model the fact that initially it is common knowledge 
that the network is in some
set ${\cal N}$ by analyzing global computation in the context 
$\gamma^{GC,{\cal N}}$ such that all the networks
encoded in the initial local states are in ${\cal N}$.
For example, if we assume that the network is a ring and that it is 
common knowledge that the network is a ring, then we focus only on 
contexts where the graph encoded in any initial environment state is a ring; 
we call this context $\gamma^{GC, ring}$. Similarly, if we assume that 
the topology $G$ of the network is common knowledge, then we focus on 
contexts where all initial environment states encode the same graph $G$; 
we call this context $\gamma^{GC,G}$. In general, for any property of 
the network expressed as a $Pr$ predicate on $Net$, analyzing the 
problem of global computation when $Pr(N)$ is common knowledge amounts to 
restricting attention to contexts in which the networks encoded in initial 
states $s_e$ satisfy $Pr$; we denote this environment as $\gamma^{Pr}$.
An important special case of contexts refers to  settings in which it 
is common knowledge among agents that 
no two agents are locally the same; we  call these contexts 
$\gamma^{GC,distinct}$. If in addition agents also
know their identity, as is the case of kb programs 
proposed by Fagin et al. \citeyear{FHMV,FHMV94}
in addition to properties of $\gamma^{GC}$, states also encode agents' 
identities, which formally amounts  
to two extra conditions:
\begin{itemize}
\item for all initial global states $s=(l_e, (l_i)_{i \in {\cal A}(s)}) 
\in {\cal G}_0$, $l_i$ encodes $i$
\item for any state $s$ and global action $a$, for any state 
$s'=\tau(a)(s)$, with 
$s'=(l'_e, (l'_i)_{i \in {\cal A}(s)}) \in {\cal G}_0$, $l'_i$ encodes 
$i$.
\end{itemize}
}
}
}
\commentout{
\section{Semantics of kb programs}\label{sec:apx:kb-progs}
We can use the machinery we have developed to give semantics to
formulas such as $K_{\name}\phi$.  
Define an \emph{extended interpreted system} to be a tuple
$\I = (\R,\pi,\mu)$, where $\R$ is a system, $\pi$ is an interpretation,
and $\mu$ is a naming function. As mentioned before, the statement $K_{\name}\varphi$ is subjective:   
it may be false or true, depending on the agent making the statement;
this means that we interpret $K_{\name}\varphi$ with respect to a point $(r,m)$ and an agent $i$ in $r$.
(We call a tuple $(r,m,i)$ a {\em situation}.)
Intuitively, if the agent 
that
$i$ names $\name$ is $j$, then $i$'s statement ``$\name$ knows $\varphi$'' is true
if $\varphi$ holds in all situations $j$ considers possible, i.e., in all situations where it has the same 
local state as in $(r,m)$. Notice that, since $j$ may be uncertain about his own identity, 
these are exactly the situations $(r',m',j')$ such that $j$ has the same local state in $(r,m)$ as $j'$ in $(r',m')$.
We can then define
$$\begin{array}{lll}
(\I, r,m, i) \models K_\name \phi & \mbox{ iff, for all
$j, j'$ and points $(r',m')$ such that $\mu(r(m),i,\name) = j$}\\
&\mbox{and
$r_j(m) = r'_{j'}(m')$, we have
$(\I,r',m',j') \sat \phi$.}
\end{array}
$$

We can associate with a kb program $\Pgkb$ and an
extended
interpreted system $\I = (\R,\pi,\mu)$ a 
protocol for agent $i$ denoted  $(\Pgkb)_i^{\cal I}$.
Intuitively, we evaluate the standard tests in $\Pgkb$ 
according to~$\pi$ 
and $\mu$
and evaluate the knowledge tests according to $\I$.
Formally, for each local state $\ell$ of agent $i$, we define 
$(\Pgkb)_i^\I(\ell)$ to consist of all actions $\sfa_j$ such that the
test $t_j \land k_j$ holds with respect to a tuple $(r,m,i')$ in $\I$
such that $r_{i'}(m) = \ell$ (recall that protocols can be
nondeterministic); if there is no point in $\I$ where some agent has
local state $\ell$, then $(\Pgkb)_i^\I(\ell)$ performs the null action
(which leaves the state unchanged).

A 
joint
protocol $P$ is said to {\em implement} $\Pgkb$ in 
interpreted context $(\gamma,\pi)$ 
if, by interpreting $\Pgkb$ with respect to $\Isys(P,\gamma,\pi)$,
we get back protocol $P$; i.e., if, 
for each agent $i$, we have $P_i=(\Pgkb)^{{\cal I}(P,\gamma,\pi)}_i$.
Here we seem to be implicitly assuming that all agents run the same
kb program.  
This is certainly true for the programs we give for global function
computation, and actually does not result in any loss of generality.
For example, if names are commonly known,
the actions performed by agents can depend on tests of the form ``if
your name is $\name$ then \ldots''.  
Similarly, if we have a system where some agents are senders and others
are receivers, the roles of agents can be encoded in their local states,
and tests in the program can ensure that all agents act appropriately,
despite using the same program.

In certain cases we are interested in 
joint
protocols $P$ that
satisfy  
a condition slightly weaker than implementation, first defined by 
HM.
We say that joint protocols $P$ and $P'$
are {\em equivalent in context $\gamma$},
denoted $P \approx_\gamma P'$,
$P_i(\ell) = P'_i(\ell)$ for every local state $\ell=r_i(m)$ with
$r \in \Rrep(P,\gamma)$.
We remark that if $P \approx_\gamma P'$, then it easily follows that
$\Rrep(P,\gamma) = \Rrep(P',\gamma)$: we simply show by induction on $m$
that every prefix of a run in $\Rrep(P,\gamma)$ is a prefix of a run in
$\Rrep(P',\gamma)$, and vice versa.
We say that $P$ {\em de facto implements}
$\Pgkb$ in context $\gamma$ if 
$P \approx_\gamma \Pgkb^{{\cal I}(P,\gamma,\pi)}$.
Arguably, de facto implementation suffices for most purposes, since all we
care about are the runs generated by the protocol.  We do not care about
the behavior of the protocol on local states that never arise
when we run the protocol.

\commentout{
\section{Proof of correctness for $Pg^{GC}$}
\othm{thm:kb correct}
$\Pg^{GC}$  solves the global function computation problem whenever 
possible.
That is, if $\N$ and $f$ satisfy the condition in
Theorem~\ref{thm:solvability}, then
with $\Pg^{GC}$ eventually all agents know the function value;
otherwise, no agent ever knows the function value. 
\eothm

\prf
Suppose that $\N$ and $f$ satisfy the condition in
Theorem~\ref{thm:solvability}.
Let $(\gamma,\pi)$ be an interpreted context for global function computation
such that in all 
initial states the network encoded in the environment state is in $\N$, 
let $\I=\Isys(\Pg^{GC,\pi},\gamma,\pi)$.
let $r$ be a run in $\I$,
and 
let
$N_r$ be the network associated with run~$r$.

We first show that at some point in $r$, some
agent knows $f(N_r)$.  Suppose not.
Let $r'$ be the unique run of the full-information protocol 
starting with the same initial global state as $r$.
We show by induction on $k$ that there is a time $m_k$ such that, at
time $(r,m_k)$, all the agents in $\A(r)$ have at least as much information 
about the network as they do at the beginning of round $k$ in $r'$.
The base case is immediate: we can take $m_1 = 0$ since, by assumption,
agents in $r$ and $r'$ start with the same initial states.
For the inductive step, 
suppose  that $i$ learns some
new information from $j$ in round $k$ of $r'$.
That means $j$ knew this information at the beginning of round $k$ in
$r'$ so, by the induction hypothesis,  $j$ must have known this
information by time $m_k$ in $r$.   
Thus, there is a time $m_k' \le m_k$ such that $j$ first learns this
information in run $r$ (where we take $m_k' = 0$ if $k=1$).
It follows that $(\I,r,m_k',j)\models \somenewinfo$. 

Thus, $j$ sends this information to $i$ at time $m_k'$ in $r$.  Since we
have assumed that communication is reliable, $i$ learns it by some time
$m_k''$.  Since $i$ has only finitely many neighbors and there are only
finitely many pieces of information about the network, there must be a
time in $r$ by which $i$ learns all the information that it learns by
the beginning of round $k+1$ in $r'$.  And since there are only finitely
many agents in $\A(r)$, there must be a time $m_{k+1}$ by which all the
agents in $\A(r)$ learn all the information about the  network that they
know at the beginning of round $k+1$ in $r'$.  

By Theorem~\ref{thm:solvability}, there exists a round 
$k_{\N,N_r,f}$ such that, running the full-information protocol, 
for all networks $N' \in \N$, all $i' \in V(N')$, and all $i \in V(N_r)$,
we have that if $(N_r,i) \sim_{k_{\N,N_r,f}} (N',i')$, then $f(N_r) = f(N')$. 
Suppose that $i$ is an agent in $N_r$, $r'\in \Rp(P,\gamma^{GC})$, and
$i'$ is
an agent in $N_{r'}$ such that $r_i(m_{k_{{\cal N},N_r,f}})=r'_{i'}(m')$.
A straightforward argument now shows that 
$(N_r,i) \sim_{k_{\N,N_r,f}} (N_{r'},i')$.  (Formally, we show by
induction on $k$ with a subinduction on $k'$ that if $k \le
k_{\N,N_r,f}$, $k' \le k$, and $j$ is 
an agent at distance  $k'$
 from $i$ in $N_r$, then there exists an agent $j'$ of distance $k'$ from
$i'$ in $N_{r'}$ such that  $(N_r,i) \sim_{k-k'} (N_{r'},i')$, and
similarly switching the roles of $i$, $i'$, $N_r$, and $N_{r'}$.)
It follows that $i$ knows $f(N_r)$ by time $m_{k_{{\cal N},N_r,f}}$ in
$r$, contradicting the assumption that no agent learns $f(N_r)$.

Suppose that $i$ is the first agent to learn the function value in $r$, 
and does so at time $m$ (or one of the first, if there are several
agents that learn the function value at time $m$).
We can now use the same argument as above to show that eventually all
agents learn the function value.  A formal proof proceeds by induction
on the distance of agent $j$ from $i$ in $N_r$; we omit details here.
\eprf
}%
}

\section{Counterfactual belief-based programs with names}\label{sec:apx:cbb-progs} 
The standard approach to giving semantics to counterfactuals
\cite{Lewis73,Stalnaker68} is that $\phi \RCond \psi$ is true at a point
$(r,m)$ if $\psi$ is true at all the points ``closest to'' or ``most
like'' $(r,m)$ where $\phi$ is true.  For example, suppose 
that we have
a wet match and we make a statement such as ``if the match were dry then
it would light''.  Using $\rimp$ this statement is trivially true, since
the antecedent is false.  However, with $\RCond$, we must consider the
worlds most like the actual world where the match is in fact dry and
decide whether it would light in those worlds.  If we think the match is
defective for some reason, then even if it were dry, it would not light.

To capture this intuition in the context of systems, we 
extend HM's approach so as to deal with names.  
We just briefly review the relevant details here; we encourage the
reader to consult \cite{HM98a} for more details and intuition.
Define an {\em order assignment\/} for an extended interpreted
system~$\cI=(\R,\pi,\mu)$ to be a 
function~$\lta$ that associates with every situation~$(r,m,i)$ a
partial order relation~$\lta_{(r,m,i)}$ over situations.
The partial orders must satisfy the constraint that $(r,m,i)$ is a 
minimal element of~$\lta_{(r,m,i)}$, so that there is no situation
$(r',m',i')$ such that 
$(r',m',i')\,\lta_{(r,m,i)}(r,m,i)$.  
Intuitively, $(r_1,m_1,i_1) \lta_{(r,m,i)} (r_2,m_2,i_2)$ if 
$(r_1,m_1,i_1)$ is ``closer'' to the true situation $(r,m,i)$ than
$(r_2,m_2,i_2)$.
A {\em counterfactual system} is a pair of the form 
$\cis=(\cI,\lta)$, 
where $\cI$ is an extended interpreted system and~$\lta$ is an order 
assignment for the situations in $\cI$.

Given a counterfactual system $\cis = (\I,\lta)$,
a set $A$ of situations, and a situation $(r,m,i)$, we define
the situations in $A$ that are closest to $(r,m,i)$, denoted 
$\closest(A,r,m,i)$, by taking
$$\begin{array}{ll}
\closest(A,r,m,i) = \\
\qquad\lbrace (r', m',i') \in A :
\mbox{ there is no situation } (r'',m'',i'') \in A \\
\qquad \mbox{such 
that } (r'',m'',i'') \lta_{(r,m,i)} (r',m',i') \rbrace.
\end{array}$$ 

A counterfactual formula is assigned meaning with respect to a 
counterfactual system $\cis$ by interpreting all formulas not 
involving $\RCond$ with respect to ${\cal I}$ using the earlier
definitions, and defining
$$(\cis, r,m,i)\models \varphi \RCond \psi~\mbox{iff}~ 
\mbox{for all }
(r',m',i') \in 
\closest(\intension{\phi}_{\cis}, r,m,i), ~
(\cis, r', m',i') \models \psi,$$
where 
$\intension{\phi}_{\cis}  = \{(r,m,i): (\cis,r,m,i) \sat \phi\}$; that is,
$\intension{\phi}_\cis$ consists of all situations in $\cis$ satisfying
$\phi$.

All earlier analyses of  (epistemic) properties of a protocol $P$ in a 
context $\gamma$ used the runs in $\Rrep(P,\gamma)$, that is,
the runs consistent with~$P$ in context~$\gamma$.
However, counterfactual reasoning involves events that occur on runs that
are not consistent with $P$ (for example, we may need to
counterfactually consider the run where a certain
message is not sent, although $P$ may say that it should be sent).
To support such reasoning, we need to consider runs not in
$\Rrep(P,\gamma)$.
The runs that must be added can, in  
general, depend on the type of counterfactual statements allowed in 
the logical language. Thus, for example, if we allow formulas of 
the form $\doact(i,\sfa)\RCond\psi$ for process~$i$ and action~$\sfa$, then 
we must allow, at every point of the system, a possible future in which~$i$'s
next action is~$\sfa$.
Following \cite{HM98a}, we do reasoning with respect to the
system $\Rp(\gamma)$ consisting of {\em all} 
runs compatible with $\gamma$, that is, all runs consistent with some
protocol $P'$ in context $\gamma$.

We want to define an order assignment in the system
$\Rp(\gamma)$ that ensures that the counterfactual tests in
$\Pgcb^{GC}$, which have an antecedent $\neg
\doact(\send_{\name}(\msg)$, get interpreted appropriately.
HM defined a way of doing so for counterfactual tests whose antecedent
has the form $\doact(i,{\sfa})$.  We modify their construction here.
Given a context $\gamma$, situation $(r,m,i)$ in
$\Rp(\gamma)$, action $\sfa$, and 
a deterministic
protocol $P$,%
\footnote{We restrict in this paper to deterministic protocols.  We can
generalize this definition to randomized protocols in a straightforward
way, but we do not need this generalization for the purposes of this
paper.} 
we define the closest set of situations to $(r,m,i)$ where $i$ does
\emph{not} perform action $\send_{\name}(\msg)$,
\commentout{
define 
$\close(\sfa,P,\gamma,r,m,i) =
\{(r',m,i'):$ (a) $r' \in \Rp(\gamma)$,
(b) $r'(m') = r(m')$ for all $m' \le m$, 
(c) if agent $i$ performs $\sfa$ according to $P$ in local state
$r_i(m)$, then $r' = r$
and $i = i'$,
(d) if agent~$i$ does not perform 
$\sfa$ in local state $r_i(m)$ in run $r$, 
then $r'_{i'}(m) = r_i(m)$, $i'$ 
performs $\sfa$ in local state $r_i(m)$ in run~$r'$, and follows~$P$ 
in all other local states in run $r'$,
(e) all agents other than $i'$ follow $P$ 
at all points of $r'\}$.
That is, $\close(\overline{\send_{\name}(\msg)},P,\gamma,r,m,i)$ is 
the set of situations $(r',m,i')$ where 
$(r',i')=(r,i)$ if $i$ performs
otherwise, $r'$ 
is identical to $r$ up to time $m$ and all the agents act according to
$P$ at
later times, except that 
at the local state $r'_{i'}(m) = r_i(m)$ in $r'$,
agent $i'$ who is indistinguishable from $i$ performs action~$\sfa$.%
}
$\close(\overline{\send_{\name}(\msg)},P,\gamma,r,m,i)$,  as 
$\{(r',m,i'):$ 
(a) $r' \in \Rp(\gamma)$,
(b) $r'(m') = r(m')$ for all $m' \le m$, 
(c) if agent $i$ performs some action $\send_A(\msg')$ according to $P$
in local state 
$r_i(m)$
and $\name \notin A$ or $\msg' \ne \msg$,  
or if $i$ does not perform action $\send_A(\msg')$ for any set $A$
of agents and message $\msg'$,
then $r' = r$ and $i = i'$,
(d) if agent $i$ performs $\send_A(\msg)$ according to $P$
in local state $r_i(m)$
and $\name \in A$, then 
$i$ performs $\send_{A- \{\name\}}(\msg)$ in local state $r_i(m)$ in
run~$r'$, and follows~$P$  
in all other local states in run $r'$,
(e) all agents other than $i'$ follow $P$ at all points of $r'\}$.
That is,
$\close(\overline{\send_{\name}(\msg)},P,\gamma,r,m,i)$ is 
$\{r,m,i\}$ if $i$ does not
send $\msg$ to $\name$ at the local state $r_i(m)$; otherwise 
$\close(\overline{\send_{\name}(\msg)},P,\gamma,r,m,i)$ 
is the set consisting of situations $(r',m,i')$ such that $r'$
is identical to $r$ up to time $m$ and all the agents act according to
$P$ at 
later times, except that at the local state $r'_{i'}(m) = r_i(m)$ in $r'$,
agent $i'$ who is indistinguishable from $i$ does 
not send $\msg$ to $\name$, but does send it to all other agents to
which it sent $\msg$ in $r_i(m)$.

Define an {\em order generator $\ogen$\/} to
be a function that associates with every protocol~$P$ 
an order assignment $\lta^P=\ogen(P)$ on the situations of
$\Rp(\gamma)$.
We are interested in order generators that prefer runs in which agents 
follow their protocols as closely as possible.  
An order generator~$\ogen$ for~$\gamma$ 
{\em respects protocols\/} if, for every (deterministic) protocol~$P$, 
interpreted context $\zeta = (\gamma,\pi)$
for global computation,
situation  $(r,m,i)$ in~$\Rrep(P,\gamma)$, and action $\sfa$,
$\closest(\intension{\neg \send_A(\msg)}_{\Isys(P,\zeta)},r,m,i)$ 
is a nonempty subset of
$\close(\overline{\send_\name(\msg)},P,\gamma,r,m,i)$ that 
includes $(r,m,i)$
if $(r,m,i) \in \close(\overline{\send_A(\msg)},P,\gamma,r,m,i)$.
Perhaps the most obvious order generator that respects protocols just
sets $\closest(\intension{\neg \send_\name(\msg)}_{\Isys(P,\zeta)},$ $r,m,i) = $
$\close( $ $\overline{\send_\name(\msg)},P,\gamma,r,m,i)$, although our
results hold 
if $=$ is 
replaced by $\subseteq$.

Reasoning in terms of the large set 
of runs $\Rp(\gamma)$ as opposed to 
$\Rrep(P,\gamma)$
leads to agents not knowing properties of $P$.
For example, even if, according to $P$, some agent $i$ always performs 
action $\sfa$ when in local state $l_i$, 
in $\Rp(\gamma)$ 
there are bound to 
be
runs $r$ and times $m$ such that 
$r_i(m)=l_i$, but $i$ does not perform action $\sfa$ at the point $(r,m)$. 
Thus, when we evaluate knowledge with respect to $\Rp(\gamma)$, 
$i$  no longer knows that, according to $P$, he performs $\sfa$ in state
$l_i$.  
Following HM, 
we deal with this by adding extra information to the models that allows
us to capture the agents' beliefs.  Although the agents will not {\em
know\/} they are running protocol $P$, they will {\em believe\/} that
they are.
We do this by associating with each run $r\in
\Rp(\gamma)$ a {\em rank}~$\rank(r)$, which is either a natural number
or~$\infty$, 
such that $\min_{r \in \Rp(\gamma)} \rank(r) = 0$.
Intuitively, the rank of a run defines the likelihood of the run. Runs
of rank~0 are most likely; runs of rank~1 are somewhat less likely, those
of rank~2 are even more unlikely, and so on. 
Very roughly speaking, if~$\epsilon>0$ is small, we can think of 
the runs of rank~$k$ as having probability $O(\epsilon^k)$.
We can use ranks to define a notion of belief (cf.~\cite{FrH1Full}).

Intuitively, of all the points considered possible 
by a given agent in a situation $(r,m,i)$, the ones believed to have
occurred are the ones appearing in runs of minimal rank. 
More formally, for a point $(r,m)$ define 
\[\mini^\rank(r,m)~=~\min\{\rank(r')\,|\, r'\in\Rp(\gamma)~{\rm and} 
~r'_{i'}(m')=r_i(m) \mbox{~for some $m'\ge 0$ and $i' \in \A(r')$}\}.\]
Thus, $\mini^\rank(r,m)$ is the minimal $\rank$-rank 
of runs $r'$ in which~$r_i(m)$ appears as a local state at the point $(r',m)$.

A {\em counterfactual belief system} (or just {\em cb} system for short)
is a triple of the form 
$\J=(\cI,\lta,\rank)$, where $(\cI,\lta)$ is a counterfactual system, 
and $\rank$ is a ranking function on the runs of~$\I$. 
In cb systems we can define a notion of belief.
We add the modal operator $B_\name$ to the language for each $\name \in
\Names$,
and define 
\[\begin{array}{lll}
(\I,\lta,\rank,r,m,i)\sat B_\name\phi~ \mbox{ iff, } 
 \mbox{for all
$j, j'$ and points $(r',m')$ such that $\mu(r,m,i,\name) = j$},\\
\qquad \qquad \qquad \qquad \mbox{$r_j(m) = r'_{j'}(m')$, and
$\rank(r')=\minj^\rank(r,m)$, we have
$(\I,r',m',j') \sat \phi$.}\\
\end{array}
$$

The following lemma illustrates a key feature of the definition of belief.
What distinguishes knowledge from belief is that knowledge satisfies the
{\em knowledge axiom}: $K_i \phi \rimp \phi$ is valid.   While $B_i \phi
\rimp \phi$ is not valid, it is true in runs of rank~0.

\lem\label{knowledge-axiom}{\rm \cite{HM98a}}
Suppose that $\J=(\R,\pi,\mu,\lta,\rank)$ is a cb system, $r \in
\R$, and $\rank(r) = 0$.
Then for every formula~$\phi$ and all times~$m$, we have
$(\J,r,m,i)\sat B_I\phi\rimp\phi$.
\elem

By analogy with order generators, we want a uniform way of
associating with each protocol $P$ a ranking function.  Intuitively, we
want to do this in a way that lets us recover $P$.
We say that a ranking function~$\rank$ is {\em $P$-compatible} 
(for~$\gamma$) if $\rank(r)=0$ if and only if $r\in\Rrep(P,\gamma)$.
A {\em ranking generator} for a context~$\gamma$ is a
function~$\rgen$ ascribing to every protocol~$P$ a ranking
$\rgen(P)$ on the runs of $\Rp(\gamma)$.  
A ranking generator~$\rgen$ is {\em deviation compatible\/} 
if $\rgen(P)$ is $P$-compatible for every protocol~$P$. 
An obvious example of a deviation-compatible ranking generator is the 
{\em characteristic} ranking generator $\rgen_\xi$,
where $\rgen_\xi(P)$ is the ranking that
assigns rank $0$ to every run in~$\Rrep(P,\gamma)$ and 
rank~$1$ to all other runs. 
This captures the assumption that runs of~$P$ are likely 
and all other runs are unlikely, without attempting to distinguish 
among them. 
Another deviation-compatible ranking generator is $\rgend$, 
where $\rgend(P)$ is the ranking that assigns to a run~$r$ the total number
of times that agents deviate from~$P$ in~$r$. 
Obviously, $\rgend(P)$ assigns~$r$ the rank~0 exactly if 
$r\in\Rrep(P,\gamma)$, as desired. 
Intuitively, $\rgend$ captures the assumption that not only are
deviations unlikely, but they are independent.

It remains to give semantics to the formulas 
$\exists \name' (\Calls(\name,I,\name') \land 
B_\name (\name'\mbox{'s} \phi))$
and $\exists v B_{\name}(f=v)$.  Recall that we want
$\exists \name' (\Calls(\name,I,\name') \land 
B_\name (\name'\mbox{'s} \phi))$ to be true 
at a situation $(r,m,i)$ if there exists a name $\name'$ such that the
agent $j$ that  agent  
$i$ names $\name$ calls $i$ $\name'$, and $j$ knows that $\phi$
interpreted relative to $\name'$ (i.e., $i$) holds.
More formally,
\[\begin{array}{lll}
(\I,\lta,\rank,r,m,i)\sat \exists \name' (\Calls(\name,I,\name') \land
 B_\name (\name'\mbox{'s} \phi)) \mbox{ iff, for all
$j, j'$ and points $(r',m')$ }\\
\mbox{such that $\mu(r(m),i,\name) = j$, $r_j(m) =r'_{j'}(m')$, and 
$\rank(r')=\minj^\rank(r,m)$, we have }\\
\mbox{$(\I,r',m',i) \sat \phi$.}
\end{array}
$$
Note that the semantics for $\exists \name' (\Calls(\name,I,\name') \land
B_\name (\name'\mbox{'s} \phi))$ is almost the same as that for
$B_\name \phi$.  The difference is that we evaluate $\phi$ at $(r',m')$
with respect to $i$ (the interpretation of $I$ at the situation
$(r,m,i)$), not $j'$.   
We could give semantics to a much richer
logic that allows arbitrary quantification over names, and give separate
semantics to formulas of the form $\Calls(\name,I,\name')$ 
and $\name'\mbox{'s} \phi$, but what we have done suffices for our
intended application.

The semantics of  $\exists v B_{\name}(f=v)$ is straightforward.
Recall that the value of $f$ in run $r$ is $f(N_r)$. 
We can then take $\exists v B_{\name}(f=v)$ to be true at a point
$(r,m)$ according so some agent $i$  
if all runs $\name$ believes possible 
are associated with the same function value:
\[\begin{array}{lll}
(\I,\lta,\rank,r,m,i)\sat \exists v B_\name(f=v)~ \mbox{ iff, } 
 \mbox{for all $j, j'$ and points $(r',m')$ such that $\mu(r(m),i,\name) = j$},\\
\mbox{$r_j(m) = r'_{j'}(m')$, and $\rank(r')=\minj^\rank(r,m)$, we have $f(N_r)=f(N_{r'})$.}
\end{array}
$$

With all these definitions in hand, we can define the semantics of
counterfactual belief-based programs such as $\Pgcb^{GC}$.
A {\em counterfactual belief-based program\/} (or {\em cbb program}, for
short) $\Pgcbb$ is similar to a kb  
program, except that the knowledge modalities $K_\name$ are replaced
by the belief modalities $B_\name$.
We allow counterfactuals in belief tests but, for simplicity, do not allow
counterfactuals in the standard tests.

As with kb programs, we are interested in when a protocol~$P$ 
{\em implements} a cbb program $\Pgcb$.
Again, the idea is that the protocol should act according to the high-level 
program, when the tests are evaluated in the cb
system corresponding to~$P$. 
To make this precise, given a cb system $\J=(\I,\lta,\rank)$,
an agent $i$, 
and a cbb program $\Pgcb$, let 
$(\Pgcb)^\J_i$
denote the protocol derived from $\Pgcb$ 
by using $\J$ to evaluate the belief tests.  That is, a
test in $\Pgcb$ such as $B_\name\phi$ holds at a situation $(r,m,i)$
in $\J$ if $\phi$ holds at all situations 
$(r',m',j')$ in
$\J$ such that $\mu(r(m),i,\name) = j$, $r'_{j'}(m') =r_j(m)$,  and
$\rank(r') =  \minj^\rank(r,m)$. 
Define a {\em cb context\/} to be a tuple
$(\gamma,\pi, \ogen,\sigma)$, where $(\gamma,\pi)$ is an
interpreted context with naming function $\mu_\gamma$ 
(for simplicity, we use $\mu_\gamma$ to refer to the naming function in
context $\gamma$),
$\ogen$ is an order generator for $\Rp(\gamma)$ that respects protocols,
and~$\sigma$ is a deviation-compatible ranking generator for $\gamma$.
A cb system $\J = (\I,\lta,\rank)$ {\em represents\/}
the cbb program
$\Pgcb$ in cb context $(\gamma,\pi,\ogen,\rgen)$ if 
(a) $\I = (\Rp(\gamma),\pi,\mu_\gamma)$, 
(b) $\lta=\ogen(\Pgcb^{\J})$, and 
(c) $\rank = \sigma(\Pgcb^{\J})$.
A protocol $P$ {\em implements\/} $\Pgcb$ in cb context 
$\chi = (\gamma,\pi,\ogen,\sigma)$
if $P = \Pgcb^{(\I,\ogen(P),\sigma(P))}$.
Protocol
$P$ {\em de facto implements\/} $\Pgcb$ in
$\chi$ if $P \approx_\gamma
\Pgcb^{(\I,\ogen(P),\sigma(P))}$.

\section{Proof of correctness for $\Pgcb^{GC}$}\label{sec: cb correct prf}
\othm{thm:cbb correct}
\commentout{
$\Pgcb^{GC}$  solves the global function computation problem 
whenever possible:
for all ${\cal N}$ and $f$ such that the condition in 
Theorem~\ref{thm:solvability}  is satisfied
and all protocols $P$ that de facto implement $\Pgcb^{GC}$, 
in every run $r$ of the system that represents $P$,
eventually all agents know  $f(N_r)$. 
}
If $f$ and $\N(\gamma^{GC})$ satisfy the condition
in Theorem~\ref{thm:solvability}, then
$\Pgcb^{GC}$ solves the global function computation problem 
for $f$ in all interpreted contexts $(\gamma^{GC},\pi)$
for global function computation. 
\eothm

\prf
Let $f$ and $\N$ be such that the condition in Theorem~\ref{thm:solvability}
is satisfied. 
Suppose that $\ogen$ is an order generator that respects protocols, 
$\rgen$  
is a deviation-compatible ranking generator,
$\gamma^{GC}$ is a context for global computation such that in all
initial states  
the network encoded in the environment state is in $\N$,
$\chi^{GC}$ is the  cb context $(\gamma^{GC},\pi,\ogen,\rgen)$, 
$P$ is a protocol that de facto implements $\Pgcbb^{GC}$ in $\chi^{GC}$,
$\J = (\Rp(\gamma), \pi, \mu_\gamma, \ogen(P), \rgen(P))$,
and $r \in \Rrep(P,\gamma^{GC})$.
We prove that at some point in run $r$  all agents in $N_r$ know $f(N_r)$.

We proceed much as in the proof of Theorem~\ref{thm:kb correct}; we just
highlight the differences here.  Again,
we first show that some agent in $r$ learns $f(N_r)$.  Suppose not.
Let $r'$ be the unique run of the full-information protocol in a
synchronous context starting with the same initial global state as $r$.
Again, we show by induction on $k$ that there is a time $m_k$ such that, at
the
point
$(r,m_k)$, all the agents in $\A(r)$ have at least as much information 
about the network as they do at the beginning of round $k$ in $r'$.
The base case is immediate, as before.  For the inductive step, suppose
that $i$ learns some information about the network from $j$ during round $k$.
Again, there must exist a time $m_k' \le m$ where $j$ first learns this
information in run $r$.
It follows that $(\J,r,m_k',j)\models \somenewinfo$.

Suppose that $j$ names $i$ $\name$ in $r$; that is
$\mu_\gamma(r(m_k),j,\name)=i$.  
Now either (a) $j$ believes at time $m_k'$ that, if he does not perform
a $\sf \send_{A}(\newinfo)$ action with $\name \in A$, 
$i$ will eventually learn its new information or the function value
anyway, or (b) $j$ does not believe this.  In case (b), it follows that 
$$
\begin{array}{ll}
(\J,r,m_k',j)\models & \neg B_I [\neg\doact(\send_{\name}(\newinfo))\RCond  
\Diamond ((\exists \name' (\Calls(\name,I,\name')\wedge \\
& \quad B_{\name}(\name'
~\mbox{'s}~\inewinfo))  \vee \exists v B_{\name}(f=v))].
\end{array}
$$ 
Since $P$ implements $\Pgcbb^{GC}$ in $\chi^{GC}$, in case (b), $j$
sends $i$ $\newinfo$ at time $m_k'$, so there is some round $m_k''$ by
which $i$ learns this information. 
On the other hand, 
in case (a), it must be the case that
$$
\begin{array}{ll}
(\J,r,m_k',j)\models  & B_I [\neg \doact(\send_{\name}(\newinfo))\RCond  
\Diamond (\exists \name' (\Calls(\name,I,\name')\wedge \\
& \quad B_{\name}(\name'~\mbox{'s}~\inewinfo))  \vee \exists v B_{\name}(f=v))].
\end{array}
$$ 
Since $\rgen$ is deviation compatible by assumption, and $r$ is a run
of $P$, it follows that $\rank(r) = 0$.  Thus by
Lemma~\ref{knowledge-axiom}, 
$$
\begin{array}{ll}
(\J,r,m_k',j)\models  & \neg \doact(\send_{\name}(\newinfo))\RCond  
\Diamond (\exists \name' (\Calls(\name,I,\name')\wedge \\
& \quad B_{\name}(\name'
~\mbox{'s}~\inewinfo))  \vee \exists v B_{\name}(f=v)).
\end{array}
$$ 
Since $P$ implements $\Pgcbb^{GC}$ in $\chi^{GC}$, in case (a), $j$
does not send $\newinfo$ to $i$ in round $m_k'$.  
Thus, 
$(\J,r,m_k',j)\models  \neg \doact(\send_{\name}(\newinfo))$.
It follows that 
$$(\J,r,m_k',j)\models  \exists \name' (\Calls(\name,I,\name')\wedge
B_{\name}(\name' ~\mbox{'s}~\inewinfo))  \vee \exists v B_{\name}(f=v)).$$ 
Since, by assumption, no one learns the function value in $r$, we have
that 
$$(\J,r,m_k',j)\models  \exists \name' (\Calls(\name,I,\name')\wedge
B_{\name}(\name' ~\mbox{'s}~\inewinfo)).$$
Thus, 
it
follows that $i$ must eventually learn $j$'s information in this case too.

It now follows, just as in the proof of Theorem~\ref{thm:kb correct},
that some agent learns $f(N_r)$ in $r$, and that eventually all agents
learn it.  We omit details here. \eprf

\section{Proof of Theorem~\ref{thm:le proofs}}\label{sec:le proofs}

In this section we prove Theorem~\ref{thm:le proofs}, which says that
LCR$'$, P1$'$, and P2$'$ de facto implement $\Pgcb^{GC}$.  We start by
sketching the proof for LCR$'$, and then provide a detailed proof for
P2$'$.  The proof for P1$'$ is similar and is omitted here.

\subsection{The argument for LCR$'$}

The pseudocode for LCR and LCR$'$ is given in Figures~\ref{LCRcode}
and~\ref{LCR'code} respectively.  
In the code for LCR, we 
use $\id$ to denote the agent's initial id.
We assume that 
each agent has one queue, denoted $RQ$, which
holds messages received from the right. 
The placing of
messages in the queue is controlled by the channel, not the agent.
We use $RQ = \bot$ to denote that the right queue is empty. 
We write $\val_R :=\mathit{dequeue}(RQ)$ to denote the operation of
removing the top message from the right queue and assigning it to the 
variable  $\val_R$.
If $RQ = \bot$ when a $\mathit{dequeue}$ operation is performed, then 
the agent 
waits until it is nonempty.
Each agent has a local variable $\mathit{status}$
that is initially set to $\mathit{nonleader}$ and 
is changed to 
$\mathit{leader}$ only by the agent with the maximum id in the ring
when it discovers it is the leader.
We take $\mathit{done}$ to be a binary variable that is initialized to
0 and changed to 1 after the maximum id has been computed.  
Agents keep track of the maximum id seen so far in the variable $\maxid$. 
We call a message of the form ``$M$ is the leader'' a \emph{leader
message}.  
Note that in our version of LCR, after the leader finds out that it is
the leader, it informs all the other agents of this fact.  This is not
the case for the original LCR protocol.  
We include it here for compatibility with our global function
computation protocol.  (Similar remarks hold for P2.)

\begin{figure}[tb]
$$
\begin{array}{l}
\mathit{status:=nonleader;~\maxid:=\id;~\val_R:=\perp;~done:=0}\\
\send_L(\id)\\
{\bf do~until}~ \mathit{done=1}\\ 
\quad \mathit{receive}\\
\quad {\bf if}~\mathit{RQ} \ne \bot~ {\bf then}\\
\quad \quad \val_R := \mathit{dequeue}(RQ)\\
\quad \quad {\bf if}~ (\val_R=\id) ~{\bf then}\\
\quad \quad \quad \mathit{status:=leader};~ \send_{\lefta}(\mathit{``\id~is~the~leader"});~\mathit{\done:=1}\\
\quad \quad {\bf else~if}~(\val_R>\maxid)~{\bf then}\\
\quad \quad \quad \mathit{\maxid:=\val_R};~\send_{\lefta}(\maxid)\\
\quad \quad {\bf else~if}~(\val_R\ \mbox{is a leader message})~{\bf then}\\
\quad \quad \quad \send_{\lefta}(\val_R);~\mathit{done:=1}\\
\end{array}
$$
\caption{The LCR protocol.}
\label{LCRcode}
\end{figure}

\begin{figure}[tb]
$$\begin{array}{l}
{\bf do~until}~ (\mathit{\id \in \val_R}) \land
(\mathit{sent~leader~message} \lor \maxid = \id_L)\\
\quad \mathit{receive}\\
\quad {\bf if}~\somenewinfo~{\bf then}\\
\quad \quad {\bf if}~ 
((\id \notin \val_R \land \mmax(\val_R)>\maxid) \lor
(\id \in \val_R)  ~{\bf then}~
\send_{\lefta}(\newinfo)
\end{array}
$$
\caption{The LCR$'$ protocol.}
\label{LCR'code}
\end{figure}

In the code for LCR$'$,
$\val_R$ encodes all the new information that the sender sends (and thus
is not just a single id).
Let $\mmax(\val_R)$ be the maximum id encoded in $\val_R$. 
Since an agent sends all the new information it has, there is no need
for special messages of the form ``$M$ is the leader''.  The leader can
be computed from $\val_R$ if the message has gone around the ring, which
will be the case if $\id \in \val_R$.
Moreover, if $\id \in \val_R$, an agent can also compute
whether the leader is its left neighbor, and whether 
it has earlier essentially sent an ``$M$ is the leader message'' (more
precisely, an agent can tell if it has earlier been in a state where
$\id \in \val_R$ and it sent a message).  
We take the test $\id_L = \maxid$ to be true if an agent knows that the
leader is its left neighbor (which means that a necessary condition for 
$\id_L = \maxid$ to be true is that $\id \in \val_R$); we take
\emph{sent leader message} to be true if $\id \in \val_R$ and the agent
earlier sent a message when $i \in \val_R$ was true.  
Notice that in LCR$'$ we do not explicitly set $\val_R$; $\val_R$ can be
computed from the agent's state, by looking at the new information
received.

The basic idea of the proof is simple: we must show that $\Pgcb^{GC}$
and LCR$'$ act the same at all points in a system that represent
LCR$'$.  That means showing that an agent sends a message iff it
believes that, without the message, its neighbor will not eventually
learn the information that it has or the function value.  Since LCR$'$
solves the leader election problem, when processors do not send a
message, they believe (correctly) that their neighbor will indeed learn
the function value.  So consider a situation where a processor $i$ sends a
message according to LCR$'$.  That means that either it has gotten a
message $\val_R$ such that $\val_R > \maxid$ or it has gotten a leader
message.  If it does not forward a leader message, then it is clear that
all the processors between $i$ and the leader (of which there must be at
least one) will not learn who the leader is, because no further messages
will be sent.  If $i$ has received a message with $\val_R > \maxid$,
then consider $\maxid$ is in fact the largest id.  Then it is easy to
see that $i$ will never receive any further messages, and no processor
will ever find out who the leader is.  Since this ring is consistent with
$i$'s information, $i$ does not believe that, if it does not forward the
message, 
$i$'s left 
neighbor will learn the information or learn who the
leader is.  Thus, according to $\Pgcb^{GC}$, $i$ should forward the message.
We omit the formal details of the proof here, since we do the proof for
P2$'$ (which is harder) in detail. 

\subsection{The argument for P2$'$}

We start by describing P2.
Since P2 works in bidirectional rings, rather than just having one
queue, as in LCR, in P2, each agent
has two queues, denoted $LQ$ and $RQ$, which
hold messages received from the left and right, respectively.  
While an agent is active, it processes a message from $RQ$, then $LQ$,
then $RQ$, and so on.  
The status of an agent, i.e., whether it is active, passive or the
leader, is indicated by the variable $\mathit{status}$. Initially,
$\mathit{status}$ is $\mathit{active}$.
Finally, we take $\mathit{wl}$  to be a  binary variable  that indicates
whether the agent is waiting to receive a message from its left.
When an active agent receives $\mathit{\val}_R$, %
it compares $\val_R$ to its id. If $\val_R = \id$ (which can happen only
if $i$ is active) then, 
as in the LCR protocol, $i$ declares itself to be the 
leader (by setting $\mathit{status}$ to $\mathit{leader}$), and it sends
out a message to this effect.
If $i$ is active and $\val_R > \id$, 
then $i$ becomes passive; if $\val_R < \id$, then $i$ remains active and 
sends its id to the right.
Finally, if $i$ is passive, then $i$ forwards $\val_R$ to the left.  
The situation is symmetric if $i$ receives $\val_L$.
The pseudocode for P2 is given in Figure~\ref{P2code}.

\begin{figure}[htb]
$$
\begin{array}{l}
\mathit{status:=active;~\val_L := \perp; ~\val_R := \perp}; 
~\mathit{done:=0};~\mathit{wl}=0\\
\send_{\lefta}(\id);\\
{\bf do~until}~ \mathit{done=1}\\ 
\quad  {\bf if}~ 
(RQ\neq \perp) \land (\mathit{wl}=0)~ {\bf then}\\ 
\quad \quad \val_R:=\mathit{dequeue}(RQ)\\
\quad \quad \mathit{wl} := 1\\
\quad \quad {\bf if}~ (\val_R=\id) ~{\bf then}~
\mathit{status:=leader};~
\send_{\righta}(\mathit{``\id~is~the~leader"});~\mathit{\done:=1}\\ 
\quad \quad  {\bf if}~\mathit{status = active} \land  \val_R > \id~ {\bf
then}~\mathit{status}:=\mathit{passive}\\
\quad \quad  {\bf if}~ \mathit{status = active} \land \val_R < \id~ {\bf then}~
\send_{\righta}(\id)\\
\quad \quad  {\bf if}~ \mathit{status = passive}~{\bf then}~
\send_{\lefta}(\id); 
{\bf if}~(\val_R\ \mbox{is a leader message})~{\bf then}~\mathit{done:=1}\\
\quad   {\bf if}~
(LQ\neq \perp) \land (\mathit{wl}=1)~{\bf then}\\ 
\quad \quad \val_L:=\mathit{dequeue}(LQ)\\
\quad \quad \mathit{wl}:=0\\
\quad \quad {\bf if}~ (\val_L=\id) ~{\bf then}~
\mathit{status:=leader};~
\send_{\lefta}(\mathit{``\id~is~the~leader"});~\mathit{\done:=1}\\ 
\quad \quad  {\bf if}~\mathit{status = active} \land  \val_L > \id~ {\bf
then}~\mathit{status}:=\mathit{passive}\\
\quad \quad  {\bf if}~ \mathit{status = active} \land \val_L < \id~ {\bf then}~
\send_{\lefta}(\id)\\
\quad \quad  {\bf if}~ \mathit{status = passive}~{\bf then}~
\send_{\righta}(\id); 
{\bf if}~(\val_L\ \mbox{is a leader message})~{\bf then}~\mathit{done:=1}\\
\end{array}
$$
\caption{Peterson's protocol P2.}
\label{P2code}
\end{figure}
To understand in more detail how P2 and P2$'$ work, it is  helpful to
characterize 
the order in which agents
following P2 send and process messages. 
Since P2 and P2$'$ are
identical up to the point that an agent knows the leader, the
characterization will apply equally well to P2$'$.
We can get a complete characterization despite the fact that we do not
assume synchrony, nor that messages are received in FIFO order.
As usual, we use $(a_1, \dots,a_k)^*$
to denote 0 or more  
repetitions of a sequence of actions $a_1, \dots, a_k$.
We denote the action of sending left (resp. right)
as $SL$ (resp. $SR$), and the 
action of processing from the left (resp. right) as $PL$ (resp. $PR$).

\lem\label{pro:msg order}
For all runs $r$ of P2, times $m$, and agents $i$ in $N_r$
\begin{itemize}
\item[(a)] if $i$ is active at time $m$, 
then $i$'s sequence of actions
in the time interval $[0,m)$ 
is a prefix of the sequence  (SL, PR, SR, PL)$^*$; 
\item[(b)] if $i$ is passive at time $m$,
$i$ does not yet know which agent has the maximum id,
and $i$ became passive at time
$m' \le m$ 
after processing a message from the right (resp., left), then 
$i$'s history in the time interval $[m',m]$ is a prefix of the sequence 
(PL, SR, PR, SL)$^*$
(resp., 
(PR, SL, PL, SR)$^*$). 
\end{itemize}
\elem

\prf 
We proceed by induction on the time $m$.  The result is trivially true
if $m=0$, since no actions are performed in the interval $[0,0]$.
Suppose the result is true for time $m$; we show it for time $m+1$.  If
$i$ is active at time $m+1$, then the result is immediate from the
description of P2 (since it is immediate that, as long as $i$ is
active, 
it cycles through the sequence $SL$, $PR$, $SR$, $PL$).  
So suppose that $i$ is passive at time $m+1$.  It is clear from the
description of P2 that, while $i$ is passive, $PL$ is immediately
followed by $SR$ and $PR$ is immediately followed by $SL$.  Thus, it
suffices to show that (i) if $i$ was active when it performed its last
action, and this action was $PR$, then $i$'s next action is $PL$;
(ii) if $i$ was active when it performed its last
action, and this action was $PL$, then $i$'s next action is $PR$;
(iii) if $i$ was passive when it performed its last action, and this
action was $SR$, then $i$'s next action is $PR$; and (iv) if $i$ was
passive when it performed its last action, and this 
action was $SL$, then $i$'s next action is $PL$.  The proofs of
(i)--(iv) are all essentially the same, so we just do (i) here.

Suppose that  $i$'s last action before time $m+1$ was $PR$, and then
$i$ became passive.  It is clear from the description of P2 that
$i$'s next action is either $PR$ or $PL$.  Suppose, by way of
contradiction, that  $i$ performs $PR$ at time $m+1$.  It follows from
the induction hypothesis that there must exist some $k$ such that 
$i$ performed $SR$ $k$ times and $PR$ $k+2$ times in the interval 
$[0,m+1]$.  But then the agent $R_i$ to $i$'s right performed $SL$ at
least $k+2$ times and $PL$ at most $k$ in the interval $[0,m]$.  This
contradicts the induction hypothesis.  
\eprf

Intuitively, P2 and P2$'$ act the same as long as agents do not know who
the leader is.  
In P2$'$, they will know who the leader is once they know
all the agents on the ring.  To make this latter notion precise, define
the sets $I_L(i,r,m)$ and
$I_R(i,r,m)$ of agents as follows:
$I_R(i,r,0) = I_L(i,r,0) =\{ i\}$.
If, at time $m+1$, $i$ processes
a message from its right, and this message was sent by $R_i$ at time
$m'$, then 
$$I_R(i,r,m+1)= I_R(i,r,m)\cup I_R(R_i,r,m') ~ \mbox{and}~
I_L(i,r,m+1) =I_L(i,r,m)\cup I_L(R_i,r,m') - \{ R_i\}.$$
If, at time $m+1$, $i$ processes a message from its left, and this
message was sent by $L_i$ at time $m'$, then 
$$I_L(i,r,m+1)= I_L(i,r,m)\cup I_L(L_i,r,m')~\mbox{and}~
I_R(i,r,m+1) =I_R(i,r,m)\cup I_R(L_i,r,m') -  \{L_i\}. $$
Finally, if $i$ does not process a message at time $m+1$, then 
$$I_R(i,r,m+1) = I_R(i,r,m)~\mbox{and}~I_L(i,r,m+1) =I_L(i,r,m).$$

$I_R(i,r,m)$ and $I_L(i,r,m)$ characterize the set of
agents to $i$'s right and left, respectively, that $i$ knows about at
the point $(r,m)$.  
$I_L(i,r,m)$ and $I_R(i,r,m)$ are always intervals for agents running a
full-information protocol (we prove this formally below).  Thus, agent
$i$ has \emph{heard from everybody in the ring}, 
denoted $\heardainfo$, if
$I_L(i,r,m) \cup I_R(i,r,m)$ contains all agents in the ring.
More formally, 
$(\J,r,m,i) \sat \heardainfo$ if $I_L(i,r,m) \union I_R(i,r,m)$
consists of all the agents in the network $N$ encoded in the environment
state in $(r,m)$.
Note that $\heardainfo$ may hold relative to agent $i$ without $i$
knowing it; $i$ may consider it possible that there are agents between
the rightmost agent in $I_R(i,r,m)$ and the leftmost agent in $I_L(i,r,m)$.
We define the primitive proposition $\ainfo$ to be true at 
at the point $(r,m)$ relative to $i$ 
if $I_L(i,r,m)\inter I_R(i,r,m)- \{ i\} \neq \emptyset$.
It is not difficult to show that 
$\ainfo$ is equivalent to $K_I (\heardainfo)$; thus, we say that
$i$ \emph{knows it has all the information} if $\ainfo$ holds relative
to $i$.

The pseudocode for P2$'$ while agents do not 
know that they 
have all the information is
given in Figure~\ref{P2code'}. 
(We describe what agents do when 
they know all the information at the end
of this 
section.)
Note that the pseudocode does not describe what happens if an agent is
active and $\val_R \ge id$.  Intuitively, at this point, the agent
becomes passive, but 
with P2$'$ there is no action that changes an
agent's status; rather, the status is inferred from the messages that
have been received.  (This is similar to the reason that the LCR$'$
protocol had so many fewer steps than the LCR protocol.)
Since agents running P2 perform the same actions under essentially the
same conditions as agents running P2$'$ up to the point that an agent
knows that it has all the information, Lemma~\ref{pro:msg order} also applies 
to
all runs $r$ 
of P2$'$, times $m$, and agents $i$ in $N_r$ such that 
$i$ did  not know that it had all the information at time $m-1$ in $r$.

\begin{figure}[htb]
$$
\begin{array}{l}
\send_{\lefta}(\newinfo);\\
{\bf do~until}~\ainfo\\
\quad  {\bf if}~ 
(RQ\neq \perp) \land (\mathit{wl}=0)~ {\bf then}\\ 
\quad \quad  {\bf if}~ \mathit{status = active} \land \val_R < \id~ {\bf then}~
\send_{\righta}(\newinfo)\\
\quad \quad  {\bf if}~ \mathit{status = passive}~{\bf then}~
\send_{\lefta}(\newinfo); \\
\quad   {\bf if}~
(LQ\neq \perp) \land (\mathit{wl}=1)~{\bf then}\\ 
\quad \quad  {\bf if}~ \mathit{status = active} \land \val_L < \id~ {\bf then}~
\send_{\lefta}(\newinfo)\\
\quad \quad  {\bf if}~ \mathit{status = passive}~{\bf then}~
\send_{\righta}(\newinfo); 
\end{array}
$$
\caption{The initial part of protocol P2$'$, run while agents do not
know that they have all the information.} 
\label{P2code'}
\end{figure}

We now prove a number of properties of $I_L(i,r,m)$ and $I_R(i,r,m)$
that will be useful in our analysis of P2$'$.

\lem\label{lem: i-segment}
For all runs $r$ of P2$'$ and times $m$ the following hold:
\begin{itemize}
\item[(a)]
$I_R(i,r,m)$ is an interval of agents starting with $i$ and going to the
right of $i$, and 
$I_L(i,r,m)$ is an interval of agents starting with $i$ and going to the
left of $i$. 
\item[(b)]
If, at time $m$, $i$ processes a message 
from the right sent by $R_i$ at time $m'$, 
and $R_i$ did not know that it had all the information at time $m'$,
then
  \begin{itemize}
  \item[(i)] $I_R(R_i,r,m')\supset I_R(i,r,m-1)- \{ i \}$,
             $I_R(i,r,m)\supset I_R(i,r,m-1)$, and $I_R(i,r,m)=\{ i \}
             \cup I_R(R_i,r,m')$;  and
  \item[(ii)] $I_L(i,r,m)=I_L(i,r,m-1)$.
  \end{itemize}
\item[(c)]
If, at time $m$, $i$ processes a message 
from the left sent by $L_i$ at time $m'$, 
and $L_i$ did not know that it had all the information at time $m'$,
then
  \begin{itemize}
  \item[(i)] $I_L(L_i,r,m')\supset I_L(i,r,m-1) - \{ i \}$,
             $I_L(i,r,m)\supset I_L(i,r,m-1)$,  and $I_L(i,r,m)=\{ i\}
\cup I_L(L_i,r,m')$; and
  \item[(ii)] $I_R(i,r,m)=I_R(i,r,m-1)$.
\end{itemize}
\item[(d)] If $i$ processed a message from the right in the interval
$[0,m]$, 
and $R_i$ did not know that it had all the information when it last sent 
a message to $i$,
then $$\max_{\{m' \le m: \val_R(i,r,m') \ne \bot\}}
\val_R(i,r,m')$$  
is the maximum id of the agents in $I_R(i,r,m) -\{i\}$, where
$\val_R(i,r,m')$ is the value of agent $i$'s variable $\val_R$ at the
point $(r,m')$; 
if $i$ processed a message from the left in the interval $[0,m]$, then  
$$\max_{\{m' \le m: \val_L(i,r,m') \ne \bot\}}
\val_L(i,r,m')$$ is the maximum id in $I_L(i,r,m) -\{i\}$.
\item[(e)] $i$ is active at time $m$ if and only if $i$ has the largest
id in $I_L(i,r,m)\cup I_R(i,r,m)$.
\end{itemize}
\elem

\prf
We prove all parts of the lemma simultaneously by induction on $m$.
The result is immediate if $m=0$, since $i$ is active at time 0, $i$ does
not process a message at time 0, and $I_L(i,r,0) = I_R(i,r,0) = \{i\}$.
Suppose that parts (a)--(e) hold for all 
times
$m' < m$.  We show that
they also hold at time $m$.  
They clearly hold if $i$ does not process a message at time $m$, since
in that case $I_L(i,r,m) = I_L(i,r,m-1)$ and $I_R(i,r,m) = I_R(i,r,m-1)$.
So suppose that $i$ processes a message $\msg$ from its right
at time $m$, and $\msg$ was sent by $R_i$ at
time $m'$.  (The proof is similar if $i$ receives from the left, and is
left to the reader.)
If $\msg$ is the first message received by $i$ from the right, then it
follows from Lemma~\ref{pro:msg order} that $i$ has sent no
messages to the right, and $R_i$ has sent only one message to $i$.  
Thus, $I_R(i,r,m-1) = \{i\}$.  Parts (a)--(e) now follow easily from the
induction hypothesis.

So suppose that $\msg$ is not the first message that $i$ has received from
$R_i$.
Part (a) is immediate from the induction hypothesis.
To prove part (b), let 
$m_1$ be the last time prior to $m'$ that $R_i$ sent a message, say
$\msg'$, to its left.   
It easily follows from 
Lemma~\ref{pro:msg order} 
(which, as we observed, also applies to P2$'$ while agents do not know
that they have all the information) 
that there are
times $m_2$ and $m_3$, both  in the interval $(m_1,m')$, such that $i$
received $\msg'$ at time $m_2$  and $R_i$ processed a message from its
right at  $m_3$; moreover, $i$ did  not process any  messages from the
right between time $m_2$ and $m$.  By the induction hypothesis,
$I_R(i,r,m_2)=      \{       i      \}      \cup      I_R(R_i,r,m_1)$,
$I_L(i,r,m_2)=I_L(i,r,m_2-1)$,     and    $I_R(R_i,r,m_3+1)    \supset
I_R(R_i,r,m_1)$.    Since   $m_3+  1   \le   m'$,   it  follows   that
$I_R(R_i,r,m')  \supset I_R(R_i,r,m_1)$.  Since  $i$ does  not process
any messages from its right between time 
$m_2$
and $m$, by definition,
$I_R(i,r,m-1)=I_R(i,r,m_2)$.   It follows that  $I_R(R_i,r,m') \supset
I_R(i,r,m-1)$ and that
$$\begin{array}{ll} I_R(i,r,m)  = I_R(i,r,m-1) \union  I_R(R_i,r,m') =
\{i\}  \union  I_R(R_i,r,m_1) \union  I_R(R_i,r,m')\\  = \{i\}  \union
I_R(R_i,r,m')     \supset     \{i\}     \union    I_R(R_i,r,m_1)     =
I_R(i,r,m-1).\end{array}$$
This  proves  part  (i) of  (b)  for  time  $m$.   For part  (ii),  by
definition, $I_L(i,r,m)=I_L(i,r,m-1)\cup I_L(R_i,r,m') - \{R_i\}$.  By
the  induction hypothesis,  it  easily follows  that $I_L(R_i,r,m')  -
\{R_i\} \subseteq I_L(i,r,m') \subseteq$
$I_L(i,r,m-1)$.  Thus, $I_L(i,r,m) = I_L(i,r,m-1)$.

Part (c) is  immediate, since $i$ does not process  a message from the
left at time $m$.

For the first  half of part (d), there are two  cases to consider.  If
$R_i$ was active  at the point $(r,m')$, then  the result is immediate
from  part  (e)  of  the  inductive  hypothesis.   Otherwise,  by  the
inductive hypothesis, $\val_R = \val_R(i,r,m) = \val_R(R_i,r,m')$.  By
the inductive  hypothesis, $\val_R$  is greater than  or equal  to the
maximum id  in $I_R(R_i,r,m')  - \{R_i\}$.  Since  the first  value of
$\val_R$  must  be $R_i$'s  id,  it  follows  that $$\max_{\{m'\le  m:
\val_R(i,r,m') \ne  \bot\}} \val_R(i,r,m')$$ is greater  than or equal
to the  maximum id  in $I_R(i,r,m) -  \{i\} =  
I_R(R_i,r,m')$.  Since
$\val_R(i,r,m')$  must be  an id  in $I_R(i,r,m)$,  we are  done.  The
second half  of part (d)  is immediate from the  induction hypothesis,
since $I_L(i,r,m)  = I_L(i,r,m-1)$ by  part (b), and  $\val_L(i,r,m) =
\val(i,r,m-1)$.

Finally, part (e) is immediate from the induction hypothesis if $i$ is
passive at time  $m-1$.  So suppose that $i$ is  active at time $m-1$.
By the induction hypothesis, $i$'s  id is the largest in $I_L(i,r,m-1)
\union I_R(i,r,m-1)$.
If $i$ is active at time $m$ then, by the description of P2$'$, $i$'s id
must  be  greater than  $\val_R(i,r,m)$.   Applying  part  (d) of  the
induction hypothesis and  the fact that $i$'s id is  at least as large
as all those  in $I_R(i,r,m-1)$, it follows that $i$'s  id is at least
as   large  as   $\max_{\{m'   \le  m:   \val_R(i,r,m')  \ne   \bot\}}
\val_R(i,r,m')$.  By  part (d), at time  $m$, $i$'s id is  at least as
large all  those in $I_R(i,r,m)$.  Since  $I_L(i,r,m) = I_L(i,r,m-1)$,
it  follows that  $i$'s id  is the  maximum id  in  $I_R(i,r,m) \union
I_L(i,r,m)$.  Conversely, if $i$'s id is the maximum id in $I_R(i,r,m)
\union I_L(i,r,m)$,  then by part  (d) at time  $m$, $i$'s id  must be
greater than $\val_R(i,r,m)$, and hence by the description of P2$'$, $i$
is active at $(r,m)$.  \eprf

\commentout{ \lem\label{pro:agent-order}
For all runs $r$ of P2$'$,
if  $i  \in  \A_L(r,p)$,  $j$  is  the  active  agent  in  $\A_L(r,p)$
immediately
to $i$'s left (if there is one), and $k \in \A_L(r,p) - \{i,j\}$,
and $i$,  $j$, and $k$ send their  $p$th message to the  left at times
$m_i$, $m_j$, and $m_k$ respectively, then
   \begin{itemize}
   \item[(a)] %
\lbrace  i \rbrace)  \cap  \A_L(r,p)=\emptyset$ (so  $i$  knows of  no
active agent to its right):
 \item[(b)]    if   $\id_j    >    \id_i$,   then    $I_L(i,r,m_i)\cap
I_R(j,r,m_j)=\emptyset$;
   \item[(c)]
$I_L(i,r,m_i)\cap I_R(k,r,m_k)=\emptyset$.
   \end{itemize}
The same is also true with  all occurrences of left and right (and $L$
and $R$) interchanged.  \elem }

It is  not difficult to  see that P2$'$  ensures that, for  all agents
 $i$,  $I_L(i,r,m)\cup  I_R(i,r,m)$ increases  with  time $m$.   Thus,
 eventually at least  one agent must know it  has all the information.
 (Recall that we  have not yet given the pseudocode  for P2$'$ for the
 case that an agent knows it has all the information.)
\cor\label{cor:p2}
In all runs  $r$ consistent with P2$'$, eventually  at least one agent
knows that it has all the information,
i.e.,   there  exist   an   agent   $i$  and   time   $m$  such   that
$I_L(i,r,m)\inter I_R(i,r,m) -\{ i\} \neq \emptyset$.  \ecor

We   say  that   message  $\msg$   received   by  $i$   at  time   $m$
\emph{originated with $j$ at time $m'$} if $j$ is the active agent who
first sent  $\msg$, and  $\msg$ was  sent by $j$  at time  $m'$.  More
formally,  we define  origination by  induction on  the time  $m$ that
$\msg$ was  received.  If  $\msg$ is received  by $i$ from  the right,
then $\msg$  originated with $R_i$ at  the time that $R_i$  sent it if
$R_i$
was not passive when it sent $\msg$; otherwise, if $\msg$ was received
at some time  $m'' < m$ by $R_i$, then the  message $\msg$ received by
$i$ at $m$ originated with the same  agent and at the same time as the
message  $\msg$  received  by  $R_i$  at  $m''$.   The  definition  is
analogous if $\msg$ is received by $i$ from the left.

Let $[i..j]_R$  denote the agents to  $i$'s right starting  at $i$ and
going to  $j$; similarly,  let $[i..j]_L$ denote  the agents  to $i$'s
left starting at $i$ and going to $j$.

\lem\label{lem: phases}
For all runs $r$ of P2$'$ and agents $i$, $j$ in $r$,
\begin{itemize}
\item[(a)] if  at time $m$ agent  $i$ processes a  message $\msg$ from
the right
that originated with $j$ at $m'$, $\msg$ is the $p$th message $j$ sent
left,
and no  agent in $[i  .. j]_R$ knows  that it has all  the information
when  it sends  $\msg$,  then $\msg$  is  the $p$th  message that  $i$
processes  from the  right, and  $I_R(i,r,m) =  I_R(j,r,m')  \union [i
..j]_R$.
\item[(b)] if  at time $m$ agent  $i$ processes a  message $\msg$ from
the left that originated with $j$ at $m'$,
and $\msg$ is the $p$th message $j$ sent right,
and no  agent in $[i  .. j]_L$ knows  that it has all  the information
when  it sends  $\msg$,  then $\msg$  is  the $p$th  message that  $i$
processes  from  the left  and  $I_L(i,r,m)  =  I_L(j,r,m') \union  [i
..j]_L$.
\end{itemize}
\elem

\prf
We do the proof for case (a);  the proof of (b) is similar and left to
the reader.
The proof proceeds by induction on
the number of agents in $[i..j]_R$.
Since $i  \ne j$,  there are  at least two  agents in  $[i..j]_R$.  If
there are exactly two, then $j = R_i$.
Since the  only messages that $i$  processes from the  right are those
sent by $j$, it is immediate that $\msg$ is the $p$th message $i$
processed  from  the right.   Moreover,  by  definition $I_R(i,r,m)  =
I_R(j,r,m') \union \lbrace i \rbrace = I_R(j,r,m') \union [i..j]_R$.

Now suppose that (a) holds
for  all pairs of  agents $i'$,  $j'$ such  that 
$[i'..j']_R$ consists of
$d \ge 2$ agents and $[i..j]_R$ consists of $d+1$ agents.
Let $m_{R_{i}}$  be the  time $R_i$ sends  the message $\msg$  to $i$.
Since $[i  .. j]_R$ consists  of at least  3 agents, it cannot  be the
case that  $R_i = j$.   Thus, $R_i$ was  passive when it  received the
message $\msg$.  Let $m'_{R_i}$ be the time
$R_i$ processed $\msg$.  Since $[R_i  ..  j]_R$ has $d$ agents, by the
induction hypothesis, it follows that $\msg$
was the $p$th message that $R_i$ processed from the right.  By
Lemma~\ref{pro:msg order},
prior to $m'_{R_i}$, $R_i$ sent exactly $p-1$ messages to the left.
Moreover, since $R_i$ must process $p-1$ messages from the left before
processing its $p$th message from the right, it follows from
Lemma~\ref{pro:msg order}
that $i$ must  have processed all the $p-1$ messages  $R_i$ sent to it
before $R_i$ processed  $\msg$.  It now easily follows  that $\msg$ is
the $p$th message  processed by $i$ from the  right.  By the induction
hypothesis, $I_R(R_i,r,m'_{R_i}) = I_R(j,r,m') \union
[R_i .. j]_R$.  Thus, $I_R(i,r,m) = I_R(R_i,r,m'_{R_i}) \union \{i\} =
I_R(j,r,m') \union [i .. j]_R$.  \eprf
By Lemma~\ref{pro:msg order}, we can think of P2$'$ as proceeding in
phases while agents do not know all the information.
For $p = 1, 2, 3, \ldots$, we say that in run $r$, phase $2p-1$ begins
for agent $i$ when $i$ sends left for the $p$th time and
phase $2p$  begins for agent $i$  when $i$ sends right  for the $p$th
time;  
phase $p$ 
for agent $i$
ends when phase $p+1$ begins. 
The  following lemma  provides some  constraints on  what  agents know
about which agents are active and passive.

\commentout{ [[THIS  IS WHERE WE PROVE MY  CHARACTERIZATION.  GIVE THE
FIVE POSSIBLE SETTINGS  WHERE THE FIRST PROCESS LEARNS,  AND SHOW THAT
AT  LEAST THE FIRST  TWO CAN  ARISE, BY  EXAMPLE.  THE  SECOND EXAMPLE
SHOULD SHOW  THAT THE  LEADER CAN LEARN  BEFORE GETTING BACK  ITS ID..
THEN DEFINE  P2$'$.  FINALLY, PROVE  P2$'$ DE FACTO IMPLEMENTS  THE KB
PROGRAM (USING THE TECHNIQUES YOU'VE ALREADY DEVELOPED).  }

\commentout{ \lem\label{lem:phs}
For all runs $r$ of P2$'$,
phases $p$, times $m$, and agents $i$
such that, at  time $m$ in $r$,  either $i$ does not know  that it has
all the information, or if it does, then
it first learns all the information at time $m-1$ and it is one of the
first agents to do so,
the following all hold:
\begin{itemize}
\item[(a)]
If $i$ processed a message from  its left at time $m-1$ in phase $p-1$
($p>1$) and was active at time  $m-1$, or phase $p=1$ begins for agent
$i$ at time $m$,
then all agents in $I_R(i,r,m)-\{ i\}$ are passive
at the beginning of phase $p$
and  at  most  one agent  in  $I_L(i,r,m)-\{  i\}$  is active  at  the
beginning of
phase  $p$.  Moreover,  if some  agent $j$  in $I_L(i,r,m)-\{  i\}$ is
active at the beginning of phase $p$,
then $j$ is  the agent that originated the  message $i$ processed from
the left in phase $p-1$, $\id_j<\id_i$
if $i$ is active at time $m$,
and $\id_j >\id_i$ if $i$ is passive at time $m$.
\item[(b)]
If $i$ processed  a message from its right in phase  $p$ at time $m-1$
and was passive at time  $m-1$, then $I_R(i,r,m)$ contains exactly one
active agent at the beginning of phase $p$
and $I_L(i,r,m)$ contains at most one such agent.
\item[(c)]
If $i$ processed  a message from its right in phase  $p$ at time $m-1$
and $i$ was active at time $m-1$,
then $I_R(i,r,m)-\{i \}$ contains at most one active agent
at the beginning of phase $p$.
If some agent $j$ in $I_R(i,r,m)-\{ i\}$ is active at the beginning of
phase $p$, then
$j$ originated the message $i$ processed from the right in
phase $p$,
$\id_j<\id_i$ if $i$  is active at time $m$,  and $\id_j>\id_i$ if $i$
is passive at time $m$.
No agent  in $I_L(i,r,m)  -\{ i\}$  is active when  it sends  right in
phase $p$.
\item[(d)] %
passive,
If $i$  processed a message from its  left in phase $p$  at time $m-1$
and was passive at time $m-1$,
then $I_L(i,r,m)$  contains exactly one  agent that is active  when it
sends right in phase $p$, and
$I_R(i,r,m)$ contains at most one such agent.
\end{itemize}
\elem
\prf
The fact that either $i$ does not know that it has all the information, or
it is one
it first  learns all the information at  time $m-1$ and is  one of the
first to do so ensures that, for all agents $j$
such that $j \in I_L(i,r,m)\cup I_R(i,r,m)$ and
for all times  $m'$ such that a message $i$ has  processed by time $m$
originated with $j$
at time  $m'$, it is the  case that $j$ did  not know that  it had all
the
information  at  time $m'$  in  run $r$;  that  is,  $j$ followed  the
protocol in Figure~\ref{P2code'} at time $m'$.
Moreover,  it  is also  the  case that  $i$  follows  the protocol  in
Figure~\ref{P2code'} at all times
$m' < m$ in $r$.

The proof proceeds by induction on phases.  For $p=1$,
parts (a)--(c)  are immediate from  the description of  the algorithm.
For part (d),  note that if $i$ processes from its  left at time $m-1$
in phase 1, then $I_R(i,r,m)=[ i .. R_i]_R$
and $I_L(i,r,m) = [ j..  i ]_R$,
where $j$  is the  closest agent to  $i$ on  the left that  received a
message from  the right in  phase 1 from  a process with a  smaller id
than its own.  Clearly, all
agents  in $[j  .. i]_R$  other  than $i$  and $j$  are passive  after
processing from their right in phase 1, and so
they are not active when they send right in phase 1.

Now suppose  that $p>1$ and  parts (a)--(d) all  hold for phases  1 to
$p-1$.
For part~(a), by Lemma~\ref{lem: phases},
$I_R(i,r,m)= [ i .. k ]_R \cup I_R(k,r,m')$,
where the  message that  $i$ processed from  the right in  phase $p-1$
originated with $k$ at time $m'$.
By  part  (a)   of  the  induction  hypothesis,  all   the  agents  in
$I_R(k,r,m')-\{ k\}$ were passive at the beginning of phase $p-1$, and
so are clearly still passive at the beginning of phase $p$.
It also follows from the definition of origination that all the agents
in $[i ..  k]_R$ other than  $i$ and $k$ were passive at the beginning
of phase $p-1$, and so also passive at the beginning of phase $p$.
Applying Lemma~\ref{lem: phases} again, we have that
$I_L(i,r,m)= I_L(j,r,m'') \cup [i .. j]_L$,
where  the message that  $i$ processed  from the  left in  phase $p-1$
originated with $j$ at time $m''$.
By induction hypothesis (c) for $j$, no agent in $I_L(j,r,m'')-\{j\}$
is active  when it sends  right in phase  $p-1$, and so is  not active
when it
sends  left  in  phase  $p$  either.   Again,  by  the  definition  of
origination, all  agents in  $[j .. i]_R  -\{ i,j\}$ are  passive when
they send right in phase $p-1$, and so they are passive when they send
left in phase $p$.
Moreover,
$i$ is active after processing
that originated with $j$ if and only
$\id_j <\id_i$.

For part (b), suppose that the  message $i$ received from the right in
phase  $p$  originated with  $j$  at  time  $m'$.  By  Lemma~\ref{lem:
phases}, $I_R(i,r,m)=[ i .. j]_R \cup I_R(j,r,m')$.
By part (a) applied to agent  $j$ and time $m'$, all agents in $I_R(j,
r,m')-\{ j\}$ are passive at  the beginning of phase $p$. Clearly, all
agents in  $[j ..   i]_L$ other than  $j$ and  $i$ are passive  at the
beginning of phase  $p$. It follows that the only  active agent at the
beginning   of  phase   $p$   in  $I_R(i,r,m)$   is   $j$.   Also   by
Lemma~\ref{lem: phases}, $I_L(i,r,m)=[ i ..  k]_L \cup I_L(k, r,m'')$,
the message that $i$ received  from the left in phase $p-1$ originated
with $k$ at time $m''$.
By part (c) of the induction hypothesis applied to agent $k$ and phase
$p-1$, it  follows that  no agent in  $I_L(k,r,m'')-\{ k\}$  is active
when it sends right in phase $p-1$.
Thus, all agents in $I_L(k,r,m'')-\{ k\}$ are passive at the
beginning of phase $p$, so $k$  is the only agent in $I_L(i,r,m)$ that
can be active at the beginning of phase $p$.

For  part  (c),  by  the  same  reasoning as  above  it  follows  that
$I_R(i,r,m)=[ i  .. j ]_R \cup  I_R(j,r,m')$ and $I_L(i,r,m)=[  i .. k
]_L \cup I_L(k, r,m'')$, where
$j$ and $k$ are defined as in  part (b).  By part (a) applied to agent
$j$ and phase $p$, and part (c) of the induction hypothesis applied to
agent  $k$  and  phase  $p-1$,   we  can  prove  that  all  agents  in
$I_R(j,r,m')-\{
j\}$  are passive  at the  beginning  of phase  $p$, and  no agent  in
$I_L(k, r,m'')-\{ k\}$  is active when it sends  right in phase $p-1$.
Since  $i$ is  active at  the beginning  of phase  $p$, $i$  must have
stayed active after processing the message from $k$ in phase $p-1$; it
follows that $\id_i >\id_k$.
Moreover, $k$ will process from its  right in phase $p$ a message that
originated with $i$, and  subsequently become passive.  Thus, no agent
in  $I_L(i,r,m)-\{i\}$ is  active when  it sends  right in  phase $p$.
Among the  agents in  $I_R(i,r,m)-\{ i\}$, clearly  only agent  $j$ is
active when it sends right in phase $p$.  Since $i$ processes from its
right at time $m-1$ a message
that originated with $j$, $i$ stays  active at time $m$ if and only if
$\id_i>id_j$.

Finally, for part (d), it  is not difficult to show that $I_R(i,r,m)=[
i  .. j  ]_R \cup  I_R(j,  r,m')$ and  $I_L(i,r,m)=[i ..   k ]_R  \cup
I_L(k,r,m'')$, where
where  the message  that $i$  processed from  the right  in  phase $p$
originated with $j$  at time $m'$, and the  message that $i$ processed
from the left in phase $p$ originated with $k$ at time $m''$.  By part
(c) applied to  agent $k$ and phase  $p$, we can show that  $k$ is the
only agent in  $I_L(i,r,m)$ that can be active when  it sends right in
phase $p$.
By part (a) applied  to agent $j$ and phase $p$, only  $j$, $j$ is the
only agent  in $I_R(i,r,m)$ that can  be active when it  sends left in
phase $p$; it follows that at most one agent in $I_R(i,r,m)$ is active
when it sends right in phase $p$.  \eprf }
\lem\label{lem:phs} For all  runs $r$ of P2$'$, times  $m$, and agents
$i$,
if $m> 0$, the last message that $i$ processed before time $m$ was the
$p$th message, and  no agent knows all the  information at time $m-1$,
then
\begin{itemize}
\item[(a)] if $j_1, \ldots, j_k$ are the active
agents
at time  $m$ in $I_R(i,r,m)$, listed  in order of closeness  to $i$ on
the right (so that $j_1$ is the closest active process to
$i$'s right
with $j_1 = i$  if $i$ is active, and $j_k$ is  the farthest) then (i)
$\id_{j_1} > \ldots > \id_{j_k}$, (ii) if $j_1 \ne i$, then $j_l$ will
be passive
after having  processed its $(p-l+1)$st  message, for $l =  2, \ldots,
 k$,  provided that  $j_l$  processes its  $(p-l+1)$st message  before
 knowing all the  information; (iii) if $j_1 = i$,  then $j_l$ will be
 passive after after having  processed its $(p-l+3)$rd message, for $l
 =  2,  \ldots, k$,  provided  that  $j_l$  processes its  $(p-l+3)$rd
 message before knowing all the information; and (iv) the last message
 that $i$ processed from the right originated with $j_1$.
\item[(b)] if $h_1, \ldots, h_{k'}$ are the active
agents
at time $m$ in $I_L(i,r,m)$ listed in order of closeness to $i$ on the
left, then (i) $\id_{h_1} > \ldots > \id_{h_{k'}}$,
(ii) if $h_1 \ne i$, then $h_l$ will be passive after having processed
its $(p-l+1)$st message, for $l  = 2, \ldots, k'$, provided that $h_l$
processes its $(p-l+1)$st message  before knowing all the information;
(iii) if $h_1 = i$, then $h_l$
will  be  passive  after  having processed  its  $(p-l+3)$rd  message,
provided that it processes its
$(p-l+3)$rd
message before knowing all the information; and
(iv) the last message that $i$ processed from the left originated with
$h_1$.
\end{itemize}
\elem

\prf We proceed  by induction on $m$.  The lemma  is trivially true if
$m=1$, since  $I_L(i,r,1) = I_R(i,r,1)  = \{i\}$.  If $m>1$,  then the
result is  trivially true if  $i$ does not  process a message  at time
$m-1$  (since  $I_L(i,r,m)  =  I_L(i,r,m-1)$ unless  $i$  processes  a
message from the  left at time $m-1$, and  similarly for $I_R(i,r,m)$;
and even if  some agents in $I_L(i,r,m) \union  I_R(i,r,m)$ may become
passive  between time  $m-1$ and  time  $m$, the  result continues  to
hold).  So suppose that $i$ processes  a message from the left at time
$m-1$.  Since  $I_R(i,r,m) = I_R(i,r,m-1)$,  it is immediate  from the
induction hypothesis that  part (a) continues to hold.   For part (b),
by Lemma~\ref{lem: phases}, we have that $I_L(i,r,m)= I_L(j,r,m') \cup
[i ..   j]_L$, where the message  that $i$ processed from  the left at
time $m-1$  originated with  $j$ at time  $m'$.  By the  definition of
origination, all agents  in $[i .. j]_L -\{ i,j\}$  must be passive at
time $m-1$.   Thus, the result follows immediately  from the induction
hypothesis applied to  $j$ and time $m'$, together  with the following
observations:
\begin{itemize}
\item  If $j$ originated  the message  at time  $m'$, then  it follows
easily  from Lemma~\ref{pro:msg order}  that it  is the  $p$th message
sent by $j$.  Moreover, either $I_L(j,r,m') = \{j\}$ or $I_L(j,r,m') =
I_L(j,r,m'')$, where $m''-1$ is the time that
$j$
processed its $(p-2)$nd  message (since this is the  last message that
$j$ processed from the left prior to time $m'$).
\item If  $i$ is  active at time  $m$, then  $\id_i > \id_j$,  and the
$(p+1)$st message that  $j$ processes will originate from  $i$ (if $j$
does not know  all the information before processing  the message) and
will cause $j$ to become passive.
\end{itemize}
The argument is  similar if $i$ processes a message  from the right at
time $m-1$.  \eprf

\commentout{
In the following,  if $i$ first learns that it  has all the information at
time $m+1$, after  processing, at time $m$, a  message that originated
with agent $j$, then we say that $i$ stays active if $i$ was active at
time $m$  and $\id_j\le\id_i$,  becomes passive if  $i$ was  active at
time $m$  and $\id_j>\id_i$, and,  respectively, stays passive  if $i$
was passive at time $m$.
\lem\label{lem:first to k}
In all runs of  P2$'$, if $i$ is one of the  first agents to know that
it has all the information, then
\begin{itemize}
\item[(a)] if $i$ knows that it has all the information
after processing from  the left in phase $p-1$ at time  $m$ and $i$ is
active at time $m$, or after processing from the right in phase $p$ at
time $m$, and $i$ is passive at time $m$,
then there  are at most  two active agents  at the beginning  of phase
$p$;
\item[(b)] if $i$ learns
that it has  all the information after processing from  the right in phase
$p$
at time $m$, and $i$ is active
at  time $m$,  then  there are  at  most three  active  agents at  the
beginning of phase $p$;
\item[(c)]
if $i$  learns that it has  all the information after  processing from the
left in phase $p$ at time $m$, and $i$ is passive at time $m$,
then there are  at most three active agents at  the beginning of phase
$p$, and $i$ is one of them; and
\item[(d)] if $i$ learns that  it has all the information after processing
from the left in phase $p$
at time $m$, and
$i$ becomes passive after learning this information, then there are at
most three active agents that send right in phase $p$.
\end{itemize}
\elem \prf Let
$m+1$ be the first time $i$ knows it has all the information; that is,
$m+1$   is   the   first   time  $I_R(i,r,m+1)$   and   $I_L(i,r,m+1)$
non-trivially intersect.
If $i$ is active
at time $m$
and has last processed from the left in phase $p-1$
at time $m$,
then, by Lemma~\ref{lem:phs}(a),
$I_L(i,r,m+1)\cup I_R(i,r,m+1)-\{ i\}$ contains at most one
active agent  at the beginning  of phase $p$.   If $i$ was  passive at
time  $m$  when  it  processed  from  the right  in  phase  $p$,  then
Lemma~\ref{lem:phs}(b)  ensures  that there  are  at  most two  active
agents at the beginning of phase $p$. This proves (a).

Assume now  that $i$ has processed  a message from the  right in phase
$p$, at time $m$, and that $i$ was active at time $m$.  Let $j$ be the
agent that has originated the  message $i$ has last processed from the
right.   By Lemma~\ref{lem:  phases}, $I_R(i,r,m+1)=[i  .. j  ]_R \cup
I_R(j,r,m')$, where $m'$ is the time $j$ sent left in phase $p$.
$j$ did not  know that it had all the information at  time $m'$, since $i$
is one of  the first agents to learn that it  has all the information.  We
can then apply
Lemma~\ref{lem:phs}(a)
and deduce
that no  agent in $I_R(j,r,m')-\{ j\}$  is active at  the beginning of
phase $p$.  Furthermore, all agents in  $[ i .. j ]_R$, other than $i$
and $j$,  must be passive  at the beginning  of phase $p$,  since they
forward $j$'s message.  Since $i$ is active at time $m$, then by
Remark~\ref{rem:msg order}, prior  to $m$, $i$ has sent  left in phase
$p$,   say   at    time   $m''$.    By   Lemma~\ref{lem:   i-segment},
$I_L(i,r,m+1)=I_L(i,r,m'')$,     and     by    Lemma~\ref{lem:phs}(a),
$I_L(i,r,m'')-\{i\}$  contains  at  most   one  active  agent  at  the
beginning of  phase $p$.  We can  now conclude that there  are at most
two active agents at the beginning  of phase $p$, other than $i$; this
proves (b).

In case (c),  basically the same argument as for  (b) shows that there
are at most three active agents at the beginning of phase $p$.  Assume
by way of contradiction that $i$ is \emph{not} one of them.  Since $i$
first learns  that it  has all the information  after processing  from the
left in  phase $p$, it  means that  $i$ did not  know that it  had all
the
information at the  beginning of phase $p$.  $i$  being passive at the
beginning  of  phase  $p$,  it  follows  that  $i$  forwards  its  new
information to  the left  at the beginning  of phase $p$.   When $L_i$
processes this message from $i$, $L_i$  knows as much as $i$ about the
agents to their right.  Clearly, $L_i$  knows as much as $i$ about the
agents  to their  left,  too, since  $i$  only learns  about its  left
through $L_i$.  When $i$ processes from left in phase $p$, the message
must have  been sent  by $L_i$.  Since  $i$ has all  
the
information after
processing this message,  and $L_i$ knows at least as  much as $i$, it
follows that $L_i$ knew it had  all the information when it sent right
in phase $p$.  This contradicts the  choice of $i$ as one of the first
agents to know it has all the information in phase $p$.

The argument for case (d) is similar to the one for case (b).  \eprf }

We  say  that agent  $i$  \emph{can  be the  first  to  learn all  the
information in network  $N$} if there is a run $r$  of P2$'$ such that
$N_r = N$ and, in run $r$,  $i$ knows all the information at some time
$m$ and  no agent  knows all the  information at the  point $(r,m-1)$.
Our goal is to prove that there  can be at most two agents that can be
first to learn all the information in a network $N$.%
\footnote{In all  the examples we  have constructed, there is  in fact
only  one agent  that can  be first  to learn  all the  information in
network  $N$, although  that agent  may  not be  the eventual  leader.
However, we have  not been able to prove that this  must be the case.}
To prove
this  result,  we  first   show  that,  although  we  are  considering
asynchronous  systems,  what agents  know  depends  only  on how  many
messages they have processed.

\lem\label{lem:asynchronous} If $N_r = N_{r'} = N$, no agent knows all
the information at the point $(r,m)$ or the point $(r',m')$, and agent
$i$ has processed exactly $k$  messages at both the points $(r,m)$ and
$(r',m')$,  then   $I_L(i,r,m)  =  I_L(i,r',m')$   and  $I_R(i,r,m)  =
I_R(i,r',m')$.  Moreover, the $k$th  message that $i$ processed in run
$r$ originated  with $j$ iff the  $k$th message that  $i$ processed in
run $r'$ originated with $j$.  \elem

\prf We proceed by a straightforward induction on $m+m'$.  Clearly the
result is true if $m=m'=1$.  If  $i$ does not process a message at the
point  $(r,m-1)$, then  $I_L(i,r,m) \union  I_R(i,r,m)  = I_L(i,r,m-1)
\union I_R(i,r,m-1)$,  and the result is immediate  from the induction
hypothesis; similarly,  the result follows  if $i$ does not  process a
message
at the  point $(r',m'-1)$.  Thus, we  can assume that  $i$ processes a
message at both $(r,m-1)$  and $(r',m'-1)$.  Moreover, it follows from
Lemma~\ref{pro:msg order}  that $i$ either processes from  the left at
both $(r,m-1)$ and $(r',m'-1)$ or  processes from the right at both of
these points.   Assume without loss  of generality that  $i$ processes
from the  left.  Then,  using the induction  hypothesis, we  have that
$I_R(i,r,m)   =  I_R(i,r,m-1)   =   I_R(i,r',m'-1)  =   I_R(i,r',m')$.
Moreover,
$I_L(i,r,m) =  I_L(L_i,r,m_1) \cup  \{ i\}$, where  $m_1$ is  the time
$L_i$ sent the message 
that
$i$ processes at time $m-1$ in $r$;
$I_L(i,r',m') = I_L(L_i,r',m_1')\cup \{ i\}$,
where $m_1'$ is the time
that 
$L_i$ sent the message
that
$i$ processes at time $m'-1$ in $r'$.
It follows from Lemma~\ref{pro:msg order} that we must have $k = 2k'$,
$L_i$  has  sent  $k'$  messages  left at  the  points  $(r,m_1)$  and
$(r',m_1')$, and has processed $k-1$ messages at both of these points.
By the induction hypothesis, $I_L(L_i,r,m_1) = I_L(L_i,r',m_1')$.  The
desired result follows immediately.  \eprf

\lem\label{lem:atmost} There are at most  two agents that can be first
to learn all the information in  network $N$.  If an agent that can be
first to  learn all the information  is active when it  learns all the
information, then it must be the agent with the highest id.  \elem

\prf Suppose,  by way of contradiction,  that three agents  can be the
first  to learn  all the  information,  say $i_1$,  $i_2$, and  $i_3$.
Suppose that $i^*$  is the agent in $N$ with  the highest id.  Suppose
that the  message that  $i_h$ processed which  caused it 
to
know  all the
information was the
$p_h$th message that $i_h$ processed, for $h = 1, 2, 3$.  First assume
that   $i^*   \notin  \{i_1,i_2,i_3\}$.    It   easily  follows   from
Lemma~\ref{lem:phs} that, for $h = 1,2,3$, either the
$p_h$th message  or the $(p_h-1)$st message that  $i_h$ processed must
have come from $i^*$.
Suppose that for two of $i_1$, $i_2$, or $i_3$, the message that $i_h$
processed from $i^*$ came from the right.
Suppose, without loss  of generality, that these two  agents are $i_1$
and $i_2$.  Now  a simple case analysis shows  that either $i_1$ knows
all the information before $i_2$ in all runs of P2$'$ where $N_r = N$,
or $i_2$ knows all the information before $i_1$ in all runs where $N_r
=  N$.  For  example, suppose  that the  message that  originated with
$i^*$ is the
$p_h'$th message that $i_h$ processed, for $h = 1 ,2 $; note that
$p_h'$     is    either     $p_h$     or    $p_h     -    1$.      (By
Lemma~\ref{lem:asynchronous},
$p_h'$ is same in all runs $r$ such that $N_r = N$.)  If
$p_1' > p_2'$ then it follows from Lemma~\ref{pro:msg order} that
$p_1' \ge p_2' +  2$, and it is easy to see  that $i_1$ must learn all
the information before $i_2$.  Similarly, if
$p_2' > p_1'$,  then it is easy  to see that $i_2$ must  learn all the
information before $i_1$.  Finally, suppose that
$p' = p_1' = p_2'$.  Without loss of generality,
assume that going  from $i^*$ left on the ring,  we reach $i_1$ before
$i_2$.  Then it is easy to see that if
$p_1' =  p_1$, so that  $i_1$ knows it  has all the  information after
processing the  message from  $i^*$, then $i_1$  knows it has  all the
information before $i_2$ in all runs $r$ with $N_r = N$, while if
$p_1 =  p_1' + 1$, then  $i_1$ must learn  it after $i_2$ in  all runs
(since the
$p_1$th  message processed  by  $i_1$ must  originate  with a  process
farther to the left of $i^*$  than $i_2$). Thus, it cannot be the case
that  both $i_1$  and  $i_2$ can  be  first to  learn  the message,  a
contradiction.  A similar contradiction arises if both $i_1$ and $i_2$
process $i^*$'s message from the left.

Thus, it  follows that  $i^* \in \{i_1,  i_2, i_3\}$; without  loss of
generality,
assume 
that
$i^* = i_3$.  Again, if both of $i_1$ and $i_2$ process $i^*$'s
message from the left, or both  process it from the right, then we get
a contradiction as above.
So suppose without loss of generality that $i_1$
processes  $i^*$'s  message from  the  left,  $i_2$ processes  $i^*$'s
message from the right, and
$i^*=i_3$ processes
its
$p_3$th message from the left.  Again, it is easy to show that if
$p_1\le p_3$, then in all runs $r$  with $N_r = N$, $i_1$ knows it has
all the information before $i_3 = i^*$; if
$p_1 > p_3$, then  in all runs $r$ with $N_r =  N$, $i^*$ knows it has
all the information before
$i_1$.  Either way, we have a contradiction.  \eprf

We can  now describe the remainder of 
protocol P2$'$, after an  agent
$i$ learns  all the information.  What happens depends on (a) which
agents can be first to learn all the information, and whether $i$ is one
of them; (b) whether 
$i$ is active or passive  just after learning all the information, and
(c)  whether  the  message  that  results  in  $i$  learning  all  the
information is processed  from the left or the  right.  Note that when
an agent  learns all  the information, it  can easily  determine which
agents can be first to learn all the information.
Rather than writing the pseudocode for P2$'$, we give just an English
description; we do not think that the pseudocode will be more
enlightening.    

\begin{itemize}
\item Suppose that  the only agent that can be first  to learn all the
information  is the leader.   We now  do essentially  what is  done in
Peterson's algorithm.  Suppose that
the message that  resulted in the leader learning  all the information
was processed  from the  left (if the  message was processed  from the
right, the  rest of the argument  remains the same,  replacing left by
right everywhere), the message originated with agent $i$, and was the
$p$th message processed by the leader.  We claim that after processing
the
$p$th message, all  agents other than the leader  will be passive.  If
$i$  is the  leader, this  is  almost immediate.   If $i$  is not  the
leader,  then it  follows from  Lemma~\ref{lem:phs}.  
The  leader then sends its
$(p+1)$st message to the left.  After an agent processes the leader's
$(p+1)$st message, it will then  know all the information.  We require
it to send a message to the left with all the information unless it is
the  leader's right  neighbor.   (Of  course, once  it  knows all  the
information,  the  leader's  right  neighbor  will  realize  that  the
neighbor to the  left is the leader and that  the leader already knows
all the information, so it  does not need to forward the information.)
After  this  process  is  completed,  all  the  agents  know  all  the
information.
\item Suppose that  agent $i$ is the only agent that  can know all the
information  and  $i$   is  passive  when  it  first   knows  all  the
information.  Suppose that
the message  that resulted in  $i$'s learning all the  information was
processed  from the left  (again, the  argument is  similar if  it was
processed from the right), the  message originated with agent $j$, and
was the
$p$th message processed by $i$.  It  is easy to see that $i$ must have
been active just prior to processing the
$p$th message,  for otherwise the agent  to $i$'s left  will learn all
the information before $i$.  Moreover, $i$'s
$p$th message  must have originated  with the leader (since  $i$ could
not have known about the leader  prior to receiving the message, or it
would not have been active).  Then  $i$ sends the message with all the
information back to  the leader, who forwards the  message all the way
around the ring up to the agent to $i$'s right, at which point all the
agents know all the information.

\item Suppose that two passive agents,  say $i$ and $i'$, can be first
to learn all  the information.  Again, it is not hard  to see that $i$
and  $i'$  must  have  been   active  just  before  learning  all  the
information.  If $i$ and $i'$ both first learn all the
information   after   processing    the   $p$th   message,   then   by
Lemma~\ref{lem:phs},  the  $p$th message  of  one  of  them, say  $i$,
originated with  $i^*$.  Suppose without  loss of generality  that $i$
and  $i'$ received  this  message from  the  left.  
Then  $i$ sends  a
message with all the information to the left, where it is forwarded up
to and including  $i^*$; similarly, $i'$ sends a  message to the left,
which is forwarded  up to but not including $i$.   Note that $i'$ will
also receive a  $(p+1)$st message that originates with  $i^*$ from the
right.  After  receiving this message,  $i'$ sends a message  with all
the information to the right up to but not including $i^*$.

\item Suppose that one passive agent,  say $i$, and $i^*$ can be first
to learn all the information.   If they both learn all the information
after receiving  their $p$th message,  then $i$ must have  been active
just  before  receiving the  message,  $i$'s  message originated  with
$i^*$, and $i^*$'s message either originated with $i$ or with an agent
$i'$ such that the $p$th message received by $i'$ originated with $i$,
and  $i'$  becomes  passive  after receiving  this  message.   Suppose
without loss  of generality that  the $p$th message was  received from
the left.   Then $i$ sends a  message with all the  information to the
left where it  is forwarded up to but  not including $i^*$; similarly,
$i^*$ sends a  message with all the information to  the left, where it
is  forwarded up  to but  not including  $i$.  A  straightforward case
analysis shows  that it cannot  be the case  that there exist  $p$ and
$p'$ with  $p \ne p'$ such  that $i$ learns all  the information after
receiving its $p$th message and $i^*$ learns all the information after
receiving the
$p'$th  message.   For if  $p  <  p'$, then  $i$  must  learn all  the
information before $i^*$ in all runs, and if $p' < p$, then $i^*$ must
learn all the information before $i$ in all runs.
\end{itemize}

\noindent This completes the description of P2$'$.

\commentout{ Notice that there are  six possible cases when $i$ learns
that  it  has all  information,  depending  on  whether $i$  has  last
processed  from  the left  or  from the  right,  and  whether $i$  was
passive, active and stayed active, or active and became passive.  Part
(a) of the above theorem covers  the cases when $i$ has last processed
from left  and stays active and  when $i$ has last  processed from the
right and  was passive;  part (b)  covers the case  when $i$  has last
processed  from the  right, was  active,  and either  stays active  or
becomes passive.   Parts (c) and (d)  each cover one of  the other two
remaining cases.

We are now able to characterize the network configurations at the time
agents  first  know  that   they  have  all  the  information.   These
configurations are as follows:
\begin{itemize}
\item[(A)]  Only one  agent $i$  (i.e.,  the leader)  is active  after
processing from the left in phase $p-1$,
and  only  one  process, say  $j$,  is  the  first  to learn  all  the
information This  breaks down into  two subcases: (i) $j=i$,  and (ii)
$j\neq i$.
\item[(B)]  Exactly two  agents, the  leader  $i$ and  agent $j$,  are
active after processing from the left in phase $p-1$,
and  only  one  agent $k$  is  the  first  to  learn  it has  all  the
information.   The  possible  subcases  are:  (i)  $k=i$,  (ii)  $k\in
[i.. j]_L-\{ i,j\}$, (iii) $k=j$ and (iv) $k\in [j .. i]_L-\{i,j\}$.
\item[(C)]  Exactly two agents,  $i$ and  $j$ (one  of them  being the
leader), are active after processing from the left in phase $p-1$,
 and exactly two agents, $k$ and $l$, are the first to learn they have
all the information.  The possible  subcases are: (i) $k=i$ and $l=j$,
(ii) $k\in  [i .. j  ]_L- \{ i,  j\}$ and $l=j$,  and (iii) $k  \in [i
.. j]_L-\{ i, j\}$ and $l \in [j .. i]_L-\{i,j\}$.
\item[(D)] Exactly  three agents, the  leader $i$, and agents  $j$ and
$k$, are active after processing from the left in phase $p-1$
and  only  one  agent $l$  is  the  first  to  learn  it has  all  the
information.
\item[(E)] Exactly  three agents, the  leader $i$, and agents  $j$ and
$k$, are active after processing from the left in phase $p-1$
and  exactly  two  agents  are  the  first  to  learn  they  have  all
the
information.
\item[(F)] There  are at least three  active agents that  send left in
phase $p$, at most three of them  send right in phase $p$, and at most
three agents are the first to know they have all the information.
\end{itemize}

P2$'$ can  now be described  for each of  the above cases.   Since the
descriptions are similar in spirit,  we present in detail only some of
the cases.

P2$'$(A): (i)
For all agents  $l$, other than $R_i$, when $l$ knows  that it has all
the information, $l$ sends its  new information to the left and stops.
(ii)  When  $j$ knows  that  it has 
all the 
information,  $j$  sends its  new
information both  left and  right, and then  stops.  For all  $l\in [j
.. i]_L-\{  i,j\}$, when $l$ has  all the information, $l$  sends left its
new information and  then stops.  For all $l\in  [j..  i]_R-\{ i,j\}$,
when $l$  has all  
the
information, $l$ sends  right its  new information,
unless $R_l=i$. $i$  sends no message after processing  from the right
in phase $p$.

P2$'$(B):
  (iv) When  $k$ first  knows it has  all the information, $k$  sends both
left and right its new information,  and then stops.  For all $l\in [k
.. i]_L-\{i,k\}$, when $l$ first  knows all the information, $l$ sends
left   its  new   information   and  then   stops.    For  all   $l\in
[k.. j]_R-\{j,k\}$, when  $l$ has all the information, $l$  sends left its
new information and then stops.  For all $l\in [j.. i]_R-\{ i\}$, when
$l$ has all  
the
information, $l$ sends right its  new information, unless
$R_l=i$; $u$ then stops.

P2$'$(D) and  P2$'$(E): Suppose  that there are  3 active  agents that
send left in  phase $p$: $i$ (the  leader), $j$ and $k$, and  $j \in [
i..k ]_L-\{ i,k\}$.  By  Lemma~\ref{lem:first to k}(a), no agent first
learns that it has all the information
after processing from  the left in phase $p-1$ when  it was active, or
after processing from the right in phase $p-1$ when it was passive.
By Lemma~\ref{lem:phs}(a), $i$  does not know about $k$  when it sends
left in  phase $p$; same goes  for $j$ and  $i$, and for $k$  and $j$.
$k$ processes  from the right in  phase $p$ a  message originated with
$j$; this  means that $k$ does not  learn about $i$ from  its right in
phase $p$.  Since $\id_i > \id_k$  (as $i$ is the leader), $k$ did not
know about $i$ at the beginning of phase $p$, since otherwise it would
have  been  passive.   It follows  that  $k$  does  not have  all  the
information after processing from the right in phase $p$.

We proceed  with a case  analysis on whether  $k$ stays active  or not
after processing from the right in phase $p$.  For simplicity, we show
here the case when $k$ stays active after processing from the right in
phase $p$,  i.e., $\id_j<\id_k$.  $k$  then sends right in  phase $p$.
Since $j$ was active  when it sent to the left in  phase $p$, it means
that $j$  did not know about $k$  at the beginning of  phase $p$.  $j$
does not  learn about  $k$ from  $i$ either, since  $i$ does  not know
about $k$ when  it sends left in phase $p$.  This  means that $j$ does
not have all the information  after processing from the right in phase
$p$.  We have already shown that no agent $l$ different from $i$, $j$,
and $k$  knows that it has  all the information after  processing from the
right in  phase $p$, since  $l$ is passive  at the beginning  of phase
$p$.  It follows that at most one agent may have all the information after
processing from the right in phase $p$, and that agent is $i$.
  \begin{itemize}
  \item[(i1)]  Suppose that  $i$ learns  that it  has  all the information
after processing from the right in  phase $p$.  Since $k$ did not know
about $j$ when it  sent to the left in phase $p$,  it must be that $i$
knew about  $j$ before processing from  the right in  phase $p$, i.e.,
when it  sent left in  phase $p$.  By Lemma~\ref{lem:phs}(a),  this is
only  possible if the  message $i$  processed from  the left  in phase
$p-1$ originated  with $j$.  It is  not difficult to  notice that this
ensures that there exists some  agent $l\in [k..j]_R -\{k\}$ such that
$l$ learns that it has  all the information after processing from the left
in phase $p$ the message originated with $k$.  By Lemma~\ref{lem:first
to  k}(c),  the only  possible  case is  $l=j$.   We  define P2$'$  as
follows: after processing  from the right in phase  $p$, $i$ sends its
new   information  both  left   and  right;   all  agents   $l'\in  [i
..L_j]_R-\{i,L_j\}$ send right  their new information after processing
from  the  right  in  phase  $p$,  and then  stop;  all  agents  $l\in
[i..R_j]-\{i,R_j\}$ send left  their new information, after processing
from $i$ in phase $p+1$, and then stop.
  \item[(i2)]  Suppose  that  $i$  does  not learn  that  it  has  all
the
information  after  processing  from  the  right  in  phase  $p$.   By
Lemma~\ref{lem:first to k}(c), no agent $l\in [i..k]_R-\{i,k\}$ learns
that it  has all the information after  processing from the  left in phase
$p$.  Similarly, no  agent $l\in [k..j]_R-\{ k,j\}$ knows  that it has
all  the information after  processing from  the left  in phase  $p$.  The
following cases are possible:
  \begin{itemize}
  \item[(i2.1)] Both $k$ and $j$  learn that they have all the information
after  processing from  the  left in  phase  $p$. We  define P2$'$  as
follows: after processing  from the left in phase  $p$, $k$ sends left
its new information  and stops, and $j$ sends both  left and right its
new  information  and  then  stops;  for  all  $l\in  [j..R_k]_L  \cup
[k..R_i]_L  -\{j, k\}$, after  learning all  the information  from the
left, $l$ sends left its new information and then stops; for all $l\in
[j..L_i]-\{j\}$,  after learning  all the information  from the  left, $l$
sends right  its new  information and then  stops; after  learning all the 
information from the left, $i$ stops.
  \item[(i2.2)]  $k$   learns  that  it  has   all the  information  after
processing from the left in phase  $p$, but $j$ does not. Then clearly
$i$ will learn  that it has all the information  after processing from the
left  in  phase   $p$  the  message  originated  with   $k$,  and,  by
Lemma~\ref{lem:first to k}(c),  no agent other than $i$  and $k$ learn
that they have all the information after processing from the left in phase
$p$.  We  define P2$'$ as follows:  after learning that  they have all the
information, both $i$ and $k$ send left their new information and then
stop; for all  $l\in [i..R_k]-\{i, R_k\}$, after learning  that it has
all the information (from  the right), $l$ sends left  its new information
and then  stops; for all  $l\in [k..R_i]-\{ k, R_i\}$,  after learning
that it has  all the information (from its right), $l$  sends left its new
information and then stops.
  \item[(i2.3)]  $j$   learns  that  it  has   all  the information  after
processing from the left in phase  $p$, but $k$ does not. According to
P2$'$, $j$ will then send both  right and left its new information and
stop; for  all $l\in [j..L_i]-\{j\}$,  after learning that it  has all
information (from the  left), $l$ sends right its  new information and
then  stops; when $i$  learns that  it has  all the information  (from the
left), $i$ stops;  for all $l\in [j..R_i]-\{ j\}$,  $l$ sends left its
new information after  learning that it has all  the information (from the
right), and  then stops; finally,  $R_i$ stops after learning  that it
has all the information.
  \item[(i2.4)]  $j$  and  $k$  do   not  learn  that  they  have  all the 
information after  processing from  the left in  phase $p$. It  is not
difficult to see that $i$ will  know that it has all the information after
processing from  the left  in phase $p$,  having heard about  $k$ from
both  left and  right. $i$  then sends  left its  new  information and
stops; all  other agents  in the ring  will learn all  the information
after processing  from the right in  phase $p+1$, send  left their new
information (with the exception of $R_i$), and then stop.
  \end{itemizne}
\end{itemize}
}
Having completed the description of P2$'$, we can finally prove that
P2$'$ de facto implements $\Pgcb^{GC}$ in contexts where (i) all
networks are bidirectional rings and (ii) agents have distinct
identifiers.   
Let $(\gamma^{\mathit{br,u}},\pi)$ denote  the interpreted context for
global computation
where  the initial  states  are the  bidirectional  rings with  unique
identifiers. Suppose that $\ogen$  is an order generator that respects
protocols,
$\rkgen$  is a  deviation-compatible ranking  function, and  $\J  =$ $
(\Rp(\gamma^{\mathit{br,u}}),$  $  \pi,  \mu_{\gamma^{\mathit{br,u}}},
\ogen(\mathit{P2}'), \rgen(\mathit{P2}'))$
is the  interpreted system  corresponding to P2$'$  in the  cb context
$\chi^{\mathit{br,u}}=(\gamma^{\mathit{br,u}},\pi,\ogen,\rgen)$.
Proving that P2$'$ de facto  implements $\Pgcb^{GC}$ in the cb context
$\chi^{\mathit{br,u}}$       amounts       to       showing       that
$P2'_i(\ell)={\Pgcb^{GC}}^{\J}_i(\ell)$ for every local state
$\ell$       such      that       there       exists      $r       \in
\Rrep(P2',\gamma^{\mathit{ur,u}})$ and $m$  such that $\ell = r_i(m)$.
That is, for all $r \in \Rrep(P2',\gamma^{\mathit{ur,u}})$ and times $m$, we
must show that
$P2'_i(r_i(m))=\sfa$
iff $(\J,r,m,i)\models \varphi_{\sfa}$,
where  $\varphi_{\sfa}$  is the  precondition  in $\Pgcb$  for
action $\sfa$.

\commentout{ We can then prove  \lem\label{lem:p2} For all runs $r$ of
P2$'$ in  the context  $\gamma^{\mathit{br,u}}$, times $m$  and agents
$i$   in   $N_r$,   if   $(\J,r,m,i)\models   (\neg   \ainfo)$,   then
$P2'_i(r_i(m))={\Pgcb^{GC}}^{\J}_i(r_i(m))$.  \elem }
\commentout{ The  above lemma  proves part (c)  of Theorem~\ref{thm:le
proofs}, since clearly no agent has all the information
prior to  the last phase of  the algorithm.  We can  show however that
there  are bidirectional  rings  with distinct  identifiers for  which
P2$'$ follows $\Pgcb^{GC}$  up to the last phase, but  not in the last
phase, too.  That is, there  exists an agent that has all the information,
believes that  sending the new information  to left (or  right) is not
necessary, but according  to P2$'$ it should send  the new information
anyway.

We  focus now on  changing P2$'$  such that  the resulting  program de
facto  implements $\Pgcb^{GC}$.   Again, this  modification is  not as
straightforward as in the case of LCR$'$.  The idea is to ensure that,
once  an agent  has all  the  information, it  does not  send the  new
information  unnecessarily.  One  problem  with P2$'$  is  that it  is
possible  for $\id$,  $\val_L$ and  $\val_R$ to  all be  distinct, and
still the  agent to have all the information  and to know that  it has all
the
information.  Perhaps the  tests an agent performs to  check if it has
all  the information  can  be   refined  so  that  the  resulting  program
implements  $\Pgcb^{GC}$.  One  possibility is  to ensure  that agents
keep  track  of  the  last  {\em  two}  messages  received  from  each
direction.  If $\val'_R$ (resp. $\val'_L$) is the last but one message
received from right (resp. left),  then we can enforce that $i$ checks
if it has  all the information not just by  verifying whether $\id=\val_R$
or     $\val_L=\val_R$    holds,     but     also    $\val'_L=\val_R$,
$\val'_L=\val'_R$,  or $\val_R=\val'_R$.   A  similar modification  is
made  for the  case when  $i$ decides  whether to  send right  or not.
However, keeping  track of  the last two  messages received  from each
direction is  still not enough  to implement $\Pgcb^{GC}$. Here  is an
example where  agents running our suggested  algorithm eventually have
all the information and still $\val_L$, $\val'_L$, $\val_R$, $\val'_R$
and $\id_i$ are all distinct.

\xam\label{xam:counter2}      Consider     the      following     ring
(counterclockwise):
$$[ i_3=200, 2, 3 ,1, 180, 2, 3, 1, 190, 2, 3, 1, 170, 2, 3, 1, u=400,
2, 3, 1, 80, 2, 3, 1, 90, 2, 3, 1, 70,$$
$$ 2,3,  1, p=100, 2,  3, 1, 80',2,  3, 1, 90',2,  3, 1, 70',2,  3, 1,
j_3=300, 2, 3, 1, 180', 2, 3, 1, v=190',$$
$$  170', 180', 160'].$$  The ring  is quite  regular, except  for the
interval from $j_3$ to $i_3$, to  the left of $j_3$, i.e., $(j_3 \dots
i_3)_L$.   In particular,  $v=190'$  plays an  important role.   After
everybody processes right in phase 1, all 2 and 1 become passive; also
160$'$ and  170$'$ become passive.  After everybody  processes left in
phase   1,   all  3   become   passive,   plus   180$'$.   $v$   knows
$\left[180'\dots  170'  \dots  v=190'  \dots  1\right]$,  $i_3$  knows
$\left[3  \dots  2  \dots  i_3=200  \dots  160'\right]$,  $j_3$  knows
$\left[3\dots 2\dots  j_3=300\dots 1\right]$, $p$  knows $\left[3\dots
2\dots  p=100\dots  1\right]$,  and  finally $u$  knows  $\left[3\dots
2\dots 400=u\dots 1\right]$.

After everybody  processes right  in phase 2,  the active  agents are:
$i_3=200$, 190, u=400, 90, p=100, 90$'$, $j_3=300$, and $v=190'$.  $v$
knows  $\left[180'\dots  170'\dots  v=190'\dots 1\dots  3\dots  2\dots
180'\dots  1\right]$, $i_3$  knows  $\left[3\dots 2\dots  i_3=200\dots
160'\dots 180'\dots 170'\dots 190'\dots 1\right]$.  $j_3$ knows
$$\left[3\dots  2\dots  j_3=300\dots  1\dots  3\dots  2\dots  70'\dots
1\right],$$  $p$ knows $\left[3\dots  2\dots p=100\dots  1\dots 3\dots
2\dots 70\dots  1\right]$, $u$ knows  $$\left[3\dots 2\dots u=400\dots
1\dots  3\dots  2\dots 170\dots  1\right],$$  190 knows  $\left[3\dots
2\dots 190\dots 1\dots 3\dots 2\dots 180\dots 1\right]$, 90 knows
$$\left[3\dots 2\dots 90\dots 1\dots 3\dots 2\dots 80\dots 1\right],$$
and finally  90$'$ knows  $\left[3\dots 2\dots 90'\dots  1\dots 3\dots
2\dots 80'\dots 1\right]$.

After  agents process  left in  phase 2,  the only  active  agents are
$i_3$, $j_3$,  $p$ and $u$.  $u$  is active after  processing right in
phase 3  (from agent $i_3$), and  will have all  the information after
processing left  in phase 3 (from $j_3$),  even if $\val_L(=id(j_3))$,
$\val'_L$, $\val_R(=id(i_3))$, $\val'_R$ and $\id_u$ are all distinct.
\exam Based on the construction above, we conjecture that, maintaining
the order in which messages are  processed as in P2$'$, agents need to
keep track of {\em all}  the messages received from both directions in
order to implement $\Pgcb^{GC}$.

We next consider a different modification of P2$'$ that this time will
de  facto  implement $\Pgcb^{GC}$.   We  call  our modified  algorithm
P2$''$. Agents  following P2$''$  act exactly as  in P2$'$ as  long as
they do  {\em not} know that  they have all the  information.  When an
agent knows that it has all the information, i.e., when $\ainfo$ holds wrt
$i$, then $i$ simulates P2Stop, which is just like P2$'$, except that,
when  an  agent has  all the  information,  it  simply stops.   Before  we
proceed, we need to analyze  P2Stop in more in detail.  The pseudocode
for P2Stop is as follows:
$$
\begin{array}{l}
{\bf     do~until}~      \mathit{done=1}\\     \quad     {\bf     if}~
\mathit{status=active}~{\bf    then}\\   \quad    \quad    {\bf   if}~
\mathit{begin\_phase}  =  1~  {\bf  then}\\  \quad  \quad  \quad  {\bf
if}~(\neg  \ainfo)~{\bf  then}~\send_{\lefta}(\newinfo)\\ \quad  \quad
{\bf    if}~   (\mathit{begin\_phase=0})\wedge   (\mathit{wl=0})\wedge
(RQ\neq \perp)~{\bf  then}\\ \quad \quad \quad  {\bf if}~(\neg \ainfo)
\wedge  (\val_R<\id)  ~{\bf  then} ~\send_{\righta}(\newinfo)\\  \quad
{\bf  if}~\mathit{status=passive}~{\bf then}\\  \quad \quad  {\bf if}~
(\neg         \ainfo)\wedge        (LQ\neq         \perp)~        {\bf
then}~\send_{\righta}(\newinfo)\\ \quad \quad  {\bf if}~ (\neg \ainfo)
\wedge (RQ\neq \perp) ~{\bf then}~\send_{\lefta}(\newinfo). \\
\end{array}
$$  Here we  assume that  $\mathit{done}$ is  set to  1  when $\ainfo$
holds.

Clearly, with  P2Stop few agents  have all the information;  among the
agents that do not have all  the information, none of them can set the
local variable $\mathit{done}$ to 1.  If  $r$ is a run of P2Stop (on a
bidirectional ring with distinct identifiers), let $K(N_r)$ be the set
of agents  that eventually have  all the information.  For  each agent
$i$ not  in $K(N_r)$, let  $W(i,N_r)$ be the  agent $i$ is  waiting to
processes from, after $i$ has processed the last message in P2Stop.%
\footnote{It is not difficult to see  that both the set of agents that
eventually have all the information, and the direction the rest of the
agents are waiting from depend only  on the ring $N_r$, and not on the
entire run $r$.}  For example, if  $i$ is active, not in $K(N_r)$, and
the last message it processed in $r$ came from its left (resp. right),
then $W(i,N_r)=R_i$ (resp. $W(i,N_r)=L_i$).

Coming back to P2$''$, if  $i$ has all the information, then $i$ simulates
P2Stop  and determines,  for each  of its  neighbors $j$,  whether $j$
belongs to $K(N_r)$,  and if not, then whether $j$  is waiting from it
in P2Stop, i.e., if $W(j,N_r)=i$. $i$ sends its new information to $j$
only if both tests succeed.  After $i$ decides whether to send its new
information to its left or  right neighbors, and sends the information
accordingly, $i$  stops.  When $i$  has all the information and  sends its
new information to  a neighbor $j$, then part  of this new information
is the  fact that $i$  has all the  information. We can  then optimize
further and ensure that $i$ does {\em not} send its new information to
$j$ if  $i$ has got  all the information  directly from $j$.   To make
this  precise, let  $\mathit{waits\_for\_me}_{\name}$  be a  primitive
proposition that  holds wrt the  situation $(r,m,i)$ exactly  when, if
$j=\mu_{\gamma^{br,u}}(r(m),i,\name)$,  then  $j\not  \in  K(N_r)$  and
$W(j,N_r)=i$.   Let  $\mathit{all\_info\_from}_{\name}$  hold wrt  the
situation $(r,m,i)$ iff $i$ has all the information at time $m$ in $r$ and
it got all the information from $j$, i.e., $j$ had all the information
when it last sent to $i$.  Here is the pseudocode for P2$''$:
$$
\begin{array}{l}
{\bf do~until}~  \mathit{done=1}\\ \quad  {\bf if} ~\neg  \ainfo ~{\bf
then}\\  \quad  \quad  {\bf if}~  \mathit{status=active}~{\bf  then}\\
\quad \quad  \quad {\bf if}~ (\mathit{begin\_phase} =  1)~ {\bf then}~
\send_{\lefta}(\newinfo)\\    \quad     \quad    \quad    {\bf    if}~
(\mathit{begin\_phase=0})\wedge  (\mathit{wl=0})\wedge (RQ\neq \perp)~
{\bf  then}\\ \quad  \quad \quad  \quad {\bf  if}~  (\val_R<\id) ~{\bf
then}      ~\send_{\righta}(\newinfo)\\      \quad     \quad      {\bf
if}~\mathit{status=passive}~{\bf then}\\  \quad \quad \quad  {\bf if}~
(LQ\neq  \perp)  ~{\bf  then}~\send_{\righta}(\newinfo)\\ \quad  \quad
\quad {\bf if}~  (RQ\neq \perp) ~{\bf then}~\send_{\lefta}(\newinfo)\\
\quad  {\bf  else}~\\   \quad  \quad  {\bf  ~if}~A=\lbrace  \name\:|\:
\mathit{waits\_for\_me}_{\name}\wedge                              \neg
\mathit{all\_info\_from}_{\name}\rbrace             ~             {\bf
then}~\send_{A}(\newinfo).\\
\end{array}
$$ (Here we assume that $\mathit{done}$  is set to 1 after $i$ has all
the information and has sent its new information to all the agents waiting
from it in P2Stop (if any).)

\lem\label{lem:p2''}  P2$''$ de facto  implements $\Pgcb^{GC}$  is all
contexts  where (i)  all networks  are bidirectional  rings,  and (ii)
agents have distinct identifiers.  \elem

\subsection{Detailed proofs}\label{sec: cbb proofs}
\olem{lem:lcr1} For all  runs $r$ of LCR$'$, times  $m$ and agents $i$
in $N_r$, if messages are  delivered in FIFO order, then the following
all hold:
\begin{itemize}
\item[(a)] $I_R(i,r,m)$ is an interval of agents starting from $i$ and
going to the right of $i$.
\item[(b)] If, at time $m$, $i$ processes a message, then this message
was  sent  by  $R_i$  at  some  time  $m'$  and  $I_R(R_i,r,m')\supset
I_R(i,r,m-1)-\{    i\}$,    $I_R(i,r,m)\supset   I_R(i,r,m-1)$,    and
$I_R(i,r,m)=\{ i\} \cup I_R(R_i,r,m')$.
\item[(c)] If, at time $m$,  $i$ processes a message (from right), and
$i$  does not  have  all  the information  after  processing it,  then
$\max(\val_R)$ is  the id of  the rightmost agent  in $I_R(R_i,r,m')$:
$\max(\val_R)=id_{R(R_i,r,m')}$.
\item[(d)] The  first agent to have  all the information  is the agent
$i_{\max}(r)$ with the  maximum id in $N_r$.  Prior  to having all the
information, $i_{\max}(r)$ sent left only once.
\item[(e)] For any  agent $i$ other than $i_{\max}(r)$,  if $m$ is the
first      time      $i$      has      all   the   information,      then
$I_R(i,r,m-1)=[i..i_{max}(r)]_R$,  all agents  in $[i_{\max}(r)..i)_L$
already  had  all  the  information  by  time  $m$,  while   no  agent  in
$(i..i_{\max}(r))_L$ has all the information at time $m$.
\end{itemize}
\eolem \prf  We omit the  proof for (a)  and (b) as it  simply follows
from  the  description  if  LCR$'$  and  is  similar  to  the  one  in
Lemma~\ref{lem: i-segment}, which we show in detail.

We  prove (c)  by induction  on $m$.   If $m$  is the  first  time $i$
processes a  message from  $R_i$, then, as  messages are  delivered in
FIFO  order,  $m'$  was  the   first  time  $R_i$  sent  left  and  so
$I_R(R_i,r,m')=\{  R_i\}$;  then  clearly $\max(\val_R)=id_{R_i}$  and
$R_i$ is the rightmost agent in $\val_R$.  Suppose now that $m$ is not
the first time  $i$ processes a message from  $R_i$.  Let $\bar{m}$ be
the  last time,  prior  to $m$,  when  $i$ processed  from $R_i$;  let
$\bar{m}'$   be   the   time    $R_i$   sent   this   message.    Then
$I_R(i,r,m-1)=\{i\}\cup  I_R(R_i,r,\bar{m}')$  and $I_R(i,r,m)=\{  i\}
\cup  I_R(R_i,r,m')$.   As  messages  are  delivered  in  FIFO  order,
$\bar{m'}<  m'$  and so  $i$'s  new information  at  time  $m$ is  the
interval $(R(R_i,r,\bar{m}')..R(R_i,r,m')]_R$.  Since  $m'$ is not the
first time $R_i$ sent left in $r$, it is the case that, prior to $m'$,
$R_i$ processed a message from its right.  Let $m''$ be the time $R_i$
processed  right, prior  to $m'$.   Since $i$  does not  have  all the
information after processing right  at time $m$, i.e., $I_R(i,r,m)$ is
not a ring,  then $R_i$ did not have all the  information when it sent
left at time $m'$, either.   We can apply the induction hypothesis and
deduce that the  agent with the maximum id  in $R_i$'s new information
at time $m''$ is the rightmost agent in $I_R(R_i,r,m'')$.  Since $R_i$
sent left after processing from right at time $m''$, and $R_i$ did not
have all the  information after processing right at  $m''$, it must be
that the maximum  id in $R_i$'s new information was  the maximum id in
$I_R(R_i,r,m'')$.  It  follows that the  agent with the maximum  id in
$I_R(R_i,r,m'')$ is $R(R_i,r,m'')$.   Since $R_i$ processes no message
from   right  between   $m''$  and   $m'$,   it  is   the  case   that
$I_R(R_i,r,m')=I_R(R_i,r,m'')$.  Then the agent with the maximum id in
$I_R(R_i,r,m')$ is  $R(R_i,r,m')$, and  so $R(R_i,r,m')$ also  has the
maximum id in $(R(R_i,r,\bar{m}')..R(R_i,r,m')]_R$.

We  prove (d)  now. By  the description  of LCR$'$,  it is  clear that
$i_{\max}(r)$ sends left its id initially, and then skips sending left
until it has all the information, as  no maximum id in the new information
can be larger  than its own id.  Suppose by  way of contradiction that
$i_{max}(r)$  is  not  the  first  agent  in $N_r$  to  have  all  the
information.  Let  $i\neq i_{\max}(r)$  be this agent  and $m$  be the
time $i$ has all the information for the first time, i.e., $I_R(i,r,m)$ is
a ring. Then  $i_{\max}(r)$ must have sent left  at least twice, prior
to $m$. This means that $i_{\max}(r)$ already had all the information when
it  sent  last  to  $i$.   We  have reached  a  contradiction  and  so
$i_{\max}(r)$ is the  first agent to have all  the information.  It is now
straightforward  to prove  (e) by  induction on  the length  of $N_r$.
\eprf

Furthermore, we  can characterize the knowledge  that agents following
LCR$'$ have at  each point in time.  To make  this precise, let ${\cal
N}_{\mathit{ur,u}}$ be the set of all unidirectional rings with unique
identifiers;  let ${\cal  N}^u(i,r,m)$  be the  set of  unidirectional
rings  that $i$  considers possible  at time  $m$ in  run  $r$: ${\cal
N}^u(i,r,m)=\{   N\in   {\cal   N}_{\mathit{ur,u}}\:  |   \:   \exists
(r',m',i').  $  $r'_{i'}(m')=r_i(m)~\mbox{and}~N_{r'}=N\}$.  If $I$ is
an  interval of agents,  let ${\cal  N}_{ur,u}(I)$ be  the set  of all
networks in  ${\cal N}_{\mathit{ur,u}}$ that contain  the interval $I$
(all the agents in $I$, in the same order as in $I$).

\cor\label{cor:k-lcr} For all runs $r$ of LCR$'$, times $m$ and agents
$i$ in $N_r$, ${\cal N}^u(i,r,m)={\cal N}_{ur,u}(I_R(i,r,m))$.  \ecor

\olem{lem:lcr2}   For  all  runs   $r$  of   LCR$'$  in   the  context
$\gamma^{\mathit{ur,u}}$,  times  $m$ and  agents  $i$  in $N_r$,  the
following hold:
\begin{itemize}
\item[(a)]                   $(\J,r,m,i)\models                   \neg
\varphi_{\send_{\righta}(\newinfo)}$    and    if   $(\J,r,m,i)\models
\varphi_{\send_{\lefta}(\newinfo)}$,                               then
$LCR'(r_i(m))=(\send_{\lefta}(\newinfo),\noop)$.
\item[(b)]  If  $LCR'(r_i(m))=(\send_{\lefta}(\newinfo),\sfa_R)$, then
$\sfa_R=\noop$ and,  unless $i$  has all the information  at time  $m$ and
$L_i=i_{\max}(r)$,                                   $(\J,r,m,i)\models
\varphi_{\send_{\lefta}(\newinfo)}$.
\end{itemize}
\eolem \prf We prove (a) first. By Corollary~\ref{cor:lcr}, eventually
all agents in $N_r$ will  have all the information.  Furthermore, the fact
that LCR$'$ solves  the leader election problem is  publicly known, so
$i$  knows  that  eventually  all   agents  in  $N_r$  will  have  all
the information. In particular, $i$  knows that eventually $R_i$ will have
all  the information.  This  means that  for all  $r'$ run  of LCR$'$,
times $m'$ and agents $i'$ such that $r'_{i'}(m')=r_i(m)$,
$$(\J,r,m',i')\models      \Diamond       (      (\exists      \name^*
\Calls(\righta,I,\name^*)          \wedge          B_{\righta}(\name^*
~\mbox{'s}~\inewinfo)) \vee \exists v B_{\righta}(f = v)).$$ Since $i$
does  not send right  in state  $r_i(m)$ and  $r'_{i'}(m')=r_i(m)$, it
means  that $i'$  does not  send right  in state  $r'_{i'}(m')$. This,
together with the fact  that $r'$ has minimal rank (as it  is a run of
LCR$'$),         ensures         that        $\closest(\intension{\neg
\send_{\righta}(\newinfo)}_{\J},r',m',i')=\{(r',m',i')\}$.   {F}rom the
semantics of belief, we deduce that $(\J,r,m,i)\models B_I (\Diamond (
(\exists \name^*  \Calls(\righta,I,\name^*) \wedge B_{\righta}(\name^*
~\mbox{'s}~\inewinfo))  \vee \exists  v B_{\righta}(f  =  v)))$, which
implies             that            $(\J,r,m,i)\models            \neg
\varphi_{\send_{\righta}(\newinfo)}$.     The     proof    that,    if
$(\J,r,m,i)\models      \varphi_{\send_{\lefta}(\newinfo)}$,      then
$LCR'(r_i(m))=(\send_{\lefta}(\newinfo),\noop)$       follows      now
essentially identical to the corresponding proof for P2$'$.

To            prove             (b),            assume            that
$LCR'(r_i(m))=(\send_{\lefta}(\newinfo),\sfa_R)$.      Then    clearly
$\sfa_R=\noop$.   Assume  first  that   $i$  does  not  have  all  the
information at  time $m$.  Since $i$  sends left at time  $m$, and $i$
does not have all the  information, it follows from the description of
LCR$'$ that $i$ has new information and the maximum id in $I_R(i,r,m)$
is   the  id   of   an  agent   it   has  just   learned  about.    By
Lemma~\ref{lem:lcr1}(c), the agent with the maximum id in $I_R(i,r,m)$
is  $R(i,r,m)$.  Consider  the ring  $N'$ that  contains  the interval
$I_R(i,r,m)$ and  such that  there is a  link from $i$  to $R(i,r,m)$.
Then      $N'\in     {\cal     N}_{ur,u}(I_R(i,r,m))$,      and     by
Corollary~\ref{cor:k-lcr}, $i$ considers  possible that the network is
$N'$, i.e., there exist run $r'$ of LCR$'$ in $\gamma^{\mathit{ur,u}}$
with  $N_{r'}=N'$,  time  $m'$  and  agent  $i'$  in  $N'$  such  that
$r'_{i'}(m')=r_i(m)$.  In $r'$,  the agent with the maximum  id in the
ring is $R_{i}$.  Consider now  a run $\bar{r}$ just like $r'$, except
that  $i'$ does  not  send left  at  time $m'$.   Then  in $r'$  agent
$R_{i}(=i_{\max}(\bar{r}))$  will  not   process  from  left,  and  so
$i_{\max}(\bar{r})$ will never learn the whole information. (Moreover,
no   agent   will   learn   all   the information.)    This   means   that
$(\J,\bar{r},m',i')\models        \Box        ((\exists        \name^*
\Calls(\lefta,I,\name^*) \wedge (\neg  \exists v B_{\lefta}(f = v)))$.
It     follows      that     $(\J,r',m',i')\models     \neg     ((\neg
\doact(\send_{\lefta}(\newinfo)))\RCond  (\Diamond (  (\exists \name^*
\Calls(\lefta,I,\name^*)           \wedge           B_{\lefta}(\name^*
~\mbox{'s}~\inewinfo))  \vee  \exists  v  B_{\lefta}(f =  v))))$.   It
follows that $(\J,r,m,i)\models \varphi_{\send_{\lefta}(\newinfo)}$.

Consider now the  case when $i$ has all the information  at time $m$. Then
$i$ knows  which agent is  $i_{\max}(r)$.  By Lemma~\ref{lem:lcr1}(e),
$i$ also  knows that  no agent in  $(L_i..i_{\max}(r))_L$ has  all the
information  and  that all  agents  in  $[i_{\max}(r)..i)_L$ have  all
the information. In particular, $i$ knows that $L_i$ does not have all the
information, unless  $L_i = i_{\max}(r)$.  If  $L_i \neq i_{\max}(r)$,
then,  since  $i$  has  all  the information, $i$  knows  that  $L_i  \neq
i_{\max}(r)$.  Let $\bar{r}$  be a run just like  $r$, except that $i$
does not send left its  new information. Then clearly $L_i$ will never
gain any  more information than what  it already has at  time $m$, and
since $L_i$  does not have all  the information, it means  that $L_i$ will
never  have   all  the information.   We   can  write  this   formally  as
$(\J,\bar{r},m,i)\models         \Box         ((\exists        \name^*
\Calls(\lefta,I,\name^*) \wedge  \neg \exists v  B_{\lefta}(f = v)))$.
Reasoning    as   above,    it    follows   that    $(\J,r,m,i)\models
\varphi_{\send_{\lefta}(\newinfo)}$.   On the  other hand,  if  $L_i =
i_{\max}(r)$,  then  $i$ knows  that  this is  the  case;  in any  run
$\bar{r}$ just  like $r$,  except that $i$  does not send  left, $L_i$
clearly will  have all the information,  i.e., $(\J,r,m,i)\models \neg
\varphi_{\send_{\lefta}(\newinfo)}$.  \eprf }
\commentout{
\olem{pro:msg order}
For all runs $r$ of P2$'$, times $m$, and agents $i$ in $N_r$
\begin{itemize}
\item[(a)]  if $i$  is  active at  time  $m$, then  $i$'s sequence  of
actions
in the time interval $[0,m)$ is  a prefix of the sequence (SL, PR, SR,
PL)$^*$;
\item[(b)] if  $i$ is passive  at time $m$  and $i$ became  passive at
time
$m' \le  m$ after processing a  message from the  right (resp., left),
then
$i$'s  history  in the  time  interval $[m',m]$  is  a  prefix of  the
sequence (PL, SR, PR, SL)$^*$
(resp., (PR, SL, PL, SR)$^*$).
\end{itemize}
\eolem } \commentout{
\olem{lem: i-segment}
For all runs $r$ of P2$'$ and times $m$ the following hold:
\begin{itemize}
\item[(a)] $I_R(i,r,m)$ is an interval of agents starting with $i$ and
going to the  right of $i$, and $I_L(i,r,m)$ is  an interval of agents
starting with $i$ and going to the
left of $i$.
\item[(b)]
If, at time $m$, $i$ processes a message
from the right sent by $R_i$ at time $m'$, then
  \begin{itemize}
  \item[(i)]    $I_R(R_i,r,m')\supset   I_R(i,r,m-1)-   \{    i   \}$,
$I_R(i,r,m)\supset  I_R(i,r,m-1)$,   and  $I_R(i,r,m)=\{  i   \}  \cup
I_R(R_i,r,m')$; and
  \item[(ii)] $I_L(i,r,m)=I_L(i,r,m-1)$.
  \end{itemize}
\item[(c)] If, at time $m$, $i$ processes a message from the left sent
by $L_i$ at time $m'$, then
  \begin{itemize}
  \item[(i)]   $I_L(L_i,r,m')\supset   I_L(i,r,m-1)   -  \{   i   \}$,
$I_L(i,r,m)\supset   I_L(i,r,m-1)$,   and   $I_L(i,r,m)=\{  i\}   \cup
I_L(L_i,r,m')$; and
  \item[(ii)] $I_R(i,r,m)=I_R(i,r,m-1)$.
\end{itemize}
\item[(d)]  If $i$  has  processed a  message  from the  right in  the
interval
$[0,m]$,   then  $$\max_{\{m'  \le   m:  \val_R(i,r,m')   \ne  \bot\}}
\val_R(i,r,m')$$  is  the maximum  id  of  the  agents in  $I_R(i,r,m)
-\{i\}$, where  $\val_R(i,r,m')$ is the value of  agent $i$'s variable
$\val_R$ at  the point $(r,m')$; if  $i$ has processed  a message from
the left in the interval $[0,m]$, then
$$\max_{\{m' \le m: \val_L(i,r,m') \ne \bot\}} \val_L(i,r,m')$$ is the
maximum id in $I_L(i,r,m) -\{i\}$.
\item[(e)]  $i$ is  active at  time $m$  if and  only if  $i$  has the
largest
id in $I_L(i,r,m)\cup I_R(i,r,m)$.
\end{itemize}
\eolem

\prf
We prove  all parts of the  lemma simultaneously by  induction on $m$.
The result is  immediate if $m=0$, since $i$ is active  at time 0, $i$
does not process  a message at time 0, and  $I_L(i,r,0) = I_R(i,r,0) =
\{i\}$.
Assume that parts  (a)--(e) hold for all time $m' <  m$.  We show that
they also hold at time $m$.
They clearly hold if $i$ does not process a message at time $m$, since
in   that  case   $I_L(i,r,m)  =   I_L(i,r,m-1)$  and   $I_R(i,r,m)  =
I_R(i,r,m-1)$.
So suppose  that $i$ processes  a message $x$  from its right  at time
$m$, and $x$ was sent by $R_i$ at time $m'$.  (The proof is similar if
$i$ receives from the left, and is left to the reader.)
If $x$ is the first message received by $i$ from the right, then it
follows from  Lemma~\ref{pro:msg order} that $i$ has  sent no messages
to  the right,  and $R_i$  has sent  only one  message to  $i$.  Thus,
$I_R(i,r,m-1)  = \{i\}$.  Parts  (a)--(e) now  follow easily  from the
induction hypothesis.

So suppose  that $x$ is  not the first  message that $i$  has received
from $R_i$.
We first  prove part (b).   Let $m_1$ be  the last time prior  to $m'$
that $R_i$ sent a message, say $x'$, to its left.
It easily follows from Lemma~\ref{pro:msg order} that there are
times $m_2$ and $m_3$, both  in the interval $(m_1,m')$, such that $i$
received $x'$  at time  $m_2$ and $R_i$  processed a message  from its
right at  $m_3$; moreover, $i$ did  not process any  messages from the
right between time $m_2$ and $m$.  By the induction hypothesis,
$I_R(i,r,m_2)=      \{       i      \}      \cup      I_R(R_i,r,m_1)$,
$I_L(i,r,m_2)=I_L(i,r,m_2-1)$,     and    $I_R(R_i,r,m_3+1)    \supset
I_R(R_i,r,m_1)$.    Since   $m_3+  1   \le   m'$,   it  follows   that
$I_R(R_i,r,m')  \supset I_R(R_i,r,m_1)$.  Since  $i$ does  not process
any messages from its right between time $m-2$ and $m$, by definition,
$I_R(i,r,m-1)=I_R(i,r,m_2)$.   It follows that  $I_R(R_i,r,m') \supset
I_R(i,r,m-1)$ and that
$$\begin{array}{ll} I_R(i,r,m)  = I_R(i,r,m-1) \union  I_R(R_i,r,m') =
\{i\}  \union  I_R(R_i,r,m_1) \union  I_R(R_i,r,m')\\  = \{i\}  \union
I_R(R_i,r,m')     \supset     \{i\}     \union    I_R(R_i,r,m_1)     =
I_R(i,r,m-1).\end{array}$$ This proves part  (b)(i) for time $m$.  For
part (ii), by definition, $I_L(i,r,m)=I_L(i,r,m-1)\cup I_L(R_i,r,m') -
\{R_i\}$.   By  the  induction  hypothesis,  it  easily  follows  that
$I_L(R_i,r,m') - \{R_i\} \subseteq I_L(i,r,m') \subseteq$
$I_L(i,r,m-1)$.  Thus, $I_L(i,r,m) = I_L(i,r,m-1)$.

Part  (a) for  time  $m$ follows  immediately  from part  (b) and  the
induction  hypothesis.  Part  (c)  is immediate,  since  $i$ does  not
process a message from the left at time $m$.

For the first  half of part (d), there are two  cases to consider.  If
$R_i$ was active  at the point $(r,m')$, then  the result is immediate
from  part  (e)  of  the  inductive  hypothesis.   Otherwise,  by  the
inductive hypothesis, $\val_R = \val_R(i,r,m) = \val_R(R_i,r,m')$.  By
the inductive  hypothesis, $\val_R$  is greater than  or equal  to the
maximum id  in $I_R(R_i,r,m')  - \{R_i\}$.  Since  the first  value of
$\val_R$  must  be $R_i$'s  id,  it  follows  that $$\max_{\{m'\le  m:
\val_R(i,r,m') \ne  \bot\}} \val_R(i,r,m')$$ is greater  than or equal
to the  maximum id  in $I_R(i,r,m) -  \{i\} =  \I_R(R_i,r,m')$.  Since
$\val_R(i,r,m')$  must be  an id  in $I_R(i,r,m)$,  we are  done.  The
second half  of part (d)  is immediate from the  induction hypothesis,
since $I_L(i,r,m)  = I_L(i,r,m-1)$ by  part (b), and  $\val_L(i,r,m) =
\val(i,r,m-1)$.

Finally, part (e) is immediate from the induction hypothesis if $i$ is
passive at time  $m-1$.  So suppose that $i$ is  active at time $m-1$.
By the induction hypothesis, $i$'s  id is the largest in $I_L(i,r,m-1)
\union  I_R(i,r,m-1)$.  If  $i$ is  active at  time $m$  then,  by the
description of  P2$'$, $i$'s id must be  greater than $\val_R(i,r,m)$.
Applying part (d) of the  induction hypothesis and the fact that $i$'s
id is  at least as  large as all  those in $I_R(i,r,m-1)$,  it follows
that  $i$'s   id  is  at  least   as  large  as   $\max_{\{m'  \le  m:
\val_R(i,r,m') \ne \bot\}} \val_R(i,r,m')$.  By part (d), at time $m$,
$i$'s  id is  at  least as  large  all those  in $I_R(i,r,m)$.   Since
$I_L(i,r,m) = I_L(i,r,m-1)$,  it follows that $i$'s id  is the maximum
id in $I_R(i,r,m) \union I_L(i,r,m)$.   Conversely, if $i$'s id is the
maximum id in $I_R(i,r,m) \union I_L(i,r,m)$, then by part (d) at time
$m$, $i$'s id  must be greater than $\val_R(i,r,m)$,  and hence by the
description of P2$'$, $i$ is active at $(r,m)$.  \eprf }

\commentout{ By Lemma~\ref{lem: i-segment},
we can prove the following:

\cor\label{cor:k-p2} For all  runs $r$ of P2$'$, times  $m$ and agents
$i$   in  $N_r$,  ${\cal   N}^b(i,r,m)={\cal  N}_{br,u}(I_R(i,r,m)\cup
I_L(i,r,m))$.  \ecor

With these notations, we can  prove the following result, mentioned in
the informal  description of P2$'$: \rem\label{rem:kall}  For all runs
$r$ of P2$'$, times $m$ and agents $i$ in $N_r$, $(\J, r, m, i)\models
\ainfo \Leftrightarrow K_I (\heardainfo)$ holds.  \erem }
\lem\label{lem:p2}  For   all  runs  $r$  of  P2$'$   in  the  context
$\gamma^{\mathit{br,u}}$,   times  $m$,  and   agents  $i$   in  $N_r$,
$P2'_i(r_i(m))={\Pgcb^{GC}}^{\J}_i(r_i(m))$.
\elem
\commentout{
that is, if $P2'_i(r_i(m)) 
\begin{itemize}
\item[(a)] $\varphi_{\send_{\lefta}(\newinfo)}$  holds whenever $i$ is
active, sends left, and does not know that it has all the information:
$$\begin{array}{ll} (\J,r,m,i)\models & (\mathit{status=active})\wedge
(LQ\neq \perp) \wedge (\mathit{wl}=1)  \wedge (\val_L<\id) \wedge \\ &
\neg \ainfo \quad \Rightarrow \varphi_{\send_{\lefta}(\newinfo)}.
\end{array}
$$
\item[(b)] $\varphi_{\send_{\righta}(\newinfo)}$ holds whenever $i$ is
active,  sends  right,  and  does   not  know  that  it  has  all  the
information:
$$
\begin{array}{ll}
(\J,r,m,i)\models  &(\mathit{status=active})\wedge   (RQ  \neq  \perp)
\wedge  (\mathit{wl}=0) \wedge  (\val_R<\id) \wedge  \\ &  \neg \ainfo
\quad \Rightarrow \varphi_{\send_{\righta}(\newinfo)}.
\end{array}
$$
\item[(c)]         $\varphi_{\send_{\righta}(\newinfo)}$        (resp.
$\varphi_{\send_{\lefta}(\newinfo)}$)  holds whenever $i$  is passive,
sends  left (resp.   right), and  does not  know that  it has  all the
information:
$$
\begin{array}{ll}
(\J,r,m,i)\models  &(\mathit{status=passive})  \wedge  (RQ\neq  \perp)
\wedge   (\mathit{wl}=0)   \wedge  \neg   \ainfo   \\  &   \Rightarrow
\varphi_{\send_{\lefta}(\newinfo)}\\                  (\J,r,m,i)\models
&(\mathit{status=passive})\wedge (LQ\neq \perp) \wedge (\mathit{wl}=1)
\wedge        \neg        \ainfo        \\        &        \Rightarrow
\varphi_{\send_{\righta}(\newinfo)}.
\end{array}
$$
\item[(d)] In all cases
where
$i$     knows     that     it    has     all     information,
$\varphi_{\send_{\lefta}(\newinfo)}$                             (resp.
$\varphi_{\send_{\righta}(\newinfo)}$)  holds if, according  to P2$'$,
$i$ sends left (resp. right) its new information.
\item[(e)]      $\varphi_{\send_{\lefta}(\newinfo)}$      (      resp.
$\varphi_{\send_{\righta}(\newinfo)}$) does not hold when $i$ does not
send  left (resp.   right):  for $\name  \in  \lbrace \lefta,  \righta
\rbrace$,  $(\J,r,m,i)\models (\neg \doact_i(\send_{\name}(\newinfo)))
\Rightarrow \neg \varphi_{\send_{\name}(\newinfo)}.$
\end{itemize}
\elem 
}

\prf  
As we observed above, we must show that for all $r \in
\Rrep(P2',\gamma^{\mathit{br,u}})$ and times $m$, we have that
$P2'_i(r_i(m))=\sfa$ iff $(\J,r,m,i)\models \varphi_{\sfa}$.
So suppose that $P2'_i(r_i(m))=\sfa$.
The relevant actions $\sfa$ have the form $\send_{\name}(\newinfo)$,
where $\name \in \{L,R\}$.  We consider the case that $\name = L$ here;
the proof for $\name = R$ is almost identical, and left to the reader.
The precondition of $\send_{L}(\newinfo)$ is
$$\neg B_I[\neg\doact(\send_{L}(\newinfo))\RCond  
\Diamond(\exists \name' (\Calls(L,I,\name') 
\land B_{\lefta} (\name'\mbox{'s} \inewinfo)) 
\lor \exists v   B_L(f=v))].$$
Since $R$ is the unique name that $i$'s left nieghbor calls $i$ in a
ring, we have that $(\J, r,  m, i)\models \Calls(L,I,R)$.
By    the    definitions   in
Section~\ref{sec:apx:cbb-progs},     $(\J,     r,     m,     i)\models
\varphi_{\send_L(\newinfo)}$ if and only if there exists a
situation $(r',m',i')$ such that
\begin{itemize} 
 \item[(a)] $r'_{i'}(m')=r_i(m)$,
\item[(b)] $\rkgen(P2')(r')=\mini^{\rkgen(P2')}(r,m)$, and
\item[(c)]  
$(\I,r',m',i')\models  \neg [\neg\doact(\send_{L}(\newinfo))\RCond  
\Diamond(\exists \name' (\Calls(L,I,\name') 
\land B_{\lefta} (\name'\mbox{'s} \inewinfo)) 
\lor \exists v   B_L(f=v))]$, so 
there    exists    a    situation    $(r'',m'',i'') \in 
\closest(\intension{\neg\doact(\send_{\name}$
$(\newinfo))}_{\Isys(P2',\chi^{\mathit{br,u}})},     $     $r',m',i')$
such that
$$(\J,r'',m'',i'')\models  \Box (\neg
B_{\lefta}(R\mbox{'s}~\inewinfo)\wedge      \neg      \exists      v.~
B_{\lefta}(f=v)).$$
\end{itemize} 
\commentout{
Condition  (b) says  that  $r'$ is  a  run of  P2$'$  in the  context
$\gamma^{\mathit{br,u}}$, as  $r$ is  a run of  P2$'$ and  $\rkgen$ is
deviation compatible.   As computing  $f$ is equivalent  
to
learning all the information,  and $\ogen$ respects protocols, to
prove (iii) it 
suffices     to     find     a    situation     $(r'',m'',i'')$     in
$\close(\overline{\doact(\send_{\name}(\newinfo))},               P2',$
$\gamma^{\mathit{br,u}},       $      $r',m',i')$       such      that
$(\J,r'',m'',i'')\models \Box (\neg B_{\lefta}(R\mbox{'s}~\inewinfo))$
holds.
}
Thus, we must show that there exists a situation $(r',m',i')$ satisfying
conditions (a), (b), and (c) above iff
$P2'_i(r_i(m))=\send_{\lefta}(\newinfo)$. 
To prove this, we need to consider the various cases where $i$
sends left.  
\begin{itemize}
\item Case 1: 
at $(r,m)$, $i$ is active, does not
know it has all the information,
and sends its first message at time $m$. 
In this case, 
we
can  take $r'$  to  be  a run  of  P2$'$  on  the network  $\left[
i\right]$
(i.e., the network where the only agent is $i$), 
 $m'=0$, and $i'=i$,  and take  $(r'',m'',i'')$  to be  
an arbitrary
situation  in $\close(\overline{\doact(\send_{\lefta}(\newinfo))},P2',
\gamma^{\mathit{br,u}},r',m',i')$
such that $|N_{r''}| > 1$. 
In $r''$, $L_{i''}$ does not receive a message from $i''$, so will never
process any message.  It easily follows that, in $r''$, $L_{i''}$ does not
learn the content $(i'')$'s initial information, nor does it learn who
the leader is. 
\item Case 2: $i$ is active, does not know all the information, and 
does not send its first message to the  left  at
time $m$.
In this case,
$L_i$ must be passive. 
Suppose that $i$ is about to send its $k$th message left at the point
$(r,m)$.  By Lemma~\ref{pro:msg order}, $i$ must have received
$k-1$ message from $L_{i}$, so $L_{i}$ must have processed $k-1$
messages from $i$.  Moreover, $i$ considers it possible that 
$L_i$ has already sent its $k$th message left, and is waiting to process
its $k$th message 
from $i$.  
Since $i$  does not have
all the information  at time $m$, 
it is easy to see that $i$ must also consider it possible that $L_i$
does not have all the information at time $m$.  Thus,
there   exists  a  
run $r'$ such that $r_i(m) = r'_{i}(m)$ and, at the
point
$(r',m)$, $L_{i}$ does not have all the information  and is waiting to
process the $k$th message from $i$. 
Let $(r'',m'',i'')$  be an arbitrary
situation  in $\close(\overline{\doact(\send_{\lefta}(\newinfo))},P2',
\gamma^{\mathit{br,u}},r',m,i)$.
Since $i''$ does not send left at $(r'',m'')$, $L_{i''}$ will wait
forever to process a message from $i''$.
Thus, in $r''$, $L_{i''}$ never 
learns the content of $(i'')$'s $k$th message, nor does it learn who the
leader is. 

\item Case 3: 
$i$  is passive at the point $(r,m)$ and does not have
all the  information.  Since $i$ is  about to
send left  and it is passive,  $i$ must have last  processed a message
from its right; without loss of generality, assume  that $i$ 
has processed $p$  messages from its right, 
and so must have processed  $(p-1)$
messages from its left by time $m$.
It easily follows from Lemma~\ref{pro:msg order} that $p > 1$.
Suppose that the $(p-1)$st message that $i$ processed from its left
originated with $k$.
Since $i$  does  not  have  all  the
information at time $m$, $k$ did  not have all the information when it
sent this message to the right.
After receiving its $(p-1)$st message from the left, $i$ must
consider it possible that the ring is sufficiently  large that, even
after $k$ processes its $(p-1)$st message from the left, $k$ will still
not know all the information.  That is, 
there     exists    a     situation    $(r',m',i')$     with    $r'\in
\Rrep(P2',\gamma^{\mathit{br,u}})$ such  that conditions (a)  and (b)
are satisfied, and  if 
$i'$'s $(p-1)$st message from the left in $r'$ originated with $k'$,
then $k'$ does not have all the information at the point $(r',m')$,
despite have processed its $(p-1)$st message from the left by this point.
Let     $(r'',m'',i'')$ be an arbitrary situation  in
$\close(\overline{\doact(\send_{\lefta}(\newinfo))},P2',
\gamma^{\mathit{br,u}},r',m',i')$.      
Suppose that 
$(i'')$'s 
$(p-1)$st message from the left in $r''$ originated
with $k''$.  At the point $(r'',m'')$, $k''$ has already processes its
$(p-1)$st message from the left and does not have all the information
(because this was the case for the agent $k'$ corresponding to $k''$ in
$r'$). 
In $r''$, all processes between $i''$ and $k''$ are passive.  Thus,
regardless of whether $k''$ is active or passive, in $r''$, 
$k''$ and  $i''$  and all agents between them are deadlocked, because
$k''$ is waiting from a message from the right, which must pass through
$i''$, and $i''$ is 
waiting for a message from its left, which must pass through $k''$.
It easily follows that $L_{i''}$ does not learn $(i'')$'s new
information in $r''$, nor does $L_{i''}$ learn who the leader is.

\commentout{ We prove (d) by  a case analysis. Since the arguments are
similar, we show here the proof  for case (B)(iv). Recall that this is
the case when
two
agents, the leader $i$ and agent $j$,
are active after processing from the left in phase $p-1$, and
exactly one agent $k$ is the  first to learn it has all the information in
run  $r$,  and $k\in  [j..   i]_L-\{ i,j\}$.   When  $k$  has all  the
information, it sends its new  information to both left and right. For
each agent  $l\in [k..  i]_L-\{i,k\}$,  $l$ is passive and  waiting to
process  from its right  in phase  $p$. By  induction on  the distance
between $l$ and  $k$ (to $l$'s right), we can show  that $l$ will have
all the information after processing from right in phase $p$, and that
$l$ will then send exactly one more message, and to its right. It will
also follow that $i$ will  have all the information after processing right
in phase $p$.

For each agent $l\in [k.. j]_R-\{j,k\}$, $l$ is passive and waiting to
receive from its left when $k$  has all the information.  We can then show
that $l$ will have all  the information after processing from its left
in phase $p$, and will send  one more message, to its right.  In phase
$p$, $j$  processes from its right  a message that  originates at $i$;
since  $i$ is  the leader,  it means  that $j$  becomes  passive after
processing this message. $j$ is  then waiting to process from its left
in phase $p$.   This ensures that $j$ will  indeed process the message
from  its   left  in  phase   $p$,  and  subsequently  have   all  the
information. $j$ will then send right its new information, and stop.

All the agents  $l\in [j.. i]_R-\{i,j\}$ are passive  at the beginning
of phase $p$; they forward left in phase $p$ the message originated at
$i$,  towards  $j$, and  then  wait to  process  from  left. A  simple
inductive argument can be carried out to show that all agents $l$ will
have  all the  information  after  processing  from their  left  in  phase
$p$. With this, we have shown that all agents in $N_r$ eventually have
all the information.

Furthermore, since $L_i$ does not  send right its new information when
it has all  the information.  Similar reasoning for  all cases (A)-(F)
shows that  there is no redundant  communication in the  last phase of
P2$'$.  }
\item Case 4: $i$ has all the information at time $m$ in $r$.  There are
a number of subcases to consider.
We focus on one of them here, 
where  two agents, the
leader $i^*$  and $i$, are the  first to learn all  the information;
the arguments for the other cases are similar in spirit, 
and left to the reader.
We have shown that, in this  case, $i$  turns  passive when  it  learns
all  the 
information as a result of processing a  message $\msg$ that originated
with 
$i^*$, and 
that the number  of messages $i^*$ and $i$ have  processed by the time
they learn  all the information is  the same. Without loss  of generality,
assume that both $i^*$ and $i$ first learned all the information after
processing their $p$th  message from the left. 
We showed that either the  $p$th message  that  $i^*$
processed  from its  left originated with $i$, or it originated with
some agent $i'$ whose $p$th message from the left originated with  $i$.
It is easy to see that all agents other than $i^*$ and $i$ are passive
after 
they 
process their $p$th message, do not have all the information,
and are waiting to receive a message from the right.
Thus, if $i$ does not send left, then all agents to the left of $i$ up
to but not including $i^*$ are deadlocked.  Since $i$ is supposed to
send left, it cannot be the case that $L_i = i^*$.  It easily follows
that if $i$ does not send left, and $(r',m,i')$ is an arbitrary
situation  in $\close(\overline{\doact(\send_{\lefta}(\newinfo))},P2',
\gamma^{\mathit{br,u}},r,m,i)$, then $L_{i'}$ does not learn 
$(i')$'s 
new
information nor who the leader is in $r'$.
\end{itemize}
\commentout{
We  have  thus  proved  that  we can  find  a  situation  $(r',m',i')$
satisfying            conditions            (a)-(c)           whenever
$P2'_i(r_i(m))=\send_{\lefta}(\newinfo)$.   It  is  not  difficult  to
notice that no such situation exists when $i$ does {\em not} send left
its new  information: Let $(r',m',i')$ be an  arbitrary situation with
$r'\in \Rrep(P2', \gamma^{\mathit{br,u}})$ that $i$ considers possible
at  time $m$  in  run $r$  (i.e.,  such that  conditions  (a) and  (b)
hold). Then at time $m'$ in  run $r'$ agent $i'$ is executing the same
action as agent $i$ at time $m$ in $r$, and so $i'$ does not send left
at      time     $m'$      in     $r'$.       In      other     words,
$\close(\overline{\doact(\send_{\name}(\newinfo))},P2',
\gamma^{\mathit{br,u}},r',m',i')=(r',m',i')$ holds. Since in any run of $P2$'
eventually all agents learn who the leader is, and $r'$ is a run of $P2$', it means that
in run $r'$ agent $L_{i'}$ eventually learns who the leader is, and so
$(\J, r', m',i')\not \models \varphi_{\send_{\lefta}(\newinfo)}$ holds.
\commentout{
We now prove (e). Similar to the argument above we can show that there
is no  redundant communication in the  last phase of  P2$'$.  (For the
case detailed here,  notice that $i^*$ and $i$ do  not process any new
message after learning all the information.)  Assume now that $i$ does
not have  all the information at time  $m$ in $r$.  If  $(r',m',i')$ is an
arbitrary situation  that $i$  considers possible at  time $m$  in run
$r$, then  $i'$ is executing the same  action at time $m'$  in $r'$ as
$i$ in  $r'$.  If $i$  does not send  its new information  to neighbor
$\name$ at time  $m$ in $r$, then  so does $i'$ at time  $m'$ in $r'$.
Then
$\close(\overline{\doact(\send_{\name}(\newinfo))},P2',\gamma^{\mathit{br,u}},r',m',i')=(r',m',i')$.
By Corollary~\ref{cor:p2},  some agent  will eventually learn  all the
information in run $r'$. The  argument above for case (d) proved that,
subsequently, all agents will learn all the information in run $r'$.
It follows that $\name$ will  indeed have all the information in $r'$.
}
This observation completes our proof.  \eprf
}
We have shown that, for all $r \in
\Rrep(P2',\gamma^{\mathit{br,u}})$ and times $m$, if 
$P2'_i(r_i(m))=\sfa$ then $(\J,r,m,i)\models \varphi_{\sfa}$.
For the converse, suppose that $P2'_i(r_i(m))\ne \sfa$.
Again, suppose that $\sfa$ is $\send_{L}(\newinfo)$.
Let $(r',m',i')$ be a situation that $i$ considers possible
at  time $m$  in  run $r$  (i.e.,  such that  conditions  (a) and  (b)
above hold).  Since $i$ does not send left at the point $(r,m)$, $i'$
does not send left at the point $(r',m')$.  Thus, by definition,
$\close(\overline{\doact(\send_{\name}(\newinfo))},P2',
\gamma^{\mathit{br,u}},r',m',i')=\{(r',m',i')\}$.  Since $r'$ is a run
of P2$'$, and every agent eventually learns who the leader is in every
run of P2$'$, it follows that $(\J, r', m',i') \models \Diamond
B_L(f=v)$, and hence $$(\J,r,m,i)\models \neg
\doact_i(\send_{\name}(\newinfo)) 
\RCond   \Diamond(\exists \name' (\Calls(L,I,\name') 
\land B_{\lefta} (\name'\mbox{'s} \inewinfo)) 
\lor \exists v   B_L(f=v)).$$
Thus, $(\J,r,m,i) \sat \neg \varphi_{\send_{L}(\newinfo)}$.
This completes the proof.
\eprf 

\commentout{
\olem{lem:p2''}
P2$''$ de facto implements $\Pgcb^{GC}$ is all contexts where (i) all networks are
bidirectional rings, and (ii) agents have distinct identifiers.
\eolem

\prf
Let $\ogen$ be an order generator that respects protocols, and $\rkgen$ be a 
deviation compatible ranking generator, and let $\chi^{\mathit{br,u}}=(\gamma^{\mathit{br,u}},\pi,\ogen,\rkgen)$
be the corresponding counterfactual interpreted system.
Let $\J=(\Rrep(\gamma^{\mathit{br,u}}),\pi,\mu_{\gamma^{\mathit{br,u}}},\ogen(P2''),\rkgen(P2''))$.
Let $r$ be an arbitrary run of P2$''$ in $\gamma^{\mathit{br,u}}$, and $i$ an agent in 
$N_r$. 

Assume that $i$ knows that it has has all the information by time 
$m$ in $r$: $$(\J,r,m,i)\models \ainfo.$$ Let $\name$ be an arbitrary neighbor of $i$.
We proceed with a case analysis depending on whether $i$ has received
all the information directly  
from  $\name$ or not, and whether $\name$ is waiting to process from $i$
in P2Stop. 
\begin{itemize}
\item[(a)] Consider the case when $\name$ is  waiting to process from $i$ in P2Stop. i.e.,
$$(\J,r,m,i)\models \ainfo \wedge \mathit{waits\_for\_me}_{\name} \wedge 
\neg \mathit{all\_info\_from}_{\name}$$
is true.
{F}rom the description of P2$''$, it follows that  $i$ sends its new information to $\name$
at time $m$ in $r$. We now show that this is in accordance with $\Pgcb^{GC}$, i.e.,
the following holds:
$$(\J,r,m,i)\models \varphi_{\send_{\name}(\newinfo)}.$$
That is, we will  show that $(\J,r,m,i)\models \neg B_I (\neg \doact(\send_{\name}(\newinfo))
\RCond \Diamond \exists v.~B_n(f=v))$.
Let $(r',m,i)$ be a situation in 
$\closest(\intension{\neg \doact(\send_{\name}(\newinfo))}_{\Isys(P2'',\chi^{\mathit{br,u}})},
r,m,i)$. Since $\ogen$ respects protocols, and $i$ sends its new information to $\name$
at time $m$ in $r$, it follows that $r'$ is just like $r$, except that $i$ does {\em not} send
its new information to $\name$ at time $m$. 
Since $i$ knows that it has all the information at time $m$ in $r$,
and $r'_i(m)=r_i(m)$, it follows that $i$ knows it has all the information at time $m$ in $r'$, as
well. It is not difficult to show now that, since $\name$ is waiting for $i$ in P2Stop on 
$N_r$, it is also waiting for $i$ in P2Stop on $N_{r'}$, i.e., 
if $j'$ is the agent $i$ calls $\name$ in $r'$ 
(i.e., $j'=\mu_{\gamma^{\mathit{br,u}}}(r'(m),i,\name)$), then $j' \not \in K(N_{r'})$
and $W(j',N_{r'})=i$. However, in $r'$, since $i$ does not send its new information to $j$
at time $m$, $j'$ will wait forever. This, together with the fact that $j'$ does not have all
the information in P2Stop on $N_{r'}$, ensures that $j'$ will never have all the information
in P2$''$ on $N_{r'}$. More precisely, 
$(\J,r',m,i)\models \neg \Diamond \exists v.~B_{\name}(f=v)$ is true.
\item[(b)] We analyze now the case when $i$ has received all the information directly from $\name$.
Then clearly $i$ knows that $\name$ already has all the information, and so
$(\J,r,m,i)\models B_I(\exists v.~B_{\name}(f=v))$ is true. It is now straightforward to see that
$(\J,r,m,i)\models \neg \varphi_{\name}(\newinfo)$ also holds, and so $i$ acts in accordance with 
$\Pgcb^{GC}$.
\item[(c)] The interesting case is when $i$ has not received all the information from $i$, and 
$\name$  is {\em not} waiting from $i$, i.e., if $j=\mu_{\gamma^{\mathit{br,u}}}(r,m,i,\name)$, 
then $j\not \in K(N_r)$ and $W(j,N_r)\neq i$.
We now show that $(\J,r,m,i)\models \neg \varphi_{\send_{\name}(\newinfo)}$ is true.
More specifically, we show that the following holds:
$(\J,r,m,i)\models B_I((\neg \doact(\send_{\name}(\newinfo)))\RCond \Diamond \exists v.~B_{\name}(f=v))$.
Since $\rkgen$ is deviation compatible, this amounts to showing that, in any run $r'$ of P2$''$, for any time 
$m'$ and agent $i'$ such that $r'_{i'}(m')=r_i(m)$, 
$(\J,r',m',i')\models (\neg \doact(\send_{\name}(\newinfo)))\RCond \Diamond \exists v.~B_{\name}(f=v) $  is true.
Since $i$ does not send its new information to $\name$ at time $m$ in $r$, $i'$ does not send it either 
(at time $m'$ and $r'$). Intuitively, this means that we are not in a counterfactual situation here; more
formally, we can show that we are left with proving that indeed $\name$ will have all the information
in run $r'$, i.e., $(\J,r,'m,',i')\models \Diamond \exists v.~B_{\name}(f=v)$ is true.

It is not difficult to see that, since $r'_{i}(m')=r_i(m)$, $i$ knows that it has all the information at time
$m$ in $r$; since  $\name$ is not waiting for $i$ in P2Stop on $N_r$, it follows that
$\name$ is not waiting for $i'$ in 
P2Stop on $N_{r'}$ either, i.e., $W(j',N_{r'})\neq i'$. We now show that $j'$ will have all the information
from its  neighbor other than $i'$.
Essential in the proof is the fact that no two agents are waiting for each other in P2Stop:
\lem\label{lem:p2stop}
For any bidirectional ring $N$ with unique identifiers, for any two neighbors $i$ and $j$ in $N$ such that
$i,j\not \in K(N)$, if $W(i,N)=j$, then $W(j,N)\neq i$.
\elem
\prf
We will do the proof for the case when $j=R_i$, as the proof for the case $j=L_i$ is similar.
Let $r$ be an arbitrary run of P2Stop on the ring $N$, and let $m$ be the last time $i$ processes a message 
in $r$. Since $W(i,N)=R_i$, it must be that $i$ processes a message from its left at time $m$. Let $p$ be the phase
$i$ is executing at time $m$. The following cases are possible:
\begin{itemize}
\item[(i)] $i$ is active after processing from left in phase $p$:
Prior to $m$, $i$ has last sent a message to its right in phase $p$. Since $i$ is active when it sends right 
in phase $p$, it follows that $R_i$ is passive after processing left in phase $p$. Then $R_i$ is waiting to process
from right, not from its left, i.e, $W(R_i,N)\neq i$. 
\item[(ii)] $i$ becomes passive after processing from left in phase $p$: 
As in case (i), prior to time $m$, $i$ has last sent right in phase $p$, and $R_i$ must be passive after processing 
this message. Again, $R_i$ is waiting to process from its right.
\item[(iii)] $i$ was passive when it processed from left in phase $p$:
In this case, prior to $m$, $i$ has last sent to its right in phase $p-1$. $R_i$ cannot be  active after processing 
this message, since otherwise  $R_i$ would send left in phase $p$, and so $i$ would process one more message 
after time $m$, which contradicts our assumptions on $m$. It follows that $R_i$ is passive after processing from 
its left in phase $p-1$. If $R_i$ has just became passive, then $R_i$ is next waiting to process from 
its right; same is true if $R_i$ was already passive when it processed from its left in phase $p-1$.
\end{itemize}
In all the above cases, $W(R_i,N)\neq i$. 
\eprf

We now show that, for any value $d\ge 0$, for any agent $j'$ in $N_{r'}$  such that the distance between
$j'$ and the closest agent in $K(N_{r'})$ (in the direction $j'$ is waiting from) is $d$, eventually
$j'$ will have all the information in $r'$. In the following, let $k'$ be the closest agent to $j'$ in $K(N_{r'})$,
in the direction $j'$ is waiting from. 
The base case is when $d=1$.  Then, by the description of P2$''$, when
$k'$ has all the information  in $r'$
(which is bound to happen, since $k'$ is an agent in $ K(N_{r'})$), it will know that $j'$ is waiting to receive from
$k'$, and so $k'$ will send its new information to $j'$.
Assume now that for any agent $j'$ such that the distance between $j'$ and its corresponding $k'$ is at most $d(>1)$,
it is the case that eventually $j'$ will have all the information in $r'$. Let $j'$ be now an agent in $N_{r'}$
such that the distance between $j'$ and its corresponding $k'$ is $d+1$. Let $v'$ be the neighbor 
$j'$ is waiting from in P2Stop, i.e., $W(j',N_{r'})=v'$. By the definition of $k'$, $v'\not \in K(N_{r'})$, and 
by Lemma~\ref{lem:p2stop}, $v'$ is not waiting from $j'$. Then the closest agent to $v'$ in $K(N_{r'})$  is $k'$, too,
but now the distance between $v'$ and $k'$ is $d$. By our induction hypothesis, it follows that eventually
$v'$ will have all the information in $r'$. When this happens, by the description of P2$''$, $v'$ will know that
$j'$ is waiting from it, and so will send its new information to $j'$.
\end{itemize}

Finally, consider the case when $i$ does not know that it has all the information at time, i.e.,
$(\J,r,m,i)\models \neg \ainfo$ is true. It is then clear from the description of P2$'$ and P2$''$ that 
$i$ performs the same action in P2$''$ as if it were following P2$'$. 
Essentially the same proof as for Lemma~\ref{lem:p2} can be carried out to show that 
$P2''_i(r_i(m))={\Pgcb^{GC}}^{\J}_i(r_i(m))$. 
With this, our proof is complete. 
\eprf
}

\bibliography{z,joe}
\bibliographystyle{chicagor}

\end{document}

%% file: defn.tex
\newtheorem{THEOREM}{Theorem}[section]
\newenvironment{theorem}{\begin{THEOREM} \hspace{-.85em} {\bf :} }%
                        {\end{THEOREM}}
\newtheorem{LEMMA}[THEOREM]{Lemma}
\newenvironment{lemma}{\begin{LEMMA} \hspace{-.85em} {\bf :} }%
                      {\end{LEMMA}}
\newtheorem{COROLLARY}[THEOREM]{Corollary}
\newenvironment{corollary}{\begin{COROLLARY} \hspace{-.85em} {\bf :} }%
                          {\end{COROLLARY}}
\newtheorem{PROPOSITION}[THEOREM]{Proposition}
\newenvironment{proposition}{\begin{PROPOSITION} \hspace{-.85em} {\bf :} }%
                            {\end{PROPOSITION}}
\newtheorem{DEFINITION}[THEOREM]{Definition}
\newenvironment{definition}{\begin{DEFINITION} \hspace{-.85em} {\bf :} \rm}%
                            {\end{DEFINITION}}
\newtheorem{CLAIM}[THEOREM]{Claim}
\newenvironment{claim}{\begin{CLAIM} \hspace{-.85em} {\bf :} \rm}%
                            {\end{CLAIM}}
\newtheorem{EXAMPLE}[THEOREM]{Example}
\newenvironment{example}{\begin{EXAMPLE} \hspace{-.85em} {\bf :} \rm}%
                            {\end{EXAMPLE}}
\newtheorem{REMARK}[THEOREM]{Remark}
\newenvironment{remark}{\begin{REMARK} \hspace{-.85em} {\bf :} \rm}%
                            {\end{REMARK}}

\newcommand{\thm}{\begin{theorem}}
\newcommand{\lem}{\begin{lemma}}
\newcommand{\pro}{\begin{proposition}}
\newcommand{\dfn}{\begin{definition}}
\newcommand{\rem}{\begin{remark}}
\newcommand{\xam}{\begin{example}}
\newcommand{\cor}{\begin{corollary}}
\newcommand{\prf}{\noindent{\bf Proof:} }
\newcommand{\ethm}{\end{theorem}}
\newcommand{\elem}{\end{lemma}}
\newcommand{\epro}{\end{proposition}}
\newcommand{\edfn}{\bbox\end{definition}}
\newcommand{\erem}{\bbox\end{remark}}
\newcommand{\exam}{\bbox\end{example}}
\newcommand{\ecor}{\end{corollary}}
\newcommand{\eprf}{\bbox\vspace{0.1in}}
\newcommand{\beqn}{\begin{equation}}
\newcommand{\eeqn}{\end{equation}}

\newcommand{\bbox}{\vrule height7pt width4pt depth1pt}

\newcommand{\clm}{\begin{claim}}
\newcommand{\eclm}{\end{claim}}

\newcommand{\sat}{\models}

\newcommand{\rimp}{\Rightarrow}

\newcommand{\union}{\cup}
\newcommand{\inter}{\cap}

\newcommand{\IN}{\mbox{$I\!\!N$}}

\renewcommand{\phi}{\varphi}

\newcommand{\A}{{\cal A}}

\newcommand{\G}{{\cal G}}
\newcommand{\I}{{\cal I}}
\newcommand{\J}{{\cal J}}

\newcommand{\N}{{\cal N}}

\newcommand{\R}{{\cal R}}

\renewcommand{\>}{\rangle}

\newcommand{\ol}{\setlength{\itemsep}{0pt}\begin{enumerate}}
\newcommand{\eol}{\end{enumerate}\setlength{\itemsep}{-\parsep}}
\newcommand{\ul}{\setlength{\itemsep}{0pt}\begin{itemize}}
\newcommand{\dl}{\setlength{\itemsep}{0pt}\begin{description}}
\newcommand{\edl}{\end{description}\setlength{\itemsep}{-\parsep}}
\newcommand{\eul}{\end{itemize}\setlength{\itemsep}{-\parsep}}

\newcommand{\cI}{{\cal I}}

\newcommand{\cR}{{\cal R}}

\newcommand{\true}{\mbox{{\it true}}}
\newcommand{\false}{\mbox{{\it false}}}

\newcommand{\commentout}[1]{}

\newcommand{\bi}{\begin{itemize}}
\newcommand{\ei}{\end{itemize}}
\newcommand{\be}{\begin{enumerate}}
\newcommand{\ee}{\end{enumerate}}

\newcommand{\Gz}{\G_0}

%% file: disc06corr.bbl
\begin{thebibliography}{}

\bibitem[\protect\citeauthoryear{Angluin}{Angluin}{1980}]{angluin80}
Angluin, D. (1980).
\newblock Local and global properties in netwroks of processors.
\newblock In {\em Proc.~12th ACM Symp.~on Theory of Computing}, pp.\  82--93.

\bibitem[\protect\citeauthoryear{Attyia, Gorbach, and Moran}{Attyia
  et~al.}{2002}]{AGM02}
Attyia, H., A.~Gorbach, and S.~Moran (2002).
\newblock Computing in totally anonymous asynchronous shared memory systems.
\newblock {\em Information and Computation\/}~{\em 173(2)}, 162--183.

\bibitem[\protect\citeauthoryear{Attyia, Snir, and Warmuth}{Attyia
  et~al.}{1988}]{attiya88}
Attyia, H., M.~Snir, and M.~K. Warmuth (1988).
\newblock Computing on an anonymous ring.
\newblock {\em Journal of ACM\/}~{\em 35(4)}, 845--875.

\bibitem[\protect\citeauthoryear{Bellman}{Bellman}{1958}]{bellman58}
Bellman, R. (1958).
\newblock On a routing problem.
\newblock {\em Quarterly of Applied Mathematics\/}~{\em 16\/}(1), 87--90.

\bibitem[\protect\citeauthoryear{Bickford, Constable, Halpern, and
  Petride}{Bickford et~al.}{2005}]{BCHP05}
Bickford, M., R.~L. Constable, J.~Y. Halpern, and S.~Petride (2005).
\newblock Knowledge-based synthesis of distributed systems using event
  structures.
\newblock In {\em Proc.~11th Int.~Conf.~on Logic for Programming, Artificial
  Intelligence, and Reasoning (LPAR 2004)}, Lecture Notes in Computer Science,
  vol.~3452, pp.\  449--465. Springer-Verlag.

\bibitem[\protect\citeauthoryear{Chang and Roberts}{Chang and
  Roberts}{1979}]{CR79}
Chang, E. and R.~Roberts (1979).
\newblock An improved algorithm for decentralized extrema-finding in circular
  configurations of processes.
\newblock {\em Communications of the ACM\/}~{\em 22(5)}, 281--283.

\bibitem[\protect\citeauthoryear{Dwork and Moses}{Dwork and Moses}{1990}]{DM}
Dwork, C. and Y.~Moses (1990).
\newblock Knowledge and common knowledge in a {B}yzantine environment: crash
  failures.
\newblock {\em Information and Computation\/}~{\em 88\/}(2), 156--186.

\bibitem[\protect\citeauthoryear{Fagin, Halpern, Moses, and Vardi}{Fagin
  et~al.}{1995}]{FHMV}
Fagin, R., J.~Y. Halpern, Y.~Moses, and M.~Y. Vardi (1995).
\newblock {\em Reasoning about Knowledge}.
\newblock \chicagoraddresspub{Cambridge, Mass.: }MIT Press.
\newblock A revised paperback edition was published in 2003.

\bibitem[\protect\citeauthoryear{Fagin, Halpern, Moses, and Vardi}{Fagin
  et~al.}{1997}]{FHMV94}
Fagin, R., J.~Y. Halpern, Y.~Moses, and M.~Y. Vardi (1997).
\newblock Knowledge-based programs.
\newblock {\em Distributed Computing\/}~{\em 10\/}(4), 199--225.

\bibitem[\protect\citeauthoryear{Ford and Fulkerson}{Ford and
  Fulkerson}{1962}]{ford62}
Ford, L.~R. and D.~R. Fulkerson (1962).
\newblock {\em Flows in Networks}.
\newblock \chicagoraddresspub{Princeton, N. J.: }Princeton University Press.

\bibitem[\protect\citeauthoryear{Friedman and Halpern}{Friedman and
  Halpern}{1997}]{FrH1Full}
Friedman, N. and J.~Y. Halpern (1997).
\newblock Modeling belief in dynamic systems. {P}art {I}: foundations.
\newblock {\em Artificial Intelligence\/}~{\em 95\/}(2), 257--316.

\bibitem[\protect\citeauthoryear{Gallager, Humblet, and Spira}{Gallager
  et~al.}{1983}]{gallager83}
Gallager, R.~G., P.~A. Humblet, and P.~M. Spira (1983).
\newblock A distributed algorithm for minimum-weight spanning trees.
\newblock {\em ACM Trans. on Programming Languages and Systems\/}~{\em 5\/}(1),
  66--77.

\bibitem[\protect\citeauthoryear{Grove}{Grove}{1995}]{Grove95}
Grove, A.~J. (1995).
\newblock Naming and identity in epistemic logic {II}: a first-order logic for
  naming.
\newblock {\em Artificial Intelligence\/}~{\em 74\/}(2), 311--350.

\bibitem[\protect\citeauthoryear{Grove and Halpern}{Grove and
  Halpern}{1993}]{GroveH2}
Grove, A.~J. and J.~Y. Halpern (1993).
\newblock Naming and identity in epistemic logics, {P}art {I}: the
  propositional case.
\newblock {\em Journal of Logic and Computation\/}~{\em 3\/}(4), 345--378.

\bibitem[\protect\citeauthoryear{Hadzilacos}{Hadzilacos}{1987}]{Had}
Hadzilacos, V. (1987).
\newblock A knowledge-theoretic analysis of atomic commitment protocols.
\newblock In {\em Proc.~6th ACM Symp.~on Principles of Database Systems}, pp.\
  129--134.

\bibitem[\protect\citeauthoryear{Halpern and Moses}{Halpern and
  Moses}{2004}]{HM98a}
Halpern, J.~Y. and Y.~Moses (2004).
\newblock Using counterfactuals in knowledge-based programming.
\newblock {\em Distributed Computing\/}~{\em 17\/}(2), 91--106.

\bibitem[\protect\citeauthoryear{Halpern, Moses, and Waarts}{Halpern
  et~al.}{2001}]{HMW}
Halpern, J.~Y., Y.~Moses, and O.~Waarts (2001).
\newblock A characterization of eventual {B}yzantine agreement.
\newblock {\em SIAM Journal on Computing\/}~{\em 31\/}(3), 838--865.

\bibitem[\protect\citeauthoryear{Halpern and Zuck}{Halpern and Zuck}{1992}]{HZ}
Halpern, J.~Y. and L.~D. Zuck (1992).
\newblock A little knowledge goes a long way: knowledge-based derivations and
  correctness proofs for a family of protocols.
\newblock {\em Journal of the ACM\/}~{\em 39\/}(3), 449--478.

\bibitem[\protect\citeauthoryear{Johnson and Schneider}{Johnson and
  Schneider}{1985}]{johnson85}
Johnson, R.~E. and F.~B. Schneider (1985).
\newblock Symmetry and similarity in distributed systems.
\newblock In {\em Proc.~4th ACM Symp.~on Principles of Distributed Computing},
  pp.\  13--22.

\bibitem[\protect\citeauthoryear{{Le Lann}}{{Le Lann}}{1977}]{L77}
{Le Lann}, G. (1977).
\newblock Distributed systems--towards a formal approach.
\newblock In {\em IFIP Congress}, Volume~7, pp.\  155--160.

\bibitem[\protect\citeauthoryear{Lewis}{Lewis}{1973}]{Lewis73}
Lewis, D.~K. (1973).
\newblock {\em Counterfactuals}.
\newblock \chicagoraddresspub{Cambridge, Mass.: }Harvard University Press.

\bibitem[\protect\citeauthoryear{Lynch}{Lynch}{1997}]{Lyn97}
Lynch, N. (1997).
\newblock {\em Distributed Algorithms}.
\newblock \chicagoraddresspub{San Francisco: }Morgan Kaufmann.

\bibitem[\protect\citeauthoryear{Mazer and Lochovsky}{Mazer and
  Lochovsky}{1990}]{Maz}
Mazer, M.~S. and F.~H. Lochovsky (1990).
\newblock Analyzing distributed commitment by reasoning about knowledge.
\newblock Technical Report CRL 90/10, DEC-CRL.

\bibitem[\protect\citeauthoryear{Milner}{Milner}{1989}]{milner89}
Milner, R. (1989).
\newblock {\em Communication and Concurrency}.
\newblock \chicagoraddresspub{Hertfordshire: }Prentice Hall.

\bibitem[\protect\citeauthoryear{Moses and Roth}{Moses and
  Roth}{1989}]{MosesRoth}
Moses, Y. and G.~Roth (1989).
\newblock On reliable message diffusion.
\newblock In {\em Proc.~8th ACM Symp.~on Principles of Distributed Computing},
  pp.\  119--128.

\bibitem[\protect\citeauthoryear{Peterson}{Peterson}{1982}]{peterson82}
Peterson, G.~L. (1982).
\newblock An ${O}(n\log{n})$ unidirectional distributed algorithm for the
  circular extrema problem.
\newblock {\em ACM Trans. on Programming Languages and Systems\/}~{\em 4\/}(4),
  758--762.

\bibitem[\protect\citeauthoryear{Stalnaker}{Stalnaker}{1968}]{Stalnaker68}
Stalnaker, R.~C. (1968).
\newblock A semantic analysis of conditional logic.
\newblock In N.~Rescher (Ed.), {\em Studies in Logical Theory}, pp.\  98--112.
  Oxford University Press.

\bibitem[\protect\citeauthoryear{Stulp and Verbrugge}{Stulp and
  Verbrugge}{2002}]{SV02}
Stulp, F. and R.~Verbrugge (2002).
\newblock A knowledge-based algorithm for the {I}nternet protocol ({TCP}).
\newblock {\em Bulletin of Economic Research\/}~{\em 54\/}(1), 69--94.

\bibitem[\protect\citeauthoryear{Yamashita and Kameda}{Yamashita and
  Kameda}{1996}]{YK96a}
Yamashita, M. and T.~Kameda (1996).
\newblock Computing on anonymous networks. {I}. {C}haracterizing the solvable
  cases.
\newblock {\em IEEE Trans.~on Parallel and Distributed Systems\/}~{\em 7\/}(1),
  69--89.

\bibitem[\protect\citeauthoryear{Yamashita and Kameda}{Yamashita and
  Kameda}{1999}]{YK99}
Yamashita, M. and T.~Kameda (1999).
\newblock Leader election problem on networks in which processor identity
  numbers are not distinct.
\newblock {\em IEEE Trans.~on Parallel and Distributed Systems\/}~{\em
  10\/}(9), 878--887.

\end{thebibliography}
